\documentclass[authoryear,3p,times,11pt]{elsarticle}
\journal{}

\usepackage{amssymb}
\usepackage{amsmath}
\usepackage{tikz}
\usepackage{amsthm}
\usepackage{bm}
\usepackage{booktabs}

\makeatletter
\def\appendixname{Appendix}
\renewcommand\appendix{\par
  \setcounter{section}{0}%
  \setcounter{subsection}{0}%
  \setcounter{equation}{0}
  \setcounter{figure}{0}
  \gdef\thefigure{\@Alph\c@section.\arabic{figure}}%
  \gdef\thetable{\@Alph\c@section.\arabic{table}}%
  \gdef\thesection{\appendixname~\@Alph\c@section}%
  \gdef\thesubsection{\@Alph{\c@section}.\arabic{subsection}}%
  \gdef\theequation{\@Alph{\c@section}.\arabic{equation}}%
  \addtocontents{toc}{\string\let\string\numberline\string\tmptocnumberline}{}{}
}
\makeatother
\begin{document}
\begin{frontmatter}
\title{Robust estimation of precision matrices under cellwise contamination}
\author{G.\ Tarr\corref{cor1}}
\ead{gtar4178@uni.sydney.edu.au}
\author{S.\ M\"uller\corref{cor3}}
\author{N.\ C.\ Weber\corref{cor2}}
\cortext[cor1]{Corresponding author}
\address{School of Mathematics and Statistics, University of Sydney, NSW 2006, Australia}

\begin{abstract}
There is a great need for robust techniques in data mining and machine learning contexts where many standard techniques such as principal component analysis and linear discriminant analysis are inherently susceptible to outliers.  Furthermore, standard robust procedures assume that less than half the observation rows of a data matrix are contaminated, which may not be a realistic assumption when the number of observed features is large.  This work looks at the problem of estimating covariance and precision matrices under cellwise contamination.  We consider using a robust pairwise covariance matrix as an input to various regularisation routines, such as the graphical lasso, QUIC and CLIME.  To ensure the input covariance matrix is positive semidefinite, we use a method that transforms a symmetric matrix of pairwise covariances to the nearest covariance matrix.  The result is a potentially sparse precision matrix that is resilient to moderate levels of cellwise contamination.  Since this procedure is not based on subsampling it scales well as the number of variables increases. 
\end{abstract}


\end{frontmatter}

\section{Introduction}

Often the aim of data mining and statistics is to extract information about the relationships between the variables and identify any features or structure in the data. The covariance matrix, $\bm{\Sigma} = \operatorname{var}(\mathbf{y})$, where $\mathbf{y}\sim\mathbf{F}$, the distribution of the true data generating process, and its inverse, the precision matrix $\bm{\Theta} = \bm{\Sigma}^{-1}$ are fundamental components of many statistical procedures, such as principal component analysis (PCA) and linear discriminant analysis.  However, it is well known that the classical covariance matrix is inherently non-robust to outliers and suffers from distortion in its eigenstructure in high dimensions \citep{Johnstone:2001}.  This paper combines pairwise covariance matrix estimation with recent regularisation routines currently used in bioinformatics and machine learning to produce an estimated precision matrix that is robust to moderate levels of cellwise contamination.

The need for robust statistics in data mining and associated fields is well known, see \citet{Barnett:1994} for a general overview.  In particular, it is desirable for learning algorithms to be stable with respect to noisy features and unusual fluctuations in the inputs.  For example \citet{Li:2004} considers robust incremental PCA applied to multi-view face modelling and \citet{Mavroeidis:2014} consider the stability of sparse PCA in the context of feature selection in microarray gene expression data.  Other situations where robust techniques are important include speech recognition and neural networks, see \citet{Gales:2007} and \citet{Bieroza:2011}, respectively. 

In the statistics literature, robust estimation of covariance matrices has received much attention in the past, notably the minimum volume ellipsoid and minimum covariance determinant (MCD) estimators, projection type estimators and M-estimators, see \citet{Hubert:2008} for a survey. Furthermore, research into covariance matrix estimation and its applications is ongoing, see for example \citet{Filzmoser:2013} who use the MCD estimator to construct robust Mahalanobis distances to identify local multivariate outliers; \citet{Hubert:2013} who study the shape bias of a range of existing robust covariance matrix estimators; or \citet{Cator:2010,Cator:2012} who consider asymptotic expansions and establish asymptotic normality for general MCD estimators.

An alternative approach is to estimate the covariance matrix in a component-wise manner based on a robust estimator of scale as outlined by \citet{Ma:2001}.  It is well known that the resulting symmetric matrix is not guaranteed to be positive definite (PD). Methods to ensure the resulting estimator is PD have previously been explored by \citet{Rousseeuw:1993b} with notable updates in the robustness literature by \citet{Maronna:2002} and quite separately in the finance literature by \citet{Higham:2002}.  \citet{Alqallaf:2002} also proposed a pairwise approach to covariance matrix estimation by means of first Winsorising the data.  The resulting Quadrant Covariance estimate does not necessarily require a transformation to ensure the result is positive definite.

In practice, it is often the precision matrix, the inverse of the covariance matrix, that is primarily of interest.  This is the case, for example, in Gaussian graphical model selection.  As such, this paper is primarily concerned with robustly estimating the precision matrix.   While there is an obvious link between covariance matrices and precision matrices, it is not obvious that a good (robust) estimator for one results in a good estimator for the other.   We will employ robust pairwise covariance matrices as a starting point for various regularisation techniques to facilitate the estimation of robust, potentially sparse, precision matrices.

Classical robust estimators assume that contamination occurs within a restricted subset of the observation vectors, however, in recent years there has been interest in developing robust estimators that perform well under cellwise contamination.  The cellwise contamination model was initially explored in a data mining context by \citet{Alqallaf:2002} and later defined comprehensively by \citet{Alqallaf:2009}.  This form of contamination is prevalent in large, automatically generated data sets, found in data mining and bioinformatics, where there is often little quality control over the inputs.     Cellwise contamination is common in the context of missing data, however, it represents a philosophical divergence from the traditional approach to robustness.  Recent examples where the problem of cellwise contamination have arisen include, \citet{Farcomeni:2014}, \citet{VanAelst:2012} and \citet{Agostinelli:2014}.  

We perform a detailed simulation study to assess the performance of a variety of precision matrix estimators in the presence of cellwise contamination over a number of scenarios and levels of $p$ while keeping the sample size fixed.  Our results are distilled from a comprehensive range of performance indices.  We outline these indices and consider their applicability to the various scenarios in the supplementary material accompanying this article.   

We show that the pairwise nature of the covariance estimates enables the resulting precision matrix to have a higher level of robustness than when using standard robust covariance matrix estimation procedures in the presence of cellwise contamination.  This is a novel result and a significant first step towards dealing with cellwise contamination in this context.

The remainder of this paper is structured as follows.  Section \ref{scatteredcont} outlines the cellwise contamination model and highlights why standard robust techniques fail in this setting.   Sections \ref{pairwisecovmatest} and \ref{precisionmatest} outline the theory for existing pairwise covariance matrix estimation techniques and regularisation routines and we propose a new procedure which combines robust pairwise covariance matrix estimation with regularisation.  Sections \ref{MVsim} and \ref{MVsimn50} present the results of an extensive simulation study and Section \ref{MVconclusion} summarises the important findings.
 
\section{Cellwise contamination} \label{scatteredcont}

Consider a data set $\mathbf{X}\in \mathbb{R}^{n\times p}$ consisting of $n$ observations on $p$ variables.    Classically, even the most robust procedures are designed such that they only work when at most half of the rows in $\mathbf{X}$ have contamination present. 

\citet{Alqallaf:2009} formally outline the cellwise contamination model as an extension of the standard Tukey-Huber contamination model which was first introduced in the univariate location-scale setup \citep{Tukey:1962,Huber:1964}.    Consider the data generating process for the $n$ rows in $\mathbf{X}$, $\mathbf{x}_{i} = (\mathbf{I}-\mathbf{B}_{i})\mathbf{y}_{i} + \mathbf{B}_{i}\mathbf{z}_{i}$,
where $\mathbf{y}_{i}\sim \mathbf{F}$, the distribution of well-behaved data, $\mathbf{z}_{i}\sim \mathbf{G}$, some outlier generating distribution and $\mathbf{B}_{i}=\operatorname{diag}(B_1,\ldots,B_p)$ is a diagonal matrix, where $B_1,\ldots,B_p$ are Bernoulli random variables, $B_j\sim\mathcal{B}(1,\varepsilon_j)$.  When $\mathbf{y}$, $\mathbf{B}$ and $\mathbf{z}$ are independent we have a situation that is similar to the missing completely at random model, where the missingness does not depend on the values of $\mathbf{y}$, see, for example, \citet{Little:2002}.

The structure of $\mathbf{B}_{i}$ determines the contamination model.  If $B_1,\ldots,B_p$ are fully dependent, then $\mathbf{B}_{i} = U_{i}\mathbf{I}$, where $U_{i}\sim\mathcal{B}(1,\varepsilon)$, and we recover the fully dependent contamination model, the standard model on which classical robust procedures are based.  In this setting, the probability that an observation is uncontaminated, $1-\varepsilon$, is independent of the dimensionality.  Furthermore, the proportion of contaminated observations is preserved under affine equivariant transformations.
 
In contrast, if $B_1,\ldots,B_p$ are mutually independent we have the fully independent contamination model, where each element of $\mathbf{x}_{i}$ is drawn from $\mathbf{F}$ or $\mathbf{G}$ independently of the other $p-1$ elements in $\mathbf{x}_{i}$.  That is, contaminating observations occur independently at the univariate level.  In this setting, it may be be unreasonable to assume that less than half the rows have contamination.  Furthermore, if $p$ is large and there is only one outlier in an observation vector, then down-weighting the entire observation may be wasteful.

\begin{figure}[t]
\begin{center}
\input{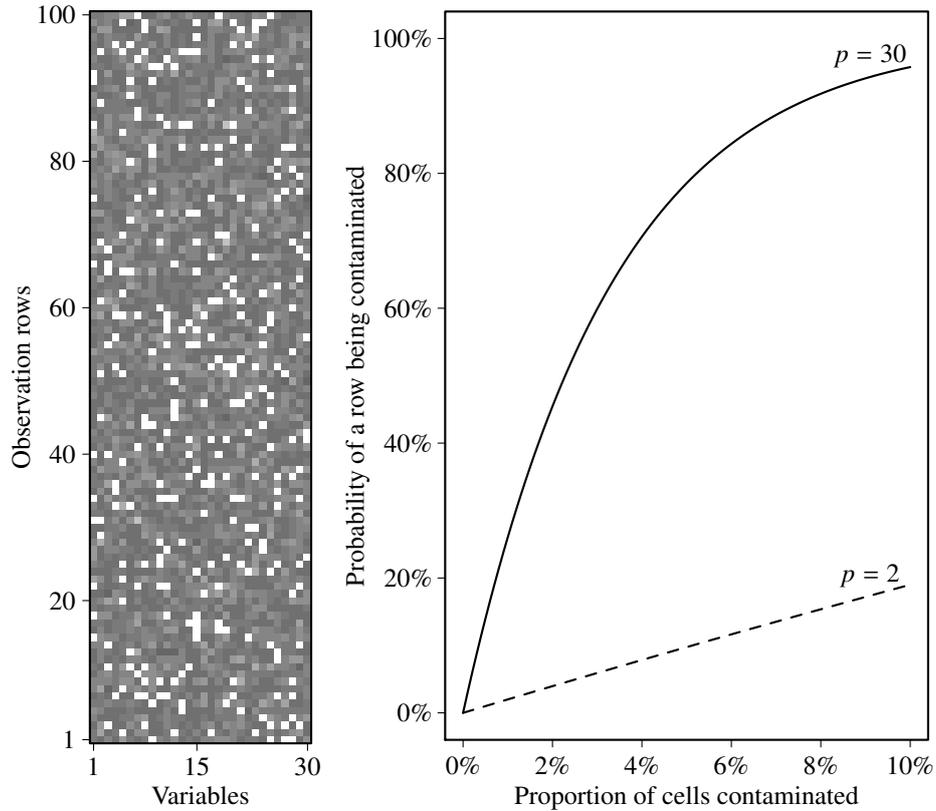}
\begin{tikzpicture}[x=1pt,y=1pt]
\definecolor[named]{fillColor}{rgb}{1.00,1.00,1.00}
\begin{scope}
\path[clip] ( 30.00, 30.00) rectangle (210.81,304.76);
\definecolor[named]{drawColor}{rgb}{0.00,0.00,0.00}

\path[draw=drawColor,line width= 0.8pt,line join=round,line cap=round] ( 36.70, 40.18) -- ( 38.37, 47.70) -- ( 40.04, 55.01) -- ( 41.72, 62.10) -- ( 43.39, 69.00) -- ( 45.07, 75.70) -- ( 46.74, 82.20) -- ( 48.42, 88.52) -- ( 50.09, 94.65) -- ( 51.76,100.61) -- ( 53.44,106.40) -- ( 55.11,112.02) -- ( 56.79,117.48) -- ( 58.46,122.78) -- ( 60.13,127.92) -- ( 61.81,132.92) -- ( 63.48,137.77) -- ( 65.16,142.48) -- ( 66.83,147.06) -- ( 68.51,151.50) -- ( 70.18,155.81) -- ( 71.85,159.99) -- ( 73.53,164.06) -- ( 75.20,168.00) -- ( 76.88,171.83) -- ( 78.55,175.55) -- ( 80.22,179.16) -- ( 81.90,182.66) -- ( 83.57,186.06) -- ( 85.25,189.36) -- ( 86.92,192.57) -- ( 88.60,195.67) -- ( 90.27,198.69) -- ( 91.94,201.62) -- ( 93.62,204.46) -- ( 95.29,207.22) -- ( 96.97,209.89) -- ( 98.64,212.49) -- (100.31,215.01) -- (101.99,217.45) -- (103.66,219.82) -- (105.34,222.13) -- (107.01,224.36) -- (108.69,226.53) -- (110.36,228.63) -- (112.03,230.67) -- (113.71,232.64) -- (115.38,234.56) -- (117.06,236.42) -- (118.73,238.23) -- (120.40,239.98) -- (122.08,241.68) -- (123.75,243.32) -- (125.43,244.92) -- (127.10,246.47) -- (128.78,247.97) -- (130.45,249.43) -- (132.12,250.84) -- (133.80,252.21) -- (135.47,253.54) -- (137.15,254.83) -- (138.82,256.08) -- (140.49,257.29) -- (142.17,258.47) -- (143.84,259.61) -- (145.52,260.71) -- (147.19,261.78) -- (148.87,262.82) -- (150.54,263.82) -- (152.21,264.80) -- (153.89,265.74) -- (155.56,266.66) -- (157.24,267.55) -- (158.91,268.41) -- (160.58,269.24) -- (162.26,270.05) -- (163.93,270.83) -- (165.61,271.59) -- (167.28,272.33) -- (168.96,273.04) -- (170.63,273.73) -- (172.30,274.40) -- (173.98,275.05) -- (175.65,275.68) -- (177.33,276.29) -- (179.00,276.88) -- (180.67,277.45) -- (182.35,278.00) -- (184.02,278.54) -- (185.70,279.06) -- (187.37,279.56) -- (189.05,280.05) -- (190.72,280.52) -- (192.39,280.98) -- (194.07,281.42) -- (195.74,281.85) -- (197.42,282.27) -- (199.09,282.67) -- (200.76,283.06) -- (202.44,283.43) -- (204.11,283.80);
\end{scope}
\begin{scope}
\path[clip] (  0.00,  0.00) rectangle (216.81,310.76);
\definecolor[named]{drawColor}{rgb}{0.00,0.00,0.00}


\path[draw=drawColor,line width= 0.4pt,line join=round,line cap=round] ( 36.70, 30.00) -- ( 36.70, 27.29);

\path[draw=drawColor,line width= 0.4pt,line join=round,line cap=round] ( 70.18, 30.00) -- ( 70.18, 27.29);

\path[draw=drawColor,line width= 0.4pt,line join=round,line cap=round] (103.66, 30.00) -- (103.66, 27.29);

\path[draw=drawColor,line width= 0.4pt,line join=round,line cap=round] (137.15, 30.00) -- (137.15, 27.29);

\path[draw=drawColor,line width= 0.4pt,line join=round,line cap=round] (170.63, 30.00) -- (170.63, 27.29);

\path[draw=drawColor,line width= 0.4pt,line join=round,line cap=round] (204.11, 30.00) -- (204.11, 27.29);

\node[text=drawColor,anchor=base,inner sep=0pt, outer sep=0pt, scale=  0.90] at ( 36.70, 18.00-1) {0\%};

\node[text=drawColor,anchor=base,inner sep=0pt, outer sep=0pt, scale=  0.90] at ( 70.18, 18.00-1) {2\%};

\node[text=drawColor,anchor=base,inner sep=0pt, outer sep=0pt, scale=  0.90] at (103.66, 18.00-1) {4\%};

\node[text=drawColor,anchor=base,inner sep=0pt, outer sep=0pt, scale=  0.90] at (137.15, 18.00-1) {6\%};

\node[text=drawColor,anchor=base,inner sep=0pt, outer sep=0pt, scale=  0.90] at (170.63, 18.00-1) {8\%};

\node[text=drawColor,anchor=base,inner sep=0pt, outer sep=0pt, scale=  0.90] at (204.11, 18.00-1) {10\%};


\path[draw=drawColor,line width= 0.4pt,line join=round,line cap=round] ( 30.00, 40.18) -- ( 27.29, 40.18);

\path[draw=drawColor,line width= 0.4pt,line join=round,line cap=round] ( 30.00, 91.06) -- ( 27.29, 91.06);

\path[draw=drawColor,line width= 0.4pt,line join=round,line cap=round] ( 30.00,141.94) -- ( 27.29,141.94);

\path[draw=drawColor,line width= 0.4pt,line join=round,line cap=round] ( 30.00,192.82) -- ( 27.29,192.82);

\path[draw=drawColor,line width= 0.4pt,line join=round,line cap=round] ( 30.00,243.70) -- ( 27.29,243.70);

\path[draw=drawColor,line width= 0.4pt,line join=round,line cap=round] ( 30.00,294.58) -- ( 27.29,294.58);

\node[text=drawColor,rotate= 90.00,anchor=east,inner sep=0pt, outer sep=0pt, scale=  0.90,rotate=270] at ( 25.20, 40.18) {0\%};

\node[text=drawColor,rotate= 90.00,anchor=east,inner sep=0pt, outer sep=0pt, scale=  0.90,rotate=270] at ( 25.20, 91.06) {20\%};

\node[text=drawColor,rotate= 90.00,anchor=east,inner sep=0pt, outer sep=0pt, scale=  0.90,rotate=270] at ( 25.20,141.94) {40\%};

\node[text=drawColor,rotate= 90.00,anchor=east,inner sep=0pt, outer sep=0pt, scale=  0.90,rotate=270] at ( 25.20,192.82) {60\%};

\node[text=drawColor,rotate= 90.00,anchor=east,inner sep=0pt, outer sep=0pt, scale=  0.90,rotate=270] at ( 25.20,243.70) {80\%};

\node[text=drawColor,rotate= 90.00,anchor=east,inner sep=0pt, outer sep=0pt, scale=  0.90,rotate=270] at ( 25.20,294.58) {100\%};

\path[draw=drawColor,line width= 0.4pt,line join=round,line cap=round,thick] ( 30.00, 30.00) -- (210.81, 30.00) -- (210.81,304.76) -- ( 30.00,304.76) -- ( 30.00, 30.00);
\end{scope}
\begin{scope}
\definecolor[named]{drawColor}{rgb}{0.00,0.00,0.00}

\node[text=drawColor,anchor=base,inner sep=0pt, outer sep=0pt, scale=  0.90] at (120.41,  2.40+3.5) {Proportion of cells contaminated};

\node[text=drawColor,rotate= 90.00,anchor=base,inner sep=0pt, outer sep=0pt, scale=  0.90] at (  9.60-10,167.38) {Probability of a row being contaminated};
\end{scope}
\begin{scope}
\path[clip] ( 30.00, 30.00) rectangle (210.81,304.76);
\definecolor[named]{drawColor}{rgb}{0.00,0.00,0.00}

\path[draw=drawColor,line width= 0.8pt,dash pattern=on 4pt off 4pt ,line join=round,line cap=round] ( 36.70, 40.18) -- ( 38.37, 40.68) -- ( 40.04, 41.19) -- ( 41.72, 41.70) -- ( 43.39, 42.21) -- ( 45.07, 42.71) -- ( 46.74, 43.22) -- ( 48.42, 43.73) -- ( 50.09, 44.23) -- ( 51.76, 44.74) -- ( 53.44, 45.24) -- ( 55.11, 45.74) -- ( 56.79, 46.25) -- ( 58.46, 46.75) -- ( 60.13, 47.25) -- ( 61.81, 47.75) -- ( 63.48, 48.25) -- ( 65.16, 48.75) -- ( 66.83, 49.25) -- ( 68.51, 49.75) -- ( 70.18, 50.25) -- ( 71.85, 50.75) -- ( 73.53, 51.25) -- ( 75.20, 51.74) -- ( 76.88, 52.24) -- ( 78.55, 52.74) -- ( 80.22, 53.23) -- ( 81.90, 53.73) -- ( 83.57, 54.22) -- ( 85.25, 54.72) -- ( 86.92, 55.21) -- ( 88.60, 55.71) -- ( 90.27, 56.20) -- ( 91.94, 56.69) -- ( 93.62, 57.18) -- ( 95.29, 57.67) -- ( 96.97, 58.16) -- ( 98.64, 58.65) -- (100.31, 59.14) -- (101.99, 59.63) -- (103.66, 60.12) -- (105.34, 60.61) -- (107.01, 61.10) -- (108.69, 61.59) -- (110.36, 62.07) -- (112.03, 62.56) -- (113.71, 63.04) -- (115.38, 63.53) -- (117.06, 64.01) -- (118.73, 64.50) -- (120.40, 64.98) -- (122.08, 65.46) -- (123.75, 65.95) -- (125.43, 66.43) -- (127.10, 66.91) -- (128.78, 67.39) -- (130.45, 67.87) -- (132.12, 68.35) -- (133.80, 68.83) -- (135.47, 69.31) -- (137.15, 69.79) -- (138.82, 70.27) -- (140.49, 70.75) -- (142.17, 71.22) -- (143.84, 71.70) -- (145.52, 72.17) -- (147.19, 72.65) -- (148.87, 73.13) -- (150.54, 73.60) -- (152.21, 74.07) -- (153.89, 74.55) -- (155.56, 75.02) -- (157.24, 75.49) -- (158.91, 75.96) -- (160.58, 76.44) -- (162.26, 76.91) -- (163.93, 77.38) -- (165.61, 77.85) -- (167.28, 78.32) -- (168.96, 78.79) -- (170.63, 79.25) -- (172.30, 79.72) -- (173.98, 80.19) -- (175.65, 80.66) -- (177.33, 81.12) -- (179.00, 81.59) -- (180.67, 82.05) -- (182.35, 82.52) -- (184.02, 82.98) -- (185.70, 83.45) -- (187.37, 83.91) -- (189.05, 84.37) -- (190.72, 84.83) -- (192.39, 85.30) -- (194.07, 85.76) -- (195.74, 86.22) -- (197.42, 86.68) -- (199.09, 87.14) -- (200.76, 87.60) -- (202.44, 88.06) -- (204.11, 88.51);

\node[anchor=south,scale=0.9] at (189.05, 84.37) {$p=2$};
\node[anchor=south,scale=0.9] at (189.05,280.05) {$p=30$};

\end{scope}

  \path
    ([shift={(-5\pgflinewidth,-3\pgflinewidth)}]current bounding box.south west)
    ([shift={( 5\pgflinewidth, 0\pgflinewidth)}]current bounding box.north east);
\end{tikzpicture}
\end{center}
\caption{On the left, a heat map of a data matrix with 30 variables and 100 observations.  10\% of the cells have been contaminated and are shown as white cells, while the uncontaminated cells are in various shades of grey.  On the right, the probability that any particular row (observations) in the data matrix will be contaminated, $1-(1-\varepsilon)^p$, over a range of $\varepsilon$, the proportion of cells affected by cellwise contamination.}
\label{cellwiseplot}
\end{figure}

If the data matrix is randomly contaminated in this elementwise manner, as the number of variables increases, the chance that more than half the rows are contaminated increases exponentially.  Formally, let $\varepsilon$ be the probability that any particular element in a data matrix is contaminated.  Assuming the contamination is randomly scattered throughout the data matrix, the probability that any particular row has no contamination is $(1-\varepsilon)^p$, which quickly decays towards zero even for small values of $\varepsilon$.  For example, if $p=30$ and $\varepsilon=0.1$, then the probability that any particular row remains uncontaminated is only 4\%.  This is demonstrated graphically in Figure \ref{cellwiseplot}.  The plot on the left shows a $100\times 30$ data matrix where 10\% of the cells have been  contaminated, the white cells.  While virtually all the rows of the data matrix have at least one contaminated element, the majority of cells remain uncontaminated in the sense that they are still real measurements from the underlying data generating process.  Even if $\varepsilon =0.03$, the probability that any particular row is uncontaminated is 40\%, however with a sample size of 100, this translates to a 98\% chance that at least half the rows are contaminated, in which case standard robust methods fail.

It is important to note that the fully independent contamination model lacks affine equivariance, in the sense that linear combinations of columns of a contaminated data set result in ``outlier propagation'' \citep{Alqallaf:2009}.  As such, affine equivariance is not an achievable outcome for any estimator in this setting.

Existing research into the problem of cellwise contamination has focussed on coordinatewise procedures, that only operate on one column at a time. \citet{Croux:2003} consider an approach based on ``alternating regressions'' using weighted $L_{1}$ regression, \citet{Maronna:2008} use a coordinatewise procedure for principal component analysis. \citet{Liu:2003} have an application involving the singular value decomposition of microarray data and \citet{DelaTorre:2001} consider cellwise contamination in the context of computer vision.  

We show that a pairwise approach is able to cope with much higher levels of cellwise contamination than existing classical robust estimators.  In the simulations in Section \ref{MVsim} we do not use the fully independent contamination model, rather, we impose restrictions on the amount of contamination in each variable.  As such the contamination is no longer strictly independent, however, the advantage is that we are able to assess the impact over various known levels of contamination in each variable.  

\section{Pairwise covariance matrix estimation} \label{pairwisecovmatest}

A pairwise approach to estimating covariance matrices in the presence of cellwise contamination has previously been explored by \citet{Alqallaf:2002} where the classical correlation coefficient was applied to a Winsorised data set.  Instead of transforming the underlying data, our approach is to take the $p(p-1)/2$ pairs of variables and robustly estimate the covariance between each pair. The primary advantage of this approach is robustness to cellwise contamination in the data set.   The main disadvantage is that the resulting symmetric matrix is not guaranteed to be positive semidefinite or affine equivariant.  However, as noted earlier, in the cellwise contamination model, affine equivariance is unachievable as there is the potential for all rows to have a contaminated cell, hence linear combinations of the rows propagate the contamination.

A simple method for turning scale estimators into covariance estimators was introduced by \citet{Gnanadesikan:1972} and brought to prominence in the context of robust estimation by \citet{Ma:2001}.  The idea is based on the identity,
\begin{equation}
\operatorname{cov}(X,Y) = \frac{1}{4\alpha\beta} \left [ \operatorname{var}( \alpha X + \beta Y) - \operatorname{var}(\alpha X -\beta Y)\right],\label{GKident}
\end{equation}
where $X$ and $Y$ are random variables.  In general, $X$ and $Y$ may have different scales, hence it is standard to let $\alpha = 1/\sqrt{{\operatorname{var}}(X)}$ and $\beta = 1/\sqrt{{\operatorname{var}}(Y)}$.  A robust covariance estimator is found by replacing the variance in \eqref{GKident} with (squared) robust scale estimators.  We will focus on the estimators $Q_n$ \citep{Rousseeuw:1993}, the $\tau$-scale as described in \citet{Maronna:2002} and an estimator that is somewhat less robust but highly efficient, $P_n$, the interquartile range of the pairwise means, and its adaptively trimmed variant $\widetilde{P}_n$ with adaptive trimming parameter $d=3$ \citep{Tarr:2012}.  We also consider the interquartile range (IQR) and the median absolute deviation from the median (MAD).   A recent discussion on the efficiency of various robust scale estimators can be found in \citet{Tarr:2012}.

Using identity \eqref{GKident}, a symmetric matrix full of pairwise covariances can easily be constructed, however, there is no guarantee that the result will be positive semidefinite.  The two methods outlined below overcome this limitation by appropriately adjusting the eigenvalues of the symmetric matrix to ensure that they are all positive, and hence ensuring a positive definite result.

\subsection{Orthogonalised Gnanadesikan Kettenring procedure}

To overcome the possible lack of positive semidefiniteness in a matrix of pairwise covariances, \citet{Maronna:2002} propose a modification based on the observation that the eigenvalues of a covariance matrix are the variances along the directions given by the respective eigenvectors.  Essentially a principal components decomposition is performed and the covariance matrix is reconstructed using robust variance estimates of the principal component vectors in place of the original eigenvalues.  This procedure is known as the Orthogonalised Gnanadesikan Kettenring (OGK) estimator.  Note that even if the original covariance matrix was already positive definite, applying the OGK procedure will not necessarily return the same matrix.  

\citet{Maronna:2002} and \citet[p.\ 207]{Maronna:2006} suggest that the OGK estimator can be improved by iterating the procedure and then using this estimate to find robust Mahalanobis distances for each observation vector.  These are then used to screen for outliers before applying the classical covariance estimator to the cleaned data, resulting in a procedure known as the reweighted OGK.  This is done in an effort to increase efficiency and to make the result ``more equivariant''.  In terms of the impact of not being affine equivariant, \citet{Maronna:2002} note that ``although the worst case may differ from the original data, for most transformations the results are very similar'' and ``the lack of equivariance is not a serious concern in our estimates''.

Regardless, neither the OGK method nor the reweighted OGK method is able to cope with cellwise contamination.  The issue of outlier propagation means that the number of contaminated principal components could easily be greater than 50\% even for small levels of cellwise contamination.  Hence, the robust variance estimates that are used in place of the eigenvalues will no longer be valid estimates -- they will be in breakdown.  Furthermore, the reweighting step will often needlessly exclude many observation vectors where there is only one contaminated cell.

\subsection{Nearest positive definite matrix procedure}
\citet{Higham:2002} considers the problem of computing the nearest positive definite (NPD) matrix to a given symmetric matrix.  The motivation stems from finance, where sample covariance matrices are constructed from vectors of stock returns, however, the problem arises when not all stocks are observed every day.  In this setting, classical covariances may be computed on a pairwise basis using data drawn only from days where both stocks have data available.  The resulting covariance matrix is not guaranteed to be PD because it has been built from inconsistent data sets.  Motivated by the same problem, \citet{Loland:2013} propose both a pseudo-likelihood and a Bayesian approach to find PD estimates of pairwise correlation matrices. However, their approach relies on expert knowledge to formulate priors for the pairwise covariances.

The NPD procedure is similar to the OGK procedure in that it performs a spectral decomposition and then updates the eigenvalues to ensure that the result is PD. However, it does not rely on linear transformations of the original dataset and hence is not affected by the ``outlier propagation'' issue associated with cellwise contamination.  Formally, for an arbitrary symmetric $p\times p$ matrix $\mathbf{A}$, the aim is to find the distance
\begin{equation}
\gamma(\mathbf{A}) = \min\{ || \mathbf{A}-\mathbf{W} ||_F : \mathbf{W} \text{ is a symmetric PD matrix}\}, \label{Higham1.1}
\end{equation}
and the resulting matrix that achieves this minimum distance.  \citet{Higham:2002} uses the Frobenius norm, $||\mathbf{B}||_F=\sqrt{\operatorname{tr}(\mathbf{B}'\mathbf{B})}$, as it is ``the easiest norm to work with for this problem and also being the natural choice from the statistical point of view''.

While \citet{Higham:2002} considers a variety of weighting mechanisms, in the simplest case without specifying any weights, the procedure is quite straightforward.  The final estimate is $\hat{\mathbf{W}} = \mathbf{E}\hat{\bm{\Lambda}}\mathbf{E}'$,
where $\mathbf{E}\bm{\Lambda}\mathbf{E}' $ is the spectral decomposition of $\mathbf{A}$, with $\bm{\Lambda} = \operatorname{diag}(\lambda_1,\ldots,\lambda_p)$ and $\hat{\bm{\Lambda}} = \operatorname{diag}(\max\{\lambda_i,\delta\})$, where $\delta$ is a small positive constant.   In contrast to the OGK procedure, if the initial symmetric matrix is already PD, then the NPD  method simply returns the original pairwise covariance matrix.  

In the presence of cellwise contamination the NPD method outperforms the OGK method.  However, the NPD method often results in estimated matrices with a number of extremely small eigenvalues which give poorly conditioned estimates, i.e.\ the condition number of these estimators is very high as is the entropy loss, which involves the log of the eigenvalues.  In general, it is not recommended to use either the OGK nor the NPD in isolation when there is cellwise contamination present.  Even in the presence of standard row-wise contamination, the NPD method is not recommended due to its propensity to return poorly conditioned estimates.

\section{Precision matrix estimation} \label{precisionmatest}

Many statistical procedures are primarily concerned with the precision matrix, the inverse of a covariance matrix, rather than the covariance matrix itself.   For example, finding Mahalanobis distances and performing linear discriminant analysis both require an estimate of $ \bm{\Theta} =  \bm{\Sigma}^{-1}$.  Finding good precision matrix estimates has been a focus of many investigators over a long period of time, the first major contribution being  \citet{Dempster:1972}.

The following routines take as an input an estimated covariance matrix and output a regularised precision matrix.  In Section \ref{MVsim} we demonstrate the advantages of using a robust pairwise covariance matrix estimate as the input to these regularisation routines.

\subsection{GLASSO}

A natural way to estimate $ \bm{\Theta}$ is by maximising the log-likelihood of the data.  With Gaussian observations, the log-likelihood takes the form,
\begin{equation}
\log | \bm{\Theta}| - \operatorname{tr} (\mathbf{S}  \bm{\Theta}),\label{LR}
\end{equation}
where $\mathbf{S}$ is an estimate of the covariance matrix of the data.  Maximising \eqref{LR} with respect to $ \bm{\Theta}$ leads to the {MLE}, $\mathbf{S}^{-1}$.  In general, $\mathbf{S}^{-1}$ will not be sparse, in the sense that it will contain no elements exactly equal to zero. Furthermore in $p>n$ situations $\mathbf{S}$ will be singular so the {MLE} cannot be computed.  \citet{Yuan:2007} consider minimising the penalised negative log-likelihood,
\begin{equation}
\operatorname{tr}(\mathbf{S} \bm{\Theta}) - \log | \bm{\Theta}| + \lambda \textstyle\sum_{i,j}|\theta_{ij}|,\label{1}
\end{equation}
over the set of PD matrices where $\lambda$ is a tuning parameter to control the amount of shrinkage. \citet{Friedman:2008} refer to this estimator as the graphical lasso (GLASSO) and note that it has two major advantages over \eqref{LR}: the solution is PD for all $\lambda>0$ even if $\mathbf{S}$ is singular, and for large values of $\lambda$ the resulting estimate, $\hat{\bm{\Theta}}$, will be sparse.

\subsection{QUIC}

The QUIC method solves the same minimisation problem as the {GLASSO}.  The improvement in speed comes from noticing that the Gaussian log-likelihood component of \eqref{1} is twice differentiable and strictly convex which lends itself to a quadratic approximation and hence faster convergence \citep{Hsieh:2011}.   On the other hand, the penalty term is convex but not differentiable and so is treated separately.   

The QUIC routine, as implemented in the R package \texttt{QUIC}, explicitly includes a step that ensures positive definiteness of the precision matrix for each iteration. Work has recently been undertaken to scale the QUIC estimator to scenarios with a million variables, see \citet{Hsieh:2013} for details. 

\subsection{CLIME}

An alternative to maximising the penalised log-likelihood is to use the constrained $\ell_{1}$ minimisation approach to sparse precision matrix estimation (CLIME), implemented in the R package \texttt{clime} \citep{Cai:2011,Cai:2012}.  The CLIME routine uses linear programming to solve the following (convex) optimisation problem,
$$ \bm{\Theta}^\star = \min | \bm{\Theta}|_{1} \quad \text{ subject to: } |\mathbf{S} \bm{\Theta} - \mathbf{I}|_{\infty}\leq \lambda,$$
where $\mathbf{S}$ is the sample covariance matrix and $|\mathbf{A}|_1 = \sum_{i,j}|a_{ij}|$ is the elementwise $\ell_{1}$ norm of a matrix, $\mathbf{A}$, and $|\mathbf{A}|_\infty = \max_{i,j}|a_{ij}|$ is the elementwise infinity norm.  No symmetry requirements are placed on $ \bm{\Theta}^\star$ so a symmetrising step is applied to obtain the final solution, $\hat{ \bm{\Theta}}$,
$$\hat\theta_{ij} = \hat\theta_{ji} = \theta_{ij}^\star\,\mathbb{I}\{|\theta_{ij}^\star|\leq |\theta_{ji}^\star|\} + \theta_{ji}^\star\,\mathbb{I}\{|\theta_{ij}^\star|> |\theta_{ji}^\star|\}.$$
Theorem 1 of \citet{Cai:2011} shows that the resulting $\hat{ \bm{\Theta}}$ is PD with high probability.

Our simulations show that there is little difference between using CLIME and QUIC -- the key point is that both appear to perform well in the presence of cellwise contamination when the input matrix is based on pairwise robust covariance estimates and it has been made PD using the NPD routine.

\section{Simulation study for $p<n$} \label{MVsim}

This section presents the results of an extensive simulation study to assess how well various robust covariance estimation techniques perform when used as an input to the regularisation routines outlined previously. 

The proposed estimator begins by finding the covariances between all $p(p-1)/2$ pairs of variables.  For the scale estimator underlying the robust covariance estimator, we consider $Q_n$, the $\tau$-scale, the {MAD} and the {IQR}.  We also consider the $P_n$ estimator and two adaptively trimmed variants $\widetilde{P}_n$, with trimming parameters $d=3$ and $d=5$, see \citet{Tarr:2012} for further details about these estimators.  The pairwise covariances are arranged in a symmetric, though not necessarily PD, matrix.  The symmetric matrix is transformed to a PD matrix using either the OGK method or the NPD method before being input into the {GLASSO}, QUIC or CLIME regularisation routines.  For comparison purposes we also include the classical covariance estimator and the MCD as initial covariance matrix estimates.

\subsection{Design}\label{design}

\begin{figure}[bt]
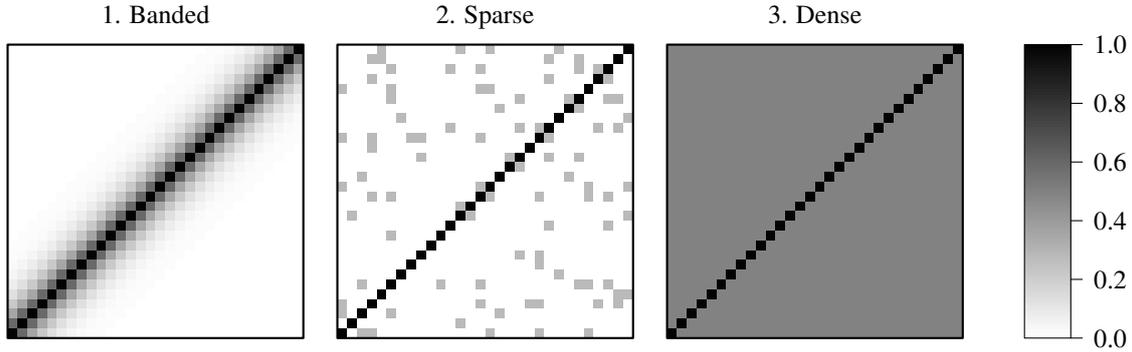

\centering
\input{T1new}\input{T2new}\input{T3new}
\caption{Heat maps of the three kinds of precision matrices used to generate the data when $p=30$.}
\label{precisionscenarios}
\end{figure}

The simulated data follows a multivariate Gaussian distribution with $n=100$ observations, $\mathcal{N} (\mathbf{0}, \bm{\Theta}^{-1})$.  The precision matrices we select as the basis for the data generating process represent a broad range of scenarios that occur in practice and are similar to those used in \citet{Cai:2011} and \citet{Hsieh:2011}.  In particular, we consider three types of precision matrices, $\bm{\Theta}$, as shown in Figure \ref{precisionscenarios} and outlined below.
\begin{enumerate}
\item\label{Banded}
\textsl{Banded precision matrices}, with elements $\theta_{ij} = 0.6^{|i-j|}$, such that the values of the entries decay the further they are from the main diagonal.
\item\label{Sparse}
\textsl{Sparse precision matrices}, with randomly allocated non-zero entries, where $ \bm{\Theta} = \mathbf{B} + \delta \mathbf{I}$ with each off diagonal entry in $\mathbf{B}$ generated independently, where $P(b_{ij} = 0.5)=0.1$ and $P(b_{ij} = 0) = 0.9$ and $\delta$ is chosen such that the condition number of the matrix equals $p$. The matrix is then standardised to have diagonal components equal to one.  This scenario will be referred to as scattered sparsity.
\item\label{Dense}
\textsl{Dense precision matrices}, where $ \bm{\Theta}$ has all off diagonal elements equal to $0.5$ and diagonal elements equal to $1$.
\end{enumerate}

The outliers were generated independently for each variable.  In our simulations we allow the number of contaminated observations within each variable to increase up to a maximum of 25 observations (out of $n=100$).  In this way we have complete control over the total number of contaminated cells.  The distribution of the outliers is a $t_{10}$ distribution scaled by either a factor of $10$ for extreme outliers or $\sqrt{10}$ for moderate outliers.  The moderate outliers are perhaps closer to what one might expect in a real data set. However, the focus here is primarily on the extreme outliers where the overwhelming majority of the unusual observations lie well outside the cloud of standard observations. The extreme nature of the outliers serves to demark clearly estimators that have effectively broken down from those that are still capable of giving ballpark correct results.  In both cases, the outliers are symmetrically distributed.

Each of the regularisation routines require a tuning parameter.  At each replication, the tuning parameter was obtained by training on a separate (uncontaminated) randomly generated data set drawn from the true data generating process.  For the training data, a sequence of precision matrices was obtained and the value of the tuning parameter corresponding to the smallest entropy loss was then used for that replication.  In practice, there was a small amount of variability in the choice of tuning parameter within each scenario and dimensionality. Furthermore, the QUIC and GLASSO routines almost always picked the same tuning parameter and the CLIME routine was free to choose slightly smaller tuning parameters.  For example, in the scattered sparsity scenario with $n=100$ and $p=60$ the training set for CLIME resulted in a tuning parameter of around $0.09$ whereas for QUIC and GLASSO it was closer to $0.12$.  In practice the tuning parameter may be chosen through cross validation, a BIC-type criterion such as in \citet{Yuan:2007}, or it can be adjusted in an ad-hoc way until a desired level of sparsity is achieved.

We performed an extensive investigation into the most appropriate way to compare estimated precision matrices in the presence of cellwise contamination. We considered a number of matrix norms, the Frobenius norm, one norm, infinity norm and spectral norm, as well as the log determinant, the condition number and the entropy loss. The definition and behaviour of these performance indices under cellwise contamination are outlined in the supplementary material.  We found that the most appropriate measure was the entropy loss defined as, $L_E( \bm{\Theta},\hat{ \bm{\Theta}})  = \operatorname{tr}( \bm{\Theta} ^{-1}\hat{ \bm{\Theta}}) - \log\det( \bm{\Theta}^{-1}\hat{ \bm{\Theta}}) - p $.

As in \citet{Lin:1985}, we report the results in terms of the percentage relative improvement in average loss (PRIAL),
$$\text{PRIAL}(\hat{ \bm{\Theta}}) = \frac{L_E( \bm{\Theta},\hat{ \bm{\Theta}}_0) - L_E( \bm{\Theta},\hat{ \bm{\Theta}})}{L_E( \bm{\Theta},\hat{ \bm{\Theta}}_0)}\times 100,$$
over $N=100$ replication of each design, where $\hat{ \bm{\Theta}}_0$ is the estimated precision matrix after a regularisation technique has been applied to the classical sample covariance matrix for uncontaminated data.  It is important to note that this is an extremely harsh benchmark to set.

\subsection{Results}

\subsubsection{No contamination}

Any good robust method should give comparable results to the classical non-robust method it is replacing when presented with a clean dataset.  Table \ref{tab:cleanPRIAL} presents the PRIAL results for the no contamination scenario.  As the PRIAL results are relative to the base case for each routine, Table \ref{tab:cleanPRIAL} cannot be used to compare the performance of the CLIME routine to the QUIC routine.

\begin{table}[b!]
\centering
\begin{tabular}{llrrrrrrr}
\toprule
 &  & \multicolumn{3}{c}{CLIME} && \multicolumn{3}{c}{QUIC} \\
\cmidrule{3-5}\cmidrule{7-9}
	&		&	$p=30$	&	$p=60$	&	$p=90$	&&	$p=30$	&	$p=60$	&	$p=90$	\\
	\midrule														
	&	OGK	&	-17.1	&	-15.1	&	-12.7	&&	-13.0	&	-9.5	&	-8.1	\\
$\tau$-scale	&	{OGKw}	&	-34.8	&	-29.3	&	-20.3	&&	-26.9	&	-15.9	&	-15.6	\\
	&	NPD	&	-31.5	&	-31.3	&	-27.6	&&	-25.3	&	-16.7	&	-15.4	\\
	\midrule														
	&	OGK	&	-16.5	&	-13.3	&	-11.7	&&	-13.6	&	-10.5	&	-9.3	\\
$Q_n$	&	{OGKw}	&	-34.7	&	-28.5	&	-12.3	&&	-27.0	&	-15.6	&	-14.6	\\
	&	NPD	&	-40.6	&	-36.9	&	-31.4	&&	-32.5	&	-21.9	&	-20.3	\\
	\midrule														
	&	OGK	&	-13.6	&	-12.7	&	-10.9	&&	-11.8	&	-9.6	&	-8.8	\\
$P_n$	&	{OGKw}	&	-33.7	&	-27.1	&	-17.5	&&	-26.2	&	-14.9	&	-14.5	\\
	&	NPD	&	-19.9	&	-18.4	&	-18.2	&&	-15.7	&	-11.4	&	-11.2	\\
	\midrule														
MCD	&		&	-53.1	&	-19.5	&	-3.7	&&	-56.2	&	-19.5	&	-3.7	\\
\bottomrule
\end{tabular}
\caption[PRIAL results when there is no contamination]{PRIAL results for the various estimators when there is no contamination present.}
\label{tab:cleanPRIAL}
\end{table}

In the uncontaminated case, the OGK method substantially outperforms the NPD method.  Overall, the methods appear to improve as the dimensionality increases, however, this is more a reflection of the deteriorating absolute performance of the baseline classical covariance matrix estimate. 

For $p=30$, the pairwise methods outperform the MCD, however the MCD method uses $\lfloor n+p+1 \rfloor /2$ observations so when $p=90$ the resulting estimator is the classical covariance estimate applied to $95$ out of a total $n=100$ observations.  Hence, it is not surprising that the PRIAL for the MCD method is so close to zero.  In fact, the MCD is not recommended for use when $n < 2p$ \citep{robustbase}.

The reweighted OGK ({OGKw}) methods essentially perform outlier detection and deletion before returning a classical covariance estimate of the cleaned data set.  The performance of these methods is broadly similar over all the various initial scale estimates.  Though not shown in Table \ref{tab:cleanPRIAL}, the {MAD} performs particularly poorly under both the OGK and the NPD corrections and would not be recommended for use.    

As would be expected, given the solid Gaussian performance of $P_n$ (see \citet{Tarr:2012}), the methods based on $P_n$ outperform those based on the $\tau$-scale and $Q_n$.  The relative deterioration in performance for the robust methods compared to the classical method is comparable to that in the simple univariate scale case.  For example, the univariate scale estimator $P_n$ has an asymptotic Gaussian relative efficiency of 86\%.

\subsubsection{Cellwise contamination}

There are a number of ways to compare and contrast the various estimators.  We consider data with $n=100$ observations from three different data generating processes, across four dimensions, $p=15$, $30$, $60$ or $90$, contaminated with either moderate or extreme outliers, as explained in Section \ref{design}. We present figures for the extreme case only but comment on both based on the results of $N=100$ replications of each design.   We implement an array of initial covariance estimation techniques and process these through the {GLASSO}, QUIC and CLIME regularisation routines.  Finally, as outlined in the supplementary material to this article, there are a number of performance indices that are considered.   This section extracts and synthesises the key results.  

\begin{figure}[p]
\centering
\begin{tikzpicture}[x=1pt,y=1pt]
\definecolor[named]{fillColor}{rgb}{1.00,1.00,1.00}
\path[use as bounding box,fill=fillColor,fill opacity=0.00] (0,0) rectangle (433.62,180.67);
\begin{scope}
\path[clip] ( 30.00, 30.00) rectangle (331.62,174.67);
\definecolor[named]{drawColor}{rgb}{0.00,0.00,0.00}

\path[draw=drawColor,line width= 0.8pt,line join=round,line cap=round] ( 41.17,169.32) --
	( 52.34,100.28) --
	( 63.51, 42.71) --
	( 74.68,  0.64) --
	( 74.87,  0.00);
\definecolor[named]{fillColor}{rgb}{0.00,0.00,0.00}

\path[draw=drawColor,line width= 0.8pt,line join=round,line cap=round,fill=fillColor] ( 41.17,169.32) circle (  1.50);

\path[draw=drawColor,line width= 0.8pt,line join=round,line cap=round,fill=fillColor] ( 52.34,100.28) circle (  1.50);

\path[draw=drawColor,line width= 0.8pt,line join=round,line cap=round,fill=fillColor] ( 63.51, 42.71) circle (  1.50);

\path[draw=drawColor,line width= 0.8pt,line join=round,line cap=round,fill=fillColor] ( 74.68,  0.64) circle (  1.50);
\end{scope}
\begin{scope}
\path[clip] (  0.00,  0.00) rectangle (460.00,180.67);
\definecolor[named]{drawColor}{rgb}{0.00,0.00,0.00}


\path[draw=drawColor,line width= 0.4pt,line join=round,line cap=round] ( 41.17, 30.00) -- ( 41.17, 32.89);

\path[draw=drawColor,line width= 0.4pt,line join=round,line cap=round] ( 97.03, 30.00) -- ( 97.03, 32.89);

\path[draw=drawColor,line width= 0.4pt,line join=round,line cap=round] (152.88, 30.00) -- (152.88, 32.89);

\path[draw=drawColor,line width= 0.4pt,line join=round,line cap=round] (208.74, 30.00) -- (208.74, 32.89);

\path[draw=drawColor,line width= 0.4pt,line join=round,line cap=round] (264.59, 30.00) -- (264.59, 32.89);

\path[draw=drawColor,line width= 0.4pt,line join=round,line cap=round] (320.45, 30.00) -- (320.45, 32.89);

\node[text=drawColor,anchor=base,inner sep=0pt, outer sep=0pt, scale = 0.90] at ( 41.17, 18.00) {0};

\node[text=drawColor,anchor=base,inner sep=0pt, outer sep=0pt, scale = 0.90] at ( 97.03, 18.00) {5};

\node[text=drawColor,anchor=base,inner sep=0pt, outer sep=0pt, scale = 0.90] at (152.88, 18.00) {10};

\node[text=drawColor,anchor=base,inner sep=0pt, outer sep=0pt, scale = 0.90] at (208.74, 18.00) {15};

\node[text=drawColor,anchor=base,inner sep=0pt, outer sep=0pt, scale = 0.90] at (264.59, 18.00) {20};

\node[text=drawColor,anchor=base,inner sep=0pt, outer sep=0pt, scale = 0.90] at (320.45, 18.00) {25};


\path[draw=drawColor,line width= 0.4pt,line join=round,line cap=round] ( 30.00, 35.36) -- ( 32.89, 35.36);

\path[draw=drawColor,line width= 0.4pt,line join=round,line cap=round] ( 30.00, 57.68) -- ( 32.89, 57.68);

\path[draw=drawColor,line width= 0.4pt,line join=round,line cap=round] ( 30.00, 80.01) -- ( 32.89, 80.01);

\path[draw=drawColor,line width= 0.4pt,line join=round,line cap=round] ( 30.00,102.34) -- ( 32.89,102.34);

\path[draw=drawColor,line width= 0.4pt,line join=round,line cap=round] ( 30.00,124.66) -- ( 32.89,124.66);

\path[draw=drawColor,line width= 0.4pt,line join=round,line cap=round] ( 30.00,146.99) -- ( 32.89,146.99);

\path[draw=drawColor,line width= 0.4pt,line join=round,line cap=round] ( 30.00,169.32) -- ( 32.89,169.32);

\node[text=drawColor,rotate= 90.00,anchor=base,inner sep=0pt, outer sep=0pt, scale = 0.90] at ( 25.20, 35.36) {-300};

\node[text=drawColor,rotate= 90.00,anchor=base,inner sep=0pt, outer sep=0pt, scale = 0.90] at ( 25.20, 80.01) {-200};

\node[text=drawColor,rotate= 90.00,anchor=base,inner sep=0pt, outer sep=0pt, scale = 0.90] at ( 25.20,124.66) {-100};

\node[text=drawColor,rotate= 90.00,anchor=base,inner sep=0pt, outer sep=0pt, scale = 0.90] at ( 25.20,169.32) {0};

\path[draw=drawColor,line width= 0.4pt,line join=round,line cap=round, thick] ( 30.00, 30.00) --
	(331.62, 30.00) --
	(331.62,174.67) --
	( 30.00,174.67) --
	( 30.00, 30.00);
\end{scope}
\begin{scope}
\path[clip] (  0.00,  0.00) rectangle (460.00,180.67);
\definecolor[named]{drawColor}{rgb}{0.00,0.00,0.00}

\node[text=drawColor,anchor=base,inner sep=0pt, outer sep=0pt, scale = 0.90] at (180.81,  2.40+5) {Percent contamination in each variable};

\node[text=drawColor,rotate= 90.00,anchor=base,inner sep=0pt, outer sep=0pt, scale = 0.90] at (  9.60,102.34) {Entropy loss {PRIAL}};
\end{scope}
\begin{scope}
\path[clip] ( 30.00, 30.00) rectangle (331.62,174.67);
\definecolor[named]{drawColor}{rgb}{0.00,0.00,0.00}

\path[draw=orange,line width= 0.8pt,dash pattern=on 4pt off 4pt ,line join=round,line cap=round] ( 41.17,157.01) --
	( 52.34,155.41) --
	( 63.51,153.41) --
	( 74.68,151.16) --
	( 85.86,148.98) --
	( 97.03,146.09) --
	(108.20,142.48) --
	(119.37,139.85) --
	(130.54,136.05) --
	(141.71,133.19) --
	(152.88,128.55) --
	(164.05,124.31) --
	(175.22,118.97) --
	(186.40,114.35) --
	(197.57,109.55) --
	(208.74,102.90) --
	(219.91, 95.38) --
	(231.08, 88.93) --
	(242.25, 82.07) --
	(253.42, 74.88) --
	(264.59, 65.94) --
	(275.76, 58.89) --
	(286.94, 49.31) --
	(298.11, 39.41) --
	(309.28, 29.75) --
	(320.45, 21.34);

\path[draw=drawColor,line width= 0.8pt,dash pattern=on 1pt off 3pt ,line join=round,line cap=round] ( 41.17,155.30) --
	( 52.34,153.87) --
	( 63.51,152.76) --
	( 74.68,150.95) --
	( 85.86,149.14) --
	( 97.03,147.01) --
	(108.20,143.98) --
	(119.37,142.17) --
	(130.54,138.97) --
	(141.71,136.20) --
	(152.88,132.42) --
	(164.05,129.01) --
	(175.22,123.96) --
	(186.40,119.98) --
	(197.57,115.45) --
	(208.74,109.48) --
	(219.91,102.33) --
	(231.08, 96.93) --
	(242.25, 89.41) --
	(253.42, 81.96) --
	(264.59, 74.19) --
	(275.76, 66.82) --
	(286.94, 57.87) --
	(298.11, 47.98) --
	(309.28, 37.76) --
	(320.45, 29.59);

\path[draw=drawColor,line width= 0.8pt,dash pattern=on 1pt off 3pt on 4pt off 3pt ,line join=round,line cap=round] ( 41.17,164.46) --
	( 52.34,127.76) --
	( 63.51, 66.70) --
	( 74.68, 15.13) --
	( 78.33,  0.00);

\path[draw=blue,line width= 0.8pt,dash pattern=on 7pt off 3pt ,line join=round,line cap=round] ( 41.17,161.17) --
	( 52.34,159.96) --
	( 63.51,158.20) --
	( 74.68,156.16) --
	( 85.86,153.96) --
	( 97.03,151.36) --
	(108.20,147.61) --
	(119.37,144.65) --
	(130.54,140.58) --
	(141.71,136.25) --
	(152.88,130.97) --
	(164.05,124.66) --
	(175.22,116.48) --
	(186.40,107.41) --
	(197.57, 96.01) --
	(208.74, 81.35) --
	(219.91, 66.21) --
	(231.08, 51.97) --
	(242.25, 32.45) --
	(253.42, 19.13) --
	(264.59,  3.02) --
	(266.88,  0.00);

\path[draw=red,line width= 0.8pt,dash pattern=on 2pt off 2pt on 6pt off 2pt ,line join=round,line cap=round] ( 41.17,154.19) --
	( 52.34,153.11) --
	( 63.51,151.51) --
	( 74.68,149.03) --
	( 85.86,147.53) --
	( 97.03,145.49) --
	(108.20,141.71) --
	(119.37,138.73) --
	(130.54,136.08) --
	(141.71,135.45) --
	(152.88,131.80) --
	(164.05,128.96) --
	(175.22,128.74) --
	(186.40,126.90) --
	(197.57,124.91) --
	(208.74,121.35) --
	(219.91,120.57) --
	(231.08,115.89) --
	(242.25,112.74) --
	(253.42,107.51) --
	(264.59,103.33) --
	(275.76, 97.66) --
	(286.94, 91.11) --
	(298.11, 83.63) --
	(309.28, 75.62) --
	(320.45, 66.75);

\path[draw=drawColor,line width= 0.8pt,line join=round,line cap=round] ( 41.17,167.66) --
	( 52.34,108.47) --
	( 63.51, 53.89) --
	( 74.68, 10.85) --
	( 77.85,  0.00);
\end{scope}
\begin{scope}
\path[clip] (  0.00,  0.00) rectangle (460.00,180.67);
\definecolor[named]{drawColor}{rgb}{0.00,0.00,0.00}
\definecolor[named]{fillColor}{rgb}{0.00,0.00,0.00}

\node[text=drawColor,anchor=base west,inner sep=0pt, outer sep=0pt, scale = 0.90] at (367.35,159.23) {$\widetilde{P}_n$ with {NPD}};
\path[draw=red,line width= 0.8pt,dash pattern=on 2pt off 2pt on 6pt off 2pt ,line join=round,line cap=round] (340.35,162.67) -- (358.35,162.67);

\node[text=drawColor,anchor=base west,inner sep=0pt, outer sep=0pt, scale = 0.90] at (367.35,147.23-5) {$Q_n$ with {NPD}};
\path[draw=drawColor,line width= 0.8pt,dash pattern=on 1pt off 3pt ,line join=round,line cap=round] (340.35,150.67-5) -- (358.35,150.67-5);

\node[text=drawColor,anchor=base west,inner sep=0pt, outer sep=0pt, scale = 0.90] at (367.35,135.23-10) {$\tau$ with {NPD}};
\path[draw=orange,line width= 0.8pt,dash pattern=on 4pt off 4pt ,line join=round,line cap=round] (340.35,138.67-10) -- (358.35,138.67-10);

\node[text=drawColor,anchor=base west,inner sep=0pt, outer sep=0pt, scale = 0.90] at (367.35,123.23-15) {$P_n$ with {NPD}};
\path[draw=blue,line width= 0.8pt,dash pattern=on 7pt off 3pt ,line join=round,line cap=round] (340.35,126.67-15) -- (358.35,126.67-15);

\node[text=drawColor,anchor=base west,inner sep=0pt, outer sep=0pt, scale = 0.90] at (367.35,111.23-20) {{MCD}};
\path[draw=drawColor,line width= 0.8pt,dash pattern=on 1pt off 3pt on 4pt off 3pt ,line join=round,line cap=round] (340.35,102.67-25) -- (358.35,102.67-25);

\node[text=drawColor,anchor=base west,inner sep=0pt, outer sep=0pt, scale = 0.90] at (367.35, 99.23-25)  {$P_n$ with {OGK}};
\path[draw=drawColor,line width= 0.8pt,line join=round,line cap=round] (340.35,114.67-20) -- (358.35,114.67-20);

\node[text=drawColor,anchor=base west,inner sep=0pt, outer sep=0pt, scale = 0.90] at (367.35, 87.23-30) {Classical};
\path[draw=drawColor,line width= 0.8pt,line join=round,line cap=round] (340.35, 90.67-30) -- (358.35, 90.67-30);
\path[draw=drawColor,line width= 0.8pt,line join=round,line cap=round,fill=fillColor] (349.35,162.67-72-30) circle (  1.50);
\node[text=drawColor,anchor=base,inner sep=0pt, outer sep=0pt, scale = 0.90] at (260.00,  160.00) {CLIME, banded, $p=90$};
\end{scope}
\end{tikzpicture}

\begin{tikzpicture}[x=1pt,y=1pt]
\definecolor[named]{fillColor}{rgb}{1.00,1.00,1.00}
\path[use as bounding box,fill=fillColor,fill opacity=0.00] (0,0) rectangle (433.62,180.67);
\begin{scope}
\path[clip] ( 30.00, 30.00) rectangle (331.62,174.67);
\definecolor[named]{drawColor}{rgb}{0.00,0.00,0.00}

\path[draw=drawColor,line width= 0.8pt,line join=round,line cap=round] ( 41.17,169.32) --
	( 52.34, 67.66) --
	( 61.99,  0.00);
\definecolor[named]{fillColor}{rgb}{0.00,0.00,0.00}

\path[draw=drawColor,line width= 0.8pt,line join=round,line cap=round,fill=fillColor] ( 41.17,169.32) circle (  1.50);

\path[draw=drawColor,line width= 0.8pt,line join=round,line cap=round,fill=fillColor] ( 52.34, 67.66) circle (  1.50);
\end{scope}
\begin{scope}
\path[clip] (  0.00,  0.00) rectangle (460.00,180.67);
\definecolor[named]{drawColor}{rgb}{0.00,0.00,0.00}


\path[draw=drawColor,line width= 0.4pt,line join=round,line cap=round] ( 41.17, 30.00) -- ( 41.17, 32.89);

\path[draw=drawColor,line width= 0.4pt,line join=round,line cap=round] ( 97.03, 30.00) -- ( 97.03, 32.89);

\path[draw=drawColor,line width= 0.4pt,line join=round,line cap=round] (152.88, 30.00) -- (152.88, 32.89);

\path[draw=drawColor,line width= 0.4pt,line join=round,line cap=round] (208.74, 30.00) -- (208.74, 32.89);

\path[draw=drawColor,line width= 0.4pt,line join=round,line cap=round] (264.59, 30.00) -- (264.59, 32.89);

\path[draw=drawColor,line width= 0.4pt,line join=round,line cap=round] (320.45, 30.00) -- (320.45, 32.89);

\node[text=drawColor,anchor=base,inner sep=0pt, outer sep=0pt, scale = 0.90] at ( 41.17, 18.00) {0};

\node[text=drawColor,anchor=base,inner sep=0pt, outer sep=0pt, scale = 0.90] at ( 97.03, 18.00) {5};

\node[text=drawColor,anchor=base,inner sep=0pt, outer sep=0pt, scale = 0.90] at (152.88, 18.00) {10};

\node[text=drawColor,anchor=base,inner sep=0pt, outer sep=0pt, scale = 0.90] at (208.74, 18.00) {15};

\node[text=drawColor,anchor=base,inner sep=0pt, outer sep=0pt, scale = 0.90] at (264.59, 18.00) {20};

\node[text=drawColor,anchor=base,inner sep=0pt, outer sep=0pt, scale = 0.90] at (320.45, 18.00) {25};


\path[draw=drawColor,line width= 0.4pt,line join=round,line cap=round] ( 30.00, 35.36) -- ( 32.89, 35.36);

\path[draw=drawColor,line width= 0.4pt,line join=round,line cap=round] ( 30.00, 57.68) -- ( 32.89, 57.68);

\path[draw=drawColor,line width= 0.4pt,line join=round,line cap=round] ( 30.00, 80.01) -- ( 32.89, 80.01);

\path[draw=drawColor,line width= 0.4pt,line join=round,line cap=round] ( 30.00,102.34) -- ( 32.89,102.34);

\path[draw=drawColor,line width= 0.4pt,line join=round,line cap=round] ( 30.00,124.66) -- ( 32.89,124.66);

\path[draw=drawColor,line width= 0.4pt,line join=round,line cap=round] ( 30.00,146.99) -- ( 32.89,146.99);

\path[draw=drawColor,line width= 0.4pt,line join=round,line cap=round] ( 30.00,169.32) -- ( 32.89,169.32);

\node[text=drawColor,rotate= 90.00,anchor=base,inner sep=0pt, outer sep=0pt, scale = 0.90] at ( 25.20, 35.36) {-300};

\node[text=drawColor,rotate= 90.00,anchor=base,inner sep=0pt, outer sep=0pt, scale = 0.90] at ( 25.20, 80.01) {-200};
\node[text=drawColor,rotate= 90.00,anchor=base,inner sep=0pt, outer sep=0pt, scale = 0.90] at ( 25.20,124.66) {-100};

\node[text=drawColor,rotate= 90.00,anchor=base,inner sep=0pt, outer sep=0pt, scale = 0.90] at ( 25.20,169.32) {0};

\path[draw=drawColor,line width= 0.4pt,line join=round,line cap=round, thick] ( 30.00, 30.00) --
	(331.62, 30.00) --
	(331.62,174.67) --
	( 30.00,174.67) --
	( 30.00, 30.00);
\end{scope}
\begin{scope}
\path[clip] (  0.00,  0.00) rectangle (460.00,180.67);
\definecolor[named]{drawColor}{rgb}{0.00,0.00,0.00}

\node[text=drawColor,anchor=base,inner sep=0pt, outer sep=0pt, scale = 0.90] at (180.81,  2.40+5) {Percent contamination in each variable};

\node[text=drawColor,rotate= 90.00,anchor=base,inner sep=0pt, outer sep=0pt, scale = 0.90] at (  9.60,102.34) {Entropy loss {PRIAL}};
\end{scope}
\begin{scope}
\path[clip] ( 30.00, 30.00) rectangle (331.62,174.67);
\definecolor[named]{drawColor}{rgb}{0.00,0.00,0.00}

\path[draw=orange,line width= 0.8pt,dash pattern=on 4pt off 4pt ,line join=round,line cap=round] ( 41.17,155.24) --
	( 52.34,151.13) --
	( 63.51,147.42) --
	( 74.68,143.86) --
	( 85.86,139.44) --
	( 97.03,133.93) --
	(108.20,130.72) --
	(119.37,125.50) --
	(130.54,119.92) --
	(141.71,115.40) --
	(152.88,108.32) --
	(164.05,103.50) --
	(175.22, 97.69) --
	(186.40, 90.29) --
	(197.57, 84.81) --
	(208.74, 78.35) --
	(219.91, 69.90) --
	(231.08, 63.61) --
	(242.25, 56.13) --
	(253.42, 45.09) --
	(264.59, 35.92) --
	(275.76, 29.14) --
	(286.94, 19.88) --
	(298.11,  7.65) --
	(305.83,  0.00);

\path[draw=drawColor,line width= 0.8pt,dash pattern=on 1pt off 3pt ,line join=round,line cap=round] ( 41.17,151.20) --
	( 52.34,148.57) --
	( 63.51,145.71) --
	( 74.68,143.96) --
	( 85.86,140.76) --
	( 97.03,137.63) --
	(108.20,133.19) --
	(119.37,129.15) --
	(130.54,124.95) --
	(141.71,120.63) --
	(152.88,115.98) --
	(164.05,109.26) --
	(175.22,103.86) --
	(186.40, 97.22) --
	(197.57, 90.86) --
	(208.74, 83.24) --
	(219.91, 75.09) --
	(231.08, 68.23) --
	(242.25, 59.76) --
	(253.42, 50.09) --
	(264.59, 37.84) --
	(275.76, 29.85) --
	(286.94, 21.29) --
	(298.11,  8.34) --
	(305.71,  0.00);

\path[draw=drawColor,line width= 0.8pt,dash pattern=on 1pt off 3pt on 4pt off 3pt ,line join=round,line cap=round] ( 41.17,163.22) --
	( 52.34,139.92) --
	( 63.51, 85.77) --
	( 74.68, 12.31) --
	( 76.81,  0.00);

\path[draw=blue,line width= 0.8pt,dash pattern=on 7pt off 3pt ,line join=round,line cap=round] ( 41.17,160.41) --
	( 52.34,157.92) --
	( 63.51,154.58) --
	( 74.68,152.34) --
	( 85.86,148.22) --
	( 97.03,143.75) --
	(108.20,138.09) --
	(119.37,131.83) --
	(130.54,125.61) --
	(141.71,119.06) --
	(152.88,110.41) --
	(164.05,101.17) --
	(175.22, 91.97) --
	(186.40, 80.29) --
	(197.57, 66.32) --
	(208.74, 54.19) --
	(219.91, 36.89) --
	(231.08, 22.45) --
	(242.25,  5.44) --
	(245.67,  0.00);

\path[draw=red,line width= 0.8pt,dash pattern=on 2pt off 2pt on 6pt off 2pt ,line join=round,line cap=round] ( 41.17,152.93) --
	( 52.34,150.14) --
	( 63.51,147.09) --
	( 74.68,144.00) --
	( 85.86,139.96) --
	( 97.03,133.99) --
	(108.20,129.17) --
	(119.37,125.14) --
	(130.54,121.44) --
	(141.71,114.39) --
	(152.88,110.25) --
	(164.05,104.22) --
	(175.22, 97.86) --
	(186.40, 95.47) --
	(197.57, 90.97) --
	(208.74, 85.27) --
	(219.91, 81.67) --
	(231.08, 79.90) --
	(242.25, 72.73) --
	(253.42, 69.00) --
	(264.59, 60.24) --
	(275.76, 56.44) --
	(286.94, 48.20) --
	(298.11, 38.29) --
	(309.28, 31.03) --
	(320.45, 21.57);

\path[draw=drawColor,line width= 0.8pt,line join=round,line cap=round] ( 41.17,145.63) --
	( 52.34,144.63) --
	( 63.51,129.42) --
	( 74.68, 94.90) --
	( 85.86, 49.17) --
	( 96.09,  0.00);
\end{scope}
\begin{scope}
\path[clip] (  0.00,  0.00) rectangle (460.00,180.67);
\definecolor[named]{drawColor}{rgb}{0.00,0.00,0.00}
\definecolor[named]{fillColor}{rgb}{0.00,0.00,0.00}

\node[text=drawColor,anchor=base west,inner sep=0pt, outer sep=0pt, scale = 0.90] at (367.35,159.23) {$\widetilde{P}_n$ with {NPD}};
\path[draw=red,line width= 0.8pt,dash pattern=on 2pt off 2pt on 6pt off 2pt ,line join=round,line cap=round] (340.35,162.67) -- (358.35,162.67);

\node[text=drawColor,anchor=base west,inner sep=0pt, outer sep=0pt, scale = 0.90] at (367.35,147.23-5) {$Q_n$ with {NPD}};
\path[draw=drawColor,line width= 0.8pt,dash pattern=on 1pt off 3pt ,line join=round,line cap=round] (340.35,150.67-5) -- (358.35,150.67-5);

\node[text=drawColor,anchor=base west,inner sep=0pt, outer sep=0pt, scale = 0.90] at (367.35,135.23-10) {$\tau$ with {NPD}};
\path[draw=orange,line width= 0.8pt,dash pattern=on 4pt off 4pt ,line join=round,line cap=round] (340.35,138.67-10) -- (358.35,138.67-10);

\node[text=drawColor,anchor=base west,inner sep=0pt, outer sep=0pt, scale = 0.90] at (367.35,123.23-15) {$P_n$ with {NPD}};
\path[draw=blue,line width= 0.8pt,dash pattern=on 7pt off 3pt ,line join=round,line cap=round] (340.35,126.67-15) -- (358.35,126.67-15);

\node[text=drawColor,anchor=base west,inner sep=0pt, outer sep=0pt, scale = 0.90] at (367.35,111.23-20) {{MCD}};
\path[draw=drawColor,line width= 0.8pt,dash pattern=on 1pt off 3pt on 4pt off 3pt ,line join=round,line cap=round] (340.35,102.67-25) -- (358.35,102.67-25);

\node[text=drawColor,anchor=base west,inner sep=0pt, outer sep=0pt, scale = 0.90] at (367.35, 99.23-25)  {$P_n$ with {OGK}};
\path[draw=drawColor,line width= 0.8pt,line join=round,line cap=round] (340.35,114.67-20) -- (358.35,114.67-20);

\node[text=drawColor,anchor=base west,inner sep=0pt, outer sep=0pt, scale = 0.90] at (367.35, 87.23-30) {Classical};
\path[draw=drawColor,line width= 0.8pt,line join=round,line cap=round] (340.35, 90.67-30) -- (358.35, 90.67-30);
\path[draw=drawColor,line width= 0.8pt,line join=round,line cap=round,fill=fillColor] (349.35,162.67-72-30) circle (  1.50);
\node[text=drawColor,anchor=base,inner sep=0pt, outer sep=0pt, scale = 0.90] at (260.00,  160.00) {CLIME, banded, $p=30$};
\end{scope}
\end{tikzpicture}

\begin{tikzpicture}[x=1pt,y=1pt]
\definecolor[named]{fillColor}{rgb}{1.00,1.00,1.00}
\path[use as bounding box,fill=fillColor,fill opacity=0.00] (0,0) rectangle (433.62,180.67);
\begin{scope}
\path[clip] ( 30.00, 30.00) rectangle (331.62,174.67);
\definecolor[named]{drawColor}{rgb}{0.00,0.00,0.00}

\path[draw=drawColor,line width= 0.8pt,line join=round,line cap=round] ( 41.17,169.32) --
	( 52.34, 31.81) --
	( 55.48,  0.00);
\definecolor[named]{fillColor}{rgb}{0.00,0.00,0.00}

\path[draw=drawColor,line width= 0.8pt,line join=round,line cap=round,fill=fillColor] ( 41.17,169.32) circle (  1.50);

\path[draw=drawColor,line width= 0.8pt,line join=round,line cap=round,fill=fillColor] ( 52.34, 31.81) circle (  1.50);
\end{scope}
\begin{scope}
\path[clip] (  0.00,  0.00) rectangle (460.00,180.67);
\definecolor[named]{drawColor}{rgb}{0.00,0.00,0.00}


\path[draw=drawColor,line width= 0.4pt,line join=round,line cap=round] ( 41.17, 30.00) -- ( 41.17, 32.89);

\path[draw=drawColor,line width= 0.4pt,line join=round,line cap=round] ( 97.03, 30.00) -- ( 97.03, 32.89);

\path[draw=drawColor,line width= 0.4pt,line join=round,line cap=round] (152.88, 30.00) -- (152.88, 32.89);

\path[draw=drawColor,line width= 0.4pt,line join=round,line cap=round] (208.74, 30.00) -- (208.74, 32.89);

\path[draw=drawColor,line width= 0.4pt,line join=round,line cap=round] (264.59, 30.00) -- (264.59, 32.89);

\path[draw=drawColor,line width= 0.4pt,line join=round,line cap=round] (320.45, 30.00) -- (320.45, 32.89);

\node[text=drawColor,anchor=base,inner sep=0pt, outer sep=0pt, scale = 0.90] at ( 41.17, 18.00) {0};

\node[text=drawColor,anchor=base,inner sep=0pt, outer sep=0pt, scale = 0.90] at ( 97.03, 18.00) {5};

\node[text=drawColor,anchor=base,inner sep=0pt, outer sep=0pt, scale = 0.90] at (152.88, 18.00) {10};

\node[text=drawColor,anchor=base,inner sep=0pt, outer sep=0pt, scale = 0.90] at (208.74, 18.00) {15};

\node[text=drawColor,anchor=base,inner sep=0pt, outer sep=0pt, scale = 0.90] at (264.59, 18.00) {20};

\node[text=drawColor,anchor=base,inner sep=0pt, outer sep=0pt, scale = 0.90] at (320.45, 18.00) {25};


\path[draw=drawColor,line width= 0.4pt,line join=round,line cap=round] ( 30.00, 35.36) -- ( 32.89, 35.36);

\path[draw=drawColor,line width= 0.4pt,line join=round,line cap=round] ( 30.00, 57.68) -- ( 32.89, 57.68);

\path[draw=drawColor,line width= 0.4pt,line join=round,line cap=round] ( 30.00, 80.01) -- ( 32.89, 80.01);

\path[draw=drawColor,line width= 0.4pt,line join=round,line cap=round] ( 30.00,102.34) -- ( 32.89,102.34);

\path[draw=drawColor,line width= 0.4pt,line join=round,line cap=round] ( 30.00,124.66) -- ( 32.89,124.66);

\path[draw=drawColor,line width= 0.4pt,line join=round,line cap=round] ( 30.00,146.99) -- ( 32.89,146.99);

\path[draw=drawColor,line width= 0.4pt,line join=round,line cap=round] ( 30.00,169.32) -- ( 32.89,169.32);

\node[text=drawColor,rotate= 90.00,anchor=base,inner sep=0pt, outer sep=0pt, scale = 0.90] at ( 25.20, 35.36) {-300};

\node[text=drawColor,rotate= 90.00,anchor=base,inner sep=0pt, outer sep=0pt, scale = 0.90] at ( 25.20, 80.01) {-200};

\node[text=drawColor,rotate= 90.00,anchor=base,inner sep=0pt, outer sep=0pt, scale = 0.90] at ( 25.20,124.66) {-100};

\node[text=drawColor,rotate= 90.00,anchor=base,inner sep=0pt, outer sep=0pt, scale = 0.90] at ( 25.20,169.32) {0};

\path[draw=drawColor,line width= 0.4pt,line join=round,line cap=round, thick] ( 30.00, 30.00) --
	(331.62, 30.00) --
	(331.62,174.67) --
	( 30.00,174.67) --
	( 30.00, 30.00);
\end{scope}
\begin{scope}
\path[clip] (  0.00,  0.00) rectangle (460.00,180.67);
\definecolor[named]{drawColor}{rgb}{0.00,0.00,0.00}

\node[text=drawColor,anchor=base,inner sep=0pt, outer sep=0pt, scale = 0.90] at (180.81,  2.40+5) {Percent contamination in each variable};

\node[text=drawColor,rotate= 90.00,anchor=base,inner sep=0pt, outer sep=0pt, scale = 0.90] at (  9.60,102.34) {Entropy loss {PRIAL}};
\end{scope}
\begin{scope}
\path[clip] ( 30.00, 30.00) rectangle (331.62,174.67);
\definecolor[named]{drawColor}{rgb}{0.00,0.00,0.00}

\path[draw=orange,line width= 0.8pt,dash pattern=on 4pt off 4pt ,line join=round,line cap=round] ( 41.17,151.95) --
	( 52.34,147.14) --
	( 63.51,141.17) --
	( 74.68,132.73) --
	( 85.86,127.46) --
	( 97.03,120.08) --
	(108.20,113.56) --
	(119.37,107.33) --
	(130.54, 95.45) --
	(141.71, 86.80) --
	(152.88, 76.97) --
	(164.05, 68.01) --
	(175.22, 54.58) --
	(186.40, 44.99) --
	(197.57, 29.71) --
	(208.74, 21.88) --
	(219.91,  6.25) --
	(225.82,  0.00);

\path[draw=drawColor,line width= 0.8pt,dash pattern=on 1pt off 3pt ,line join=round,line cap=round] ( 41.17,145.79) --
	( 52.34,142.11) --
	( 63.51,136.79) --
	( 74.68,130.22) --
	( 85.86,125.46) --
	( 97.03,117.29) --
	(108.20,113.13) --
	(119.37,106.60) --
	(130.54, 94.53) --
	(141.71, 86.19) --
	(152.88, 77.57) --
	(164.05, 67.72) --
	(175.22, 54.64) --
	(186.40, 45.06) --
	(197.57, 28.98) --
	(208.74, 19.09) --
	(219.91,  3.61) --
	(222.51,  0.00);

\path[draw=drawColor,line width= 0.8pt,dash pattern=on 1pt off 3pt on 4pt off 3pt ,line join=round,line cap=round] ( 41.17,162.62) --
	( 52.34,145.24) --
	( 63.51,108.29) --
	( 74.68, 50.07) --
	( 82.13,  0.00);

\path[draw=blue,line width= 0.8pt,dash pattern=on 7pt off 3pt ,line join=round,line cap=round] ( 41.17,156.97) --
	( 52.34,151.74) --
	( 63.51,146.41) --
	( 74.68,137.06) --
	( 85.86,130.45) --
	( 97.03,120.42) --
	(108.20,113.20) --
	(119.37,103.51) --
	(130.54, 88.93) --
	(141.71, 77.42) --
	(152.88, 63.64) --
	(164.05, 49.31) --
	(175.22, 29.28) --
	(186.40, 13.25) --
	(192.40,  0.00);

\path[draw=red,line width= 0.8pt,dash pattern=on 2pt off 2pt on 6pt off 2pt ,line join=round,line cap=round] ( 41.17,148.65) --
	( 52.34,146.03) --
	( 63.51,144.50) --
	( 74.68,140.40) --
	( 85.86,136.67) --
	( 97.03,134.84) --
	(108.20,132.86) --
	(119.37,127.42) --
	(130.54,119.71) --
	(141.71,114.33) --
	(152.88,108.67) --
	(164.05,102.05) --
	(175.22, 94.96) --
	(186.40, 86.56) --
	(197.57, 75.30) --
	(208.74, 71.41) --
	(219.91, 58.81) --
	(231.08, 48.91) --
	(242.25, 41.08) --
	(253.42, 25.93) --
	(264.59, 16.45) --
	(275.76,  3.74) --
	(278.39,  0.00);

\path[draw=drawColor,line width= 0.8pt,line join=round,line cap=round] ( 41.17,132.19) --
	( 52.34,126.31) --
	( 63.51,120.53) --
	( 74.68,113.78) --
	( 85.86,102.34) --
	( 97.03, 83.53) --
	(108.20, 49.22) --
	(119.37, 11.11) --
	(121.21,  0.00);
\end{scope}
\begin{scope}
\path[clip] (  0.00,  0.00) rectangle (460.00,180.67);
\definecolor[named]{drawColor}{rgb}{0.00,0.00,0.00}
\definecolor[named]{fillColor}{rgb}{0.00,0.00,0.00}

\node[text=drawColor,anchor=base west,inner sep=0pt, outer sep=0pt, scale = 0.90] at (367.35,159.23) {$\widetilde{P}_n$ with {NPD}};
\path[draw=red,line width= 0.8pt,dash pattern=on 2pt off 2pt on 6pt off 2pt ,line join=round,line cap=round] (340.35,162.67) -- (358.35,162.67);

\node[text=drawColor,anchor=base west,inner sep=0pt, outer sep=0pt, scale = 0.90] at (367.35,147.23-5) {$Q_n$ with {NPD}};
\path[draw=drawColor,line width= 0.8pt,dash pattern=on 1pt off 3pt ,line join=round,line cap=round] (340.35,150.67-5) -- (358.35,150.67-5);

\node[text=drawColor,anchor=base west,inner sep=0pt, outer sep=0pt, scale = 0.90] at (367.35,135.23-10) {$\tau$ with {NPD}};
\path[draw=orange,line width= 0.8pt,dash pattern=on 4pt off 4pt ,line join=round,line cap=round] (340.35,138.67-10) -- (358.35,138.67-10);

\node[text=drawColor,anchor=base west,inner sep=0pt, outer sep=0pt, scale = 0.90] at (367.35,123.23-15) {$P_n$ with {NPD}};
\path[draw=blue,line width= 0.8pt,dash pattern=on 7pt off 3pt ,line join=round,line cap=round] (340.35,126.67-15) -- (358.35,126.67-15);

\node[text=drawColor,anchor=base west,inner sep=0pt, outer sep=0pt, scale = 0.90] at (367.35,111.23-20) {{MCD}};
\path[draw=drawColor,line width= 0.8pt,dash pattern=on 1pt off 3pt on 4pt off 3pt ,line join=round,line cap=round] (340.35,102.67-25) -- (358.35,102.67-25);

\node[text=drawColor,anchor=base west,inner sep=0pt, outer sep=0pt, scale = 0.90] at (367.35, 99.23-25)  {$P_n$ with {OGK}};
\path[draw=drawColor,line width= 0.8pt,line join=round,line cap=round] (340.35,114.67-20) -- (358.35,114.67-20);

\node[text=drawColor,anchor=base west,inner sep=0pt, outer sep=0pt, scale = 0.90] at (367.35, 87.23-30) {Classical};
\path[draw=drawColor,line width= 0.8pt,line join=round,line cap=round] (340.35, 90.67-30) -- (358.35, 90.67-30);
\path[draw=drawColor,line width= 0.8pt,line join=round,line cap=round,fill=fillColor] (349.35,162.67-72-30) circle (  1.50);

\node[text=drawColor,anchor=base,inner sep=0pt, outer sep=0pt, scale = 0.90] at (260.00,  160.00) {CLIME, banded, $p=15$};
\end{scope}

\end{tikzpicture}
\caption{PRIAL results for a selection of estimators applied to data generated with a banded precision matrix with extreme outliers for $p=90$ (top), $p=30$ (middle) and $p=15$ (bottom) using the CLIME regularisation procedure.}
\label{Increasingp}
\end{figure}
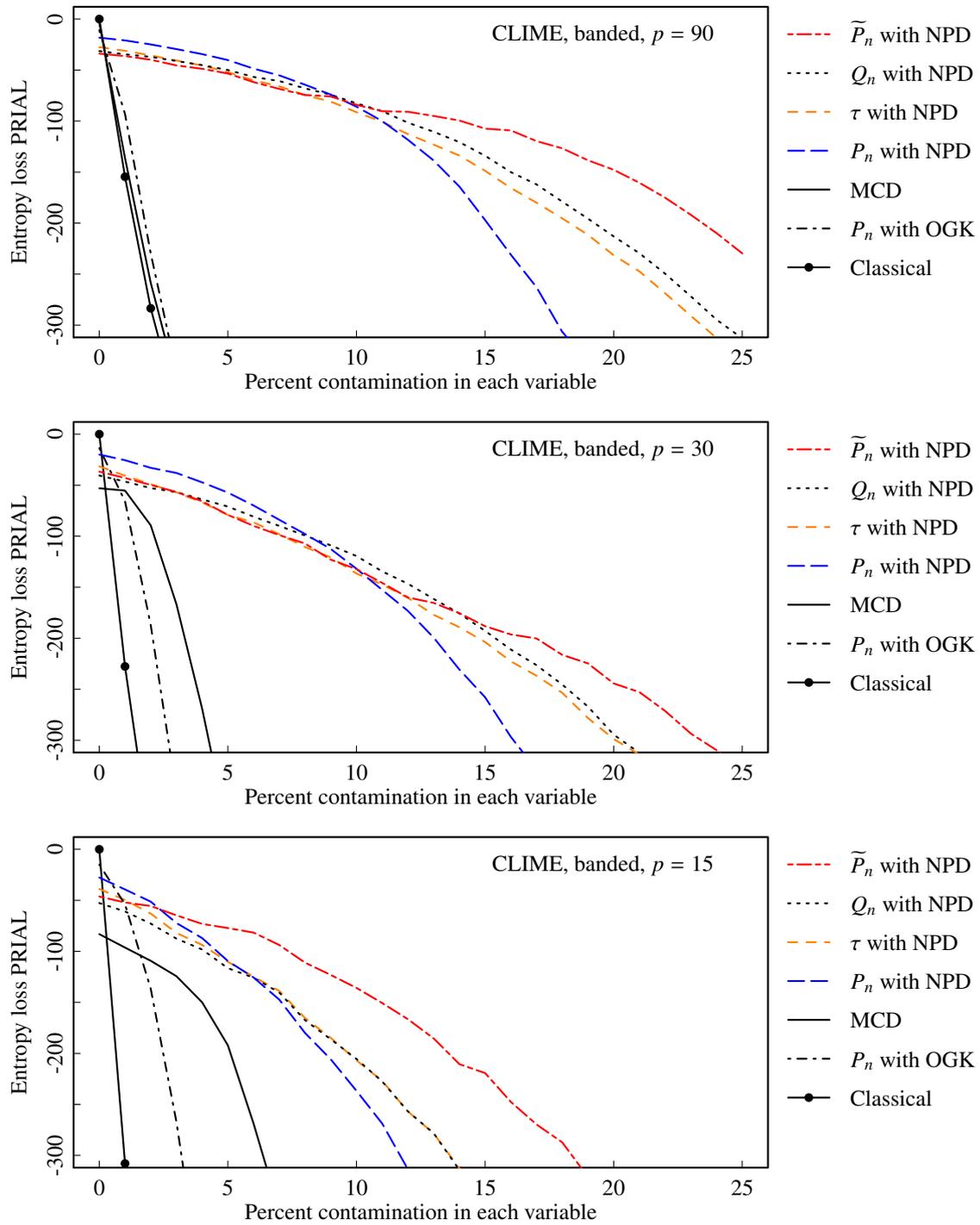

We first consider the effect of dimensionality on the performance of the various estimators.  A typical example is shown in Figure \ref{Increasingp} where we plot the PRIAL results for the precision matrix resulting from the CLIME procedure for various input covariance matrices across different amounts of extreme contamination in each variable.  The original data was generated assuming a banded precision matrix, however the trend holds true for scattered sparsity and dense precision matrices as well as for the QUIC and GLASSO procedures.

For relatively low dimensions, such as in the bottom panel of Figure \ref{Increasingp} where $p=15$, there is clearly an advantage to using the NPD method over the OGK method once there is more than a few percent of observations in each variable being contaminated.  To avoid clutter, only the OGK method with $P_n$ has been included in the plots, however, it is representative of the performance of the other scale estimators when used in conjunction with the OGK method.  

As the dimensionality increases, the OGK and the MCD methods deteriorate faster.   When $p=90$, as outlined in the previous section, the MCD method behaves like the classical method. The OGK method performs similarly poorly as outlier propagation can lead to more than half of the elements in each principle component vector being contaminated. Hence, the eigenvalues in the spectral decomposition are replaced with robust estimates of scale that may no longer be valid.

Remarkably, the NPD methods perform consistently well.  Their performance, relative to the classical method with no contamination improves as the number of variables increases.  The $P_n$ based method performs well for low levels of contamination, however once the proportion of contaminated cells is greater than 10\% it does not perform as well as the other pairwise methods due to its lower breakdown value.

It is interesting to note that the adaptively trimmed $P_n$ with adaptive trimming parameter $d=3$, $\widetilde{P}_n$, follows a somewhat different trajectory to the rest of the NPD type estimators.  It maintains a relatively high level of performance even for quite high levels of contamination.  This is due to the extreme nature of the contamination making the adaptive trimming extremely effective in identifying and excluding the errant observations.  The advantage of $\widetilde{P}_n$ is lost when the contaminating distribution has only moderately sized outliers, in which case all the NPD pairwise methods perform comparably because $\widetilde{P}_n$ does almost no trimming.

To summarise, for $p=30$, $p=60$ and $p=90$, using a pairwise method in conjunction with the NPD procedure as an input into the CLIME regularisation routine, the increase in entropy loss can be contained to less than double that of the classical method without contamination if the proportion of cellwise contamination is less than 10\%.  

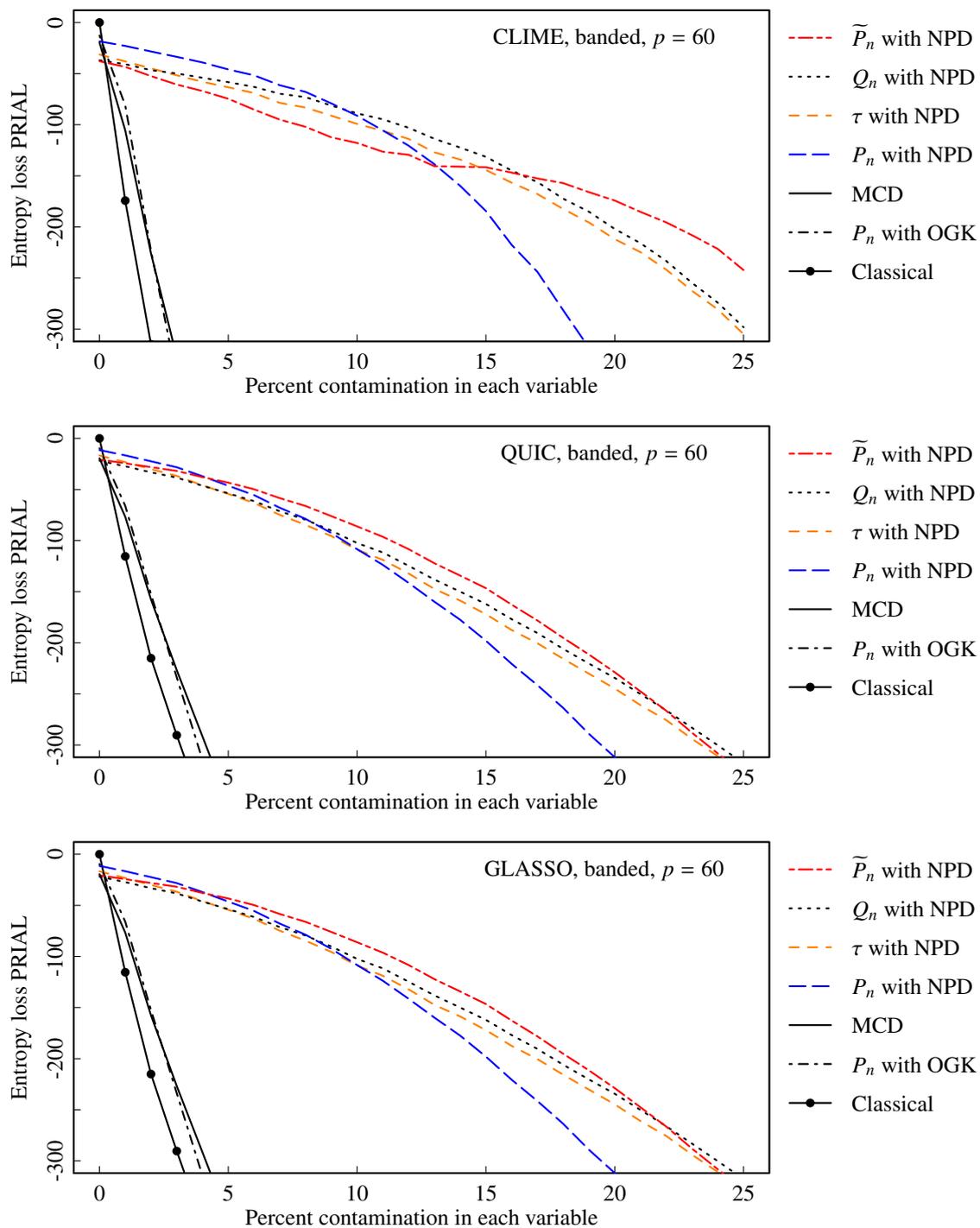
\begin{figure}[p]
\centering
\begin{tikzpicture}[x=1pt,y=1pt]
\definecolor[named]{fillColor}{rgb}{1.00,1.00,1.00}
\path[use as bounding box,fill=fillColor,fill opacity=0.00] (0,0) rectangle (433.62,180.67);
\begin{scope}
\path[clip] ( 30.00, 30.00) rectangle (331.62,174.67);
\definecolor[named]{drawColor}{rgb}{0.00,0.00,0.00}

\path[draw=drawColor,line width= 0.8pt,line join=round,line cap=round] ( 41.17,169.32) --
	( 52.34, 91.51) --
	( 63.51, 28.74) --
	( 70.36,  0.00);
\definecolor[named]{fillColor}{rgb}{0.00,0.00,0.00}

\path[draw=drawColor,line width= 0.8pt,line join=round,line cap=round,fill=fillColor] ( 41.17,169.32) circle (  1.50);

\path[draw=drawColor,line width= 0.8pt,line join=round,line cap=round,fill=fillColor] ( 52.34, 91.51) circle (  1.50);

\path[draw=drawColor,line width= 0.8pt,line join=round,line cap=round,fill=fillColor] ( 63.51, 28.74) circle (  1.50);
\end{scope}
\begin{scope}
\path[clip] (  0.00,  0.00) rectangle (460.00,180.67);
\definecolor[named]{drawColor}{rgb}{0.00,0.00,0.00}


\path[draw=drawColor,line width= 0.4pt,line join=round,line cap=round] ( 41.17, 30.00) -- ( 41.17, 32.89);

\path[draw=drawColor,line width= 0.4pt,line join=round,line cap=round] ( 97.03, 30.00) -- ( 97.03, 32.89);

\path[draw=drawColor,line width= 0.4pt,line join=round,line cap=round] (152.88, 30.00) -- (152.88, 32.89);

\path[draw=drawColor,line width= 0.4pt,line join=round,line cap=round] (208.74, 30.00) -- (208.74, 32.89);

\path[draw=drawColor,line width= 0.4pt,line join=round,line cap=round] (264.59, 30.00) -- (264.59, 32.89);

\path[draw=drawColor,line width= 0.4pt,line join=round,line cap=round] (320.45, 30.00) -- (320.45, 32.89);

\node[text=drawColor,anchor=base,inner sep=0pt, outer sep=0pt, scale = 0.90] at ( 41.17, 18.00) {0};

\node[text=drawColor,anchor=base,inner sep=0pt, outer sep=0pt, scale = 0.90] at ( 97.03, 18.00) {5};

\node[text=drawColor,anchor=base,inner sep=0pt, outer sep=0pt, scale = 0.90] at (152.88, 18.00) {10};

\node[text=drawColor,anchor=base,inner sep=0pt, outer sep=0pt, scale = 0.90] at (208.74, 18.00) {15};

\node[text=drawColor,anchor=base,inner sep=0pt, outer sep=0pt, scale = 0.90] at (264.59, 18.00) {20};

\node[text=drawColor,anchor=base,inner sep=0pt, outer sep=0pt, scale = 0.90] at (320.45, 18.00) {25};


\path[draw=drawColor,line width= 0.4pt,line join=round,line cap=round] ( 30.00, 35.36) -- ( 32.89, 35.36);

\path[draw=drawColor,line width= 0.4pt,line join=round,line cap=round] ( 30.00, 57.68) -- ( 32.89, 57.68);

\path[draw=drawColor,line width= 0.4pt,line join=round,line cap=round] ( 30.00, 80.01) -- ( 32.89, 80.01);

\path[draw=drawColor,line width= 0.4pt,line join=round,line cap=round] ( 30.00,102.34) -- ( 32.89,102.34);

\path[draw=drawColor,line width= 0.4pt,line join=round,line cap=round] ( 30.00,124.66) -- ( 32.89,124.66);

\path[draw=drawColor,line width= 0.4pt,line join=round,line cap=round] ( 30.00,146.99) -- ( 32.89,146.99);

\path[draw=drawColor,line width= 0.4pt,line join=round,line cap=round] ( 30.00,169.32) -- ( 32.89,169.32);

\node[text=drawColor,rotate= 90.00,anchor=base,inner sep=0pt, outer sep=0pt, scale = 0.90] at ( 25.20, 35.36) {-300};

\node[text=drawColor,rotate= 90.00,anchor=base,inner sep=0pt, outer sep=0pt, scale = 0.90] at ( 25.20, 80.01) {-200};

\node[text=drawColor,rotate= 90.00,anchor=base,inner sep=0pt, outer sep=0pt, scale = 0.90] at ( 25.20,124.66) {-100};

\node[text=drawColor,rotate= 90.00,anchor=base,inner sep=0pt, outer sep=0pt, scale = 0.90] at ( 25.20,169.32) {0};

\path[draw=drawColor,line width= 0.4pt,line join=round,line cap=round, thick] ( 30.00, 30.00) --
	(331.62, 30.00) --
	(331.62,174.67) --
	( 30.00,174.67) --
	( 30.00, 30.00);
\end{scope}
\begin{scope}
\path[clip] (  0.00,  0.00) rectangle (460.00,180.67);
\definecolor[named]{drawColor}{rgb}{0.00,0.00,0.00}

\node[text=drawColor,anchor=base,inner sep=0pt, outer sep=0pt, scale = 0.90] at (180.81,  2.40+5) {Percent contamination in each variable};

\node[text=drawColor,rotate= 90.00,anchor=base,inner sep=0pt, outer sep=0pt, scale = 0.90] at (  9.60,102.34) {Entropy loss {PRIAL}};
\end{scope}
\begin{scope}
\path[clip] ( 30.00, 30.00) rectangle (331.62,174.67);
\definecolor[named]{drawColor}{rgb}{0.00,0.00,0.00}

\path[draw=orange,line width= 0.8pt,dash pattern=on 4pt off 4pt ,line join=round,line cap=round] ( 41.17,155.33) --
	( 52.34,152.33) --
	( 63.51,149.30) --
	( 74.68,146.25) --
	( 85.86,143.42) --
	( 97.03,141.02) --
	(108.20,138.51) --
	(119.37,134.21) --
	(130.54,132.16) --
	(141.71,128.60) --
	(152.88,125.04) --
	(164.05,121.99) --
	(175.22,118.46) --
	(186.40,112.59) --
	(197.57,109.46) --
	(208.74,104.90) --
	(219.91, 99.31) --
	(231.08, 94.37) --
	(242.25, 87.83) --
	(253.42, 81.94) --
	(264.59, 74.68) --
	(275.76, 68.98) --
	(286.94, 61.35) --
	(298.11, 52.07) --
	(309.28, 43.98) --
	(320.45, 33.17);

\path[draw=drawColor,line width= 0.8pt,dash pattern=on 1pt off 3pt ,line join=round,line cap=round] ( 41.17,152.82) --
	( 52.34,151.12) --
	( 63.51,148.50) --
	( 74.68,147.06) --
	( 85.86,145.19) --
	( 97.03,143.26) --
	(108.20,141.20) --
	(119.37,138.19) --
	(130.54,136.72) --
	(141.71,133.03) --
	(152.88,129.64) --
	(164.05,127.03) --
	(175.22,123.34) --
	(186.40,118.52) --
	(197.57,114.68) --
	(208.74,110.63) --
	(219.91,104.61) --
	(231.08, 99.62) --
	(242.25, 92.35) --
	(253.42, 86.65) --
	(264.59, 79.12) --
	(275.76, 72.90) --
	(286.94, 65.10) --
	(298.11, 55.57) --
	(309.28, 46.89) --
	(320.45, 36.19);

\path[draw=drawColor,line width= 0.8pt,dash pattern=on 1pt off 3pt on 4pt off 3pt ,line join=round,line cap=round] ( 41.17,163.63) --
	( 52.34,133.10) --
	( 63.51, 69.63) --
	( 74.68, 14.27) --
	( 77.68,  0.00);

\path[draw=blue,line width= 0.8pt,dash pattern=on 7pt off 3pt ,line join=round,line cap=round] ( 41.17,161.09) --
	( 52.34,159.09) --
	( 63.51,156.67) --
	( 74.68,154.32) --
	( 85.86,151.87) --
	( 97.03,148.82) --
	(108.20,146.25) --
	(119.37,141.92) --
	(130.54,139.00) --
	(141.71,133.98) --
	(152.88,128.60) --
	(164.05,122.18) --
	(175.22,115.51) --
	(186.40,107.42) --
	(197.57, 97.96) --
	(208.74, 86.94) --
	(219.91, 72.17) --
	(231.08, 60.37) --
	(242.25, 43.61) --
	(253.42, 27.21) --
	(264.59,  9.20) --
	(271.60,  0.00);

\path[draw=red,line width= 0.8pt,dash pattern=on 2pt off 2pt on 6pt off 2pt ,line join=round,line cap=round] ( 41.17,152.41) --
	( 52.34,149.94) --
	( 63.51,145.91) --
	( 74.68,142.23) --
	( 85.86,139.39) --
	( 97.03,136.00) --
	(108.20,131.26) --
	(119.37,126.89) --
	(130.54,123.74) --
	(141.71,119.16) --
	(152.88,116.70) --
	(164.05,112.86) --
	(175.22,111.53) --
	(186.40,106.55) --
	(197.57,106.34) --
	(208.74,106.08) --
	(219.91,103.74) --
	(231.08,101.19) --
	(242.25, 99.14) --
	(253.42, 95.20) --
	(264.59, 91.54) --
	(275.76, 86.58) --
	(286.94, 81.98) --
	(298.11, 76.39) --
	(309.28, 70.39) --
	(320.45, 61.15);

\path[draw=drawColor,line width= 0.8pt,line join=round,line cap=round] ( 41.17,160.60) --
	( 52.34,122.20) --
	( 63.51, 68.40) --
	( 74.68, 23.25) --
	( 80.89,  0.00);
\end{scope}
\begin{scope}
\path[clip] (  0.00,  0.00) rectangle (460.00,180.67);
\definecolor[named]{drawColor}{rgb}{0.00,0.00,0.00}
\definecolor[named]{fillColor}{rgb}{0.00,0.00,0.00}

\node[text=drawColor,anchor=base west,inner sep=0pt, outer sep=0pt, scale = 0.90] at (367.35,159.23) {$\widetilde{P}_n$ with {NPD}};
\path[draw=red,line width= 0.8pt,dash pattern=on 2pt off 2pt on 6pt off 2pt ,line join=round,line cap=round] (340.35,162.67) -- (358.35,162.67);

\node[text=drawColor,anchor=base west,inner sep=0pt, outer sep=0pt, scale = 0.90] at (367.35,147.23-5) {$Q_n$ with {NPD}};
\path[draw=drawColor,line width= 0.8pt,dash pattern=on 1pt off 3pt ,line join=round,line cap=round] (340.35,150.67-5) -- (358.35,150.67-5);

\node[text=drawColor,anchor=base west,inner sep=0pt, outer sep=0pt, scale = 0.90] at (367.35,135.23-10) {$\tau$ with {NPD}};
\path[draw=orange,line width= 0.8pt,dash pattern=on 4pt off 4pt ,line join=round,line cap=round] (340.35,138.67-10) -- (358.35,138.67-10);

\node[text=drawColor,anchor=base west,inner sep=0pt, outer sep=0pt, scale = 0.90] at (367.35,123.23-15) {$P_n$ with {NPD}};
\path[draw=blue,line width= 0.8pt,dash pattern=on 7pt off 3pt ,line join=round,line cap=round] (340.35,126.67-15) -- (358.35,126.67-15);

\node[text=drawColor,anchor=base west,inner sep=0pt, outer sep=0pt, scale = 0.90] at (367.35,111.23-20) {{MCD}};
\path[draw=drawColor,line width= 0.8pt,dash pattern=on 1pt off 3pt on 4pt off 3pt ,line join=round,line cap=round] (340.35,102.67-25) -- (358.35,102.67-25);

\node[text=drawColor,anchor=base west,inner sep=0pt, outer sep=0pt, scale = 0.90] at (367.35, 99.23-25)  {$P_n$ with {OGK}};
\path[draw=drawColor,line width= 0.8pt,line join=round,line cap=round] (340.35,114.67-20) -- (358.35,114.67-20);

\node[text=drawColor,anchor=base west,inner sep=0pt, outer sep=0pt, scale = 0.90] at (367.35, 87.23-30) {Classical};
\path[draw=drawColor,line width= 0.8pt,line join=round,line cap=round] (340.35, 90.67-30) -- (358.35, 90.67-30);
\path[draw=drawColor,line width= 0.8pt,line join=round,line cap=round,fill=fillColor] (349.35,162.67-72-30) circle (  1.50);
\node[text=drawColor,anchor=base,inner sep=0pt, outer sep=0pt, scale = 0.90] at (260.00,  160.00) {CLIME, banded, $p=60$};
\end{scope}
\end{tikzpicture}

\begin{tikzpicture}[x=1pt,y=1pt]
\definecolor[named]{fillColor}{rgb}{1.00,1.00,1.00}
\path[use as bounding box,fill=fillColor,fill opacity=0.00] (0,0) rectangle (433.62,180.67);
\begin{scope}
\path[clip] ( 30.00, 30.00) rectangle (331.62,174.67);
\definecolor[named]{drawColor}{rgb}{0.00,0.00,0.00}

\path[draw=drawColor,line width= 0.8pt,line join=round,line cap=round] ( 41.17,169.32) --
	( 52.34,117.76) --
	( 63.51, 73.29) --
	( 74.68, 39.65) --
	( 85.86,  7.94) --
	( 88.82,  0.00);
\definecolor[named]{fillColor}{rgb}{0.00,0.00,0.00}

\path[draw=drawColor,line width= 0.8pt,line join=round,line cap=round,fill=fillColor] ( 41.17,169.32) circle (  1.50);

\path[draw=drawColor,line width= 0.8pt,line join=round,line cap=round,fill=fillColor] ( 52.34,117.76) circle (  1.50);

\path[draw=drawColor,line width= 0.8pt,line join=round,line cap=round,fill=fillColor] ( 63.51, 73.29) circle (  1.50);

\path[draw=drawColor,line width= 0.8pt,line join=round,line cap=round,fill=fillColor] ( 74.68, 39.65) circle (  1.50);

\path[draw=drawColor,line width= 0.8pt,line join=round,line cap=round,fill=fillColor] ( 85.86,  7.94) circle (  1.50);
\end{scope}
\begin{scope}
\path[clip] (  0.00,  0.00) rectangle (460.00,180.67);
\definecolor[named]{drawColor}{rgb}{0.00,0.00,0.00}


\path[draw=drawColor,line width= 0.4pt,line join=round,line cap=round] ( 41.17, 30.00) -- ( 41.17, 32.89);

\path[draw=drawColor,line width= 0.4pt,line join=round,line cap=round] ( 97.03, 30.00) -- ( 97.03, 32.89);

\path[draw=drawColor,line width= 0.4pt,line join=round,line cap=round] (152.88, 30.00) -- (152.88, 32.89);

\path[draw=drawColor,line width= 0.4pt,line join=round,line cap=round] (208.74, 30.00) -- (208.74, 32.89);

\path[draw=drawColor,line width= 0.4pt,line join=round,line cap=round] (264.59, 30.00) -- (264.59, 32.89);

\path[draw=drawColor,line width= 0.4pt,line join=round,line cap=round] (320.45, 30.00) -- (320.45, 32.89);

\node[text=drawColor,anchor=base,inner sep=0pt, outer sep=0pt, scale = 0.90] at ( 41.17, 18.00) {0};

\node[text=drawColor,anchor=base,inner sep=0pt, outer sep=0pt, scale = 0.90] at ( 97.03, 18.00) {5};

\node[text=drawColor,anchor=base,inner sep=0pt, outer sep=0pt, scale = 0.90] at (152.88, 18.00) {10};

\node[text=drawColor,anchor=base,inner sep=0pt, outer sep=0pt, scale = 0.90] at (208.74, 18.00) {15};

\node[text=drawColor,anchor=base,inner sep=0pt, outer sep=0pt, scale = 0.90] at (264.59, 18.00) {20};

\node[text=drawColor,anchor=base,inner sep=0pt, outer sep=0pt, scale = 0.90] at (320.45, 18.00) {25};


\path[draw=drawColor,line width= 0.4pt,line join=round,line cap=round] ( 30.00, 35.36) -- ( 32.89, 35.36);

\path[draw=drawColor,line width= 0.4pt,line join=round,line cap=round] ( 30.00, 57.68) -- ( 32.89, 57.68);

\path[draw=drawColor,line width= 0.4pt,line join=round,line cap=round] ( 30.00, 80.01) -- ( 32.89, 80.01);

\path[draw=drawColor,line width= 0.4pt,line join=round,line cap=round] ( 30.00,102.34) -- ( 32.89,102.34);

\path[draw=drawColor,line width= 0.4pt,line join=round,line cap=round] ( 30.00,124.66) -- ( 32.89,124.66);

\path[draw=drawColor,line width= 0.4pt,line join=round,line cap=round] ( 30.00,146.99) -- ( 32.89,146.99);

\path[draw=drawColor,line width= 0.4pt,line join=round,line cap=round] ( 30.00,169.32) -- ( 32.89,169.32);

\node[text=drawColor,rotate= 90.00,anchor=base,inner sep=0pt, outer sep=0pt, scale = 0.90] at ( 25.20, 35.36) {-300};

\node[text=drawColor,rotate= 90.00,anchor=base,inner sep=0pt, outer sep=0pt, scale = 0.90] at ( 25.20, 80.01) {-200};

\node[text=drawColor,rotate= 90.00,anchor=base,inner sep=0pt, outer sep=0pt, scale = 0.90] at ( 25.20,124.66) {-100};

\node[text=drawColor,rotate= 90.00,anchor=base,inner sep=0pt, outer sep=0pt, scale = 0.90] at ( 25.20,169.32) {0};

\path[draw=drawColor,line width= 0.4pt,line join=round,line cap=round, thick] ( 30.00, 30.00) --
	(331.62, 30.00) --
	(331.62,174.67) --
	( 30.00,174.67) --
	( 30.00, 30.00);
\end{scope}
\begin{scope}
\path[clip] (  0.00,  0.00) rectangle (460.00,180.67);
\definecolor[named]{drawColor}{rgb}{0.00,0.00,0.00}

\node[text=drawColor,anchor=base,inner sep=0pt, outer sep=0pt, scale = 0.90] at (180.81,  2.40+5) {Percent contamination in each variable};

\node[text=drawColor,rotate= 90.00,anchor=base,inner sep=0pt, outer sep=0pt, scale = 0.90] at (  9.60,102.34) {Entropy loss {PRIAL}};
\end{scope}
\begin{scope}
\path[clip] ( 30.00, 30.00) rectangle (331.62,174.67);
\definecolor[named]{drawColor}{rgb}{0.00,0.00,0.00}

\path[draw=orange,line width= 0.8pt,dash pattern=on 4pt off 4pt ,line join=round,line cap=round] ( 41.17,161.84) --
	( 52.34,158.94) --
	( 63.51,155.89) --
	( 74.68,152.92) --
	( 85.86,148.80) --
	( 97.03,145.05) --
	(108.20,141.10) --
	(119.37,135.90) --
	(130.54,131.52) --
	(141.71,126.63) --
	(152.88,120.67) --
	(164.05,116.22) --
	(175.22,110.29) --
	(186.40,103.51) --
	(197.57, 98.44) --
	(208.74, 92.47) --
	(219.91, 85.58) --
	(231.08, 79.81) --
	(242.25, 73.00) --
	(253.42, 66.50) --
	(264.59, 60.05) --
	(275.76, 52.76) --
	(286.94, 46.23) --
	(298.11, 38.03) --
	(309.28, 30.50) --
	(320.45, 22.86);

\path[draw=drawColor,line width= 0.8pt,dash pattern=on 1pt off 3pt ,line join=round,line cap=round] ( 41.17,159.54) --
	( 52.34,157.14) --
	( 63.51,154.48) --
	( 74.68,152.18) --
	( 85.86,148.62) --
	( 97.03,145.30) --
	(108.20,141.88) --
	(119.37,137.46) --
	(130.54,133.80) --
	(141.71,129.00) --
	(152.88,123.59) --
	(164.05,119.45) --
	(175.22,113.62) --
	(186.40,107.64) --
	(197.57,102.35) --
	(208.74, 96.96) --
	(219.91, 90.31) --
	(231.08, 84.26) --
	(242.25, 77.35) --
	(253.42, 70.92) --
	(264.59, 64.68) --
	(275.76, 57.60) --
	(286.94, 50.48) --
	(298.11, 42.89) --
	(309.28, 35.24) --
	(320.45, 27.35);

\path[draw=drawColor,line width= 0.8pt,dash pattern=on 1pt off 3pt on 4pt off 3pt ,line join=round,line cap=round] ( 41.17,165.02) --
	( 52.34,139.85) --
	( 63.51,101.25) --
	( 74.68, 64.96) --
	( 85.86, 28.81) --
	( 95.25,  0.00);

\path[draw=blue,line width= 0.8pt,dash pattern=on 7pt off 3pt ,line join=round,line cap=round] ( 41.17,164.23) --
	( 52.34,161.89) --
	( 63.51,159.30) --
	( 74.68,156.70) --
	( 85.86,152.76) --
	( 97.03,148.63) --
	(108.20,144.53) --
	(119.37,139.00) --
	(130.54,134.14) --
	(141.71,127.97) --
	(152.88,120.91) --
	(164.05,114.04) --
	(175.22,106.17) --
	(186.40, 97.95) --
	(197.57, 90.07) --
	(208.74, 80.89) --
	(219.91, 70.80) --
	(231.08, 61.50) --
	(242.25, 51.55) --
	(253.42, 40.20) --
	(264.59, 30.05) --
	(275.76, 18.35) --
	(286.94,  8.31) --
	(295.19,  0.00);

\path[draw=red,line width= 0.8pt,dash pattern=on 2pt off 2pt on 6pt off 2pt ,line join=round,line cap=round] ( 41.17,160.09) --
	( 52.34,158.43) --
	( 63.51,156.75) --
	( 74.68,155.10) --
	( 85.86,152.38) --
	( 97.03,149.98) --
	(108.20,147.09) --
	(119.37,143.20) --
	(130.54,139.77) --
	(141.71,135.35) --
	(152.88,130.86) --
	(164.05,126.31) --
	(175.22,120.91) --
	(186.40,114.75) --
	(197.57,109.36) --
	(208.74,103.86) --
	(219.91, 96.65) --
	(231.08, 89.74) --
	(242.25, 82.16) --
	(253.42, 74.88) --
	(264.59, 67.13) --
	(275.76, 58.77) --
	(286.94, 50.32) --
	(298.11, 40.71) --
	(309.28, 31.68) --
	(320.45, 22.61);

\path[draw=drawColor,line width= 0.8pt,line join=round,line cap=round] ( 41.17,160.62) --
	( 52.34,135.06) --
	( 63.51, 98.92) --
	( 74.68, 68.09) --
	( 85.86, 39.06) --
	( 97.03,  9.81) --
	(102.11,  0.00);
\end{scope}
\begin{scope}
\path[clip] (  0.00,  0.00) rectangle (460.00,180.67);
\definecolor[named]{drawColor}{rgb}{0.00,0.00,0.00}
\definecolor[named]{fillColor}{rgb}{0.00,0.00,0.00}

\node[text=drawColor,anchor=base west,inner sep=0pt, outer sep=0pt, scale = 0.90] at (367.35,159.23) {$\widetilde{P}_n$ with {NPD}};
\path[draw=red,line width= 0.8pt,dash pattern=on 2pt off 2pt on 6pt off 2pt ,line join=round,line cap=round] (340.35,162.67) -- (358.35,162.67);

\node[text=drawColor,anchor=base west,inner sep=0pt, outer sep=0pt, scale = 0.90] at (367.35,147.23-5) {$Q_n$ with {NPD}};
\path[draw=drawColor,line width= 0.8pt,dash pattern=on 1pt off 3pt ,line join=round,line cap=round] (340.35,150.67-5) -- (358.35,150.67-5);

\node[text=drawColor,anchor=base west,inner sep=0pt, outer sep=0pt, scale = 0.90] at (367.35,135.23-10) {$\tau$ with {NPD}};
\path[draw=orange,line width= 0.8pt,dash pattern=on 4pt off 4pt ,line join=round,line cap=round] (340.35,138.67-10) -- (358.35,138.67-10);

\node[text=drawColor,anchor=base west,inner sep=0pt, outer sep=0pt, scale = 0.90] at (367.35,123.23-15) {$P_n$ with {NPD}};
\path[draw=blue,line width= 0.8pt,dash pattern=on 7pt off 3pt ,line join=round,line cap=round] (340.35,126.67-15) -- (358.35,126.67-15);

\node[text=drawColor,anchor=base west,inner sep=0pt, outer sep=0pt, scale = 0.90] at (367.35,111.23-20) {{MCD}};
\path[draw=drawColor,line width= 0.8pt,dash pattern=on 1pt off 3pt on 4pt off 3pt ,line join=round,line cap=round] (340.35,102.67-25) -- (358.35,102.67-25);

\node[text=drawColor,anchor=base west,inner sep=0pt, outer sep=0pt, scale = 0.90] at (367.35, 99.23-25)  {$P_n$ with {OGK}};
\path[draw=drawColor,line width= 0.8pt,line join=round,line cap=round] (340.35,114.67-20) -- (358.35,114.67-20);

\node[text=drawColor,anchor=base west,inner sep=0pt, outer sep=0pt, scale = 0.90] at (367.35, 87.23-30) {Classical};
\path[draw=drawColor,line width= 0.8pt,line join=round,line cap=round] (340.35, 90.67-30) -- (358.35, 90.67-30);
\path[draw=drawColor,line width= 0.8pt,line join=round,line cap=round,fill=fillColor] (349.35,162.67-72-30) circle (  1.50);
\node[text=drawColor,anchor=base,inner sep=0pt, outer sep=0pt, scale = 0.90] at (260.00,  160.00) {QUIC, banded, $p=60$};
\end{scope}
\end{tikzpicture}

\begin{tikzpicture}[x=1pt,y=1pt]
\definecolor[named]{fillColor}{rgb}{1.00,1.00,1.00}
\path[use as bounding box,fill=fillColor,fill opacity=0.00] (0,0) rectangle (433.62,180.67);
\begin{scope}
\path[clip] ( 30.00, 30.00) rectangle (331.62,174.67);
\definecolor[named]{drawColor}{rgb}{0.00,0.00,0.00}

\path[draw=drawColor,line width= 0.8pt,line join=round,line cap=round] ( 41.17,169.32) --
	( 52.34,117.76) --
	( 63.51, 73.29) --
	( 74.68, 39.65) --
	( 85.86,  7.94) --
	( 88.82,  0.00);
\definecolor[named]{fillColor}{rgb}{0.00,0.00,0.00}

\path[draw=drawColor,line width= 0.8pt,line join=round,line cap=round,fill=fillColor] ( 41.17,169.32) circle (  1.50);

\path[draw=drawColor,line width= 0.8pt,line join=round,line cap=round,fill=fillColor] ( 52.34,117.76) circle (  1.50);

\path[draw=drawColor,line width= 0.8pt,line join=round,line cap=round,fill=fillColor] ( 63.51, 73.29) circle (  1.50);

\path[draw=drawColor,line width= 0.8pt,line join=round,line cap=round,fill=fillColor] ( 74.68, 39.65) circle (  1.50);

\path[draw=drawColor,line width= 0.8pt,line join=round,line cap=round,fill=fillColor] ( 85.86,  7.94) circle (  1.50);
\end{scope}
\begin{scope}
\path[clip] (  0.00,  0.00) rectangle (460.00,180.67);
\definecolor[named]{drawColor}{rgb}{0.00,0.00,0.00}


\path[draw=drawColor,line width= 0.4pt,line join=round,line cap=round] ( 41.17, 30.00) -- ( 41.17, 32.89);

\path[draw=drawColor,line width= 0.4pt,line join=round,line cap=round] ( 97.03, 30.00) -- ( 97.03, 32.89);

\path[draw=drawColor,line width= 0.4pt,line join=round,line cap=round] (152.88, 30.00) -- (152.88, 32.89);

\path[draw=drawColor,line width= 0.4pt,line join=round,line cap=round] (208.74, 30.00) -- (208.74, 32.89);

\path[draw=drawColor,line width= 0.4pt,line join=round,line cap=round] (264.59, 30.00) -- (264.59, 32.89);

\path[draw=drawColor,line width= 0.4pt,line join=round,line cap=round] (320.45, 30.00) -- (320.45, 32.89);

\node[text=drawColor,anchor=base,inner sep=0pt, outer sep=0pt, scale = 0.90] at ( 41.17, 18.00) {0};

\node[text=drawColor,anchor=base,inner sep=0pt, outer sep=0pt, scale = 0.90] at ( 97.03, 18.00) {5};

\node[text=drawColor,anchor=base,inner sep=0pt, outer sep=0pt, scale = 0.90] at (152.88, 18.00) {10};

\node[text=drawColor,anchor=base,inner sep=0pt, outer sep=0pt, scale = 0.90] at (208.74, 18.00) {15};

\node[text=drawColor,anchor=base,inner sep=0pt, outer sep=0pt, scale = 0.90] at (264.59, 18.00) {20};

\node[text=drawColor,anchor=base,inner sep=0pt, outer sep=0pt, scale = 0.90] at (320.45, 18.00) {25};


\path[draw=drawColor,line width= 0.4pt,line join=round,line cap=round] ( 30.00, 35.36) -- ( 32.89, 35.36);

\path[draw=drawColor,line width= 0.4pt,line join=round,line cap=round] ( 30.00, 57.68) -- ( 32.89, 57.68);

\path[draw=drawColor,line width= 0.4pt,line join=round,line cap=round] ( 30.00, 80.01) -- ( 32.89, 80.01);

\path[draw=drawColor,line width= 0.4pt,line join=round,line cap=round] ( 30.00,102.34) -- ( 32.89,102.34);

\path[draw=drawColor,line width= 0.4pt,line join=round,line cap=round] ( 30.00,124.66) -- ( 32.89,124.66);

\path[draw=drawColor,line width= 0.4pt,line join=round,line cap=round] ( 30.00,146.99) -- ( 32.89,146.99);

\path[draw=drawColor,line width= 0.4pt,line join=round,line cap=round] ( 30.00,169.32) -- ( 32.89,169.32);

\node[text=drawColor,rotate= 90.00,anchor=base,inner sep=0pt, outer sep=0pt, scale = 0.90] at ( 25.20, 35.36) {-300};

\node[text=drawColor,rotate= 90.00,anchor=base,inner sep=0pt, outer sep=0pt, scale = 0.90] at ( 25.20, 80.01) {-200};

\node[text=drawColor,rotate= 90.00,anchor=base,inner sep=0pt, outer sep=0pt, scale = 0.90] at ( 25.20,124.66) {-100};

\node[text=drawColor,rotate= 90.00,anchor=base,inner sep=0pt, outer sep=0pt, scale = 0.90] at ( 25.20,169.32) {0};

\path[draw=drawColor,line width= 0.4pt,line join=round,line cap=round, thick] ( 30.00, 30.00) --
	(331.62, 30.00) --
	(331.62,174.67) --
	( 30.00,174.67) --
	( 30.00, 30.00);
\end{scope}
\begin{scope}
\path[clip] (  0.00,  0.00) rectangle (460.00,180.67);
\definecolor[named]{drawColor}{rgb}{0.00,0.00,0.00}

\node[text=drawColor,anchor=base,inner sep=0pt, outer sep=0pt, scale = 0.90] at (180.81,  2.40+5) {Percent contamination in each variable};

\node[text=drawColor,rotate= 90.00,anchor=base,inner sep=0pt, outer sep=0pt, scale = 0.90] at (  9.60,102.34) {Entropy loss {PRIAL}};
\end{scope}
\begin{scope}
\path[clip] ( 30.00, 30.00) rectangle (331.62,174.67);
\definecolor[named]{drawColor}{rgb}{0.00,0.00,0.00}

\path[draw=orange,line width= 0.8pt,dash pattern=on 4pt off 4pt ,line join=round,line cap=round] ( 41.17,161.84) --
	( 52.34,158.94) --
	( 63.51,155.89) --
	( 74.68,152.92) --
	( 85.86,148.80) --
	( 97.03,145.05) --
	(108.20,141.10) --
	(119.37,135.90) --
	(130.54,131.52) --
	(141.71,126.63) --
	(152.88,120.67) --
	(164.05,116.22) --
	(175.22,110.29) --
	(186.40,103.51) --
	(197.57, 98.44) --
	(208.74, 92.47) --
	(219.91, 85.58) --
	(231.08, 79.81) --
	(242.25, 73.00) --
	(253.42, 66.50) --
	(264.59, 60.05) --
	(275.76, 52.76) --
	(286.94, 46.23) --
	(298.11, 38.03) --
	(309.28, 30.50) --
	(320.45, 22.86);

\path[draw=drawColor,line width= 0.8pt,dash pattern=on 1pt off 3pt ,line join=round,line cap=round] ( 41.17,159.54) --
	( 52.34,157.14) --
	( 63.51,154.48) --
	( 74.68,152.18) --
	( 85.86,148.62) --
	( 97.03,145.30) --
	(108.20,141.88) --
	(119.37,137.46) --
	(130.54,133.80) --
	(141.71,129.00) --
	(152.88,123.59) --
	(164.05,119.45) --
	(175.22,113.62) --
	(186.40,107.64) --
	(197.57,102.35) --
	(208.74, 96.96) --
	(219.91, 90.31) --
	(231.08, 84.26) --
	(242.25, 77.35) --
	(253.42, 70.92) --
	(264.59, 64.68) --
	(275.76, 57.60) --
	(286.94, 50.48) --
	(298.11, 42.89) --
	(309.28, 35.24) --
	(320.45, 27.35);

\path[draw=drawColor,line width= 0.8pt,dash pattern=on 1pt off 3pt on 4pt off 3pt ,line join=round,line cap=round] ( 41.17,165.02) --
	( 52.34,139.85) --
	( 63.51,101.25) --
	( 74.68, 64.96) --
	( 85.86, 28.81) --
	( 95.25,  0.00);

\path[draw=blue,line width= 0.8pt,dash pattern=on 7pt off 3pt ,line join=round,line cap=round] ( 41.17,164.23) --
	( 52.34,161.89) --
	( 63.51,159.30) --
	( 74.68,156.70) --
	( 85.86,152.76) --
	( 97.03,148.63) --
	(108.20,144.53) --
	(119.37,139.00) --
	(130.54,134.14) --
	(141.71,127.97) --
	(152.88,120.91) --
	(164.05,114.04) --
	(175.22,106.17) --
	(186.40, 97.95) --
	(197.57, 90.07) --
	(208.74, 80.89) --
	(219.91, 70.80) --
	(231.08, 61.50) --
	(242.25, 51.55) --
	(253.42, 40.20) --
	(264.59, 30.05) --
	(275.76, 18.35) --
	(286.94,  8.31) --
	(295.19,  0.00);

\path[draw=red,line width= 0.8pt,dash pattern=on 2pt off 2pt on 6pt off 2pt ,line join=round,line cap=round] ( 41.17,160.09) --
	( 52.34,158.43) --
	( 63.51,156.75) --
	( 74.68,155.10) --
	( 85.86,152.38) --
	( 97.03,149.98) --
	(108.20,147.09) --
	(119.37,143.20) --
	(130.54,139.77) --
	(141.71,135.35) --
	(152.88,130.86) --
	(164.05,126.31) --
	(175.22,120.91) --
	(186.40,114.75) --
	(197.57,109.36) --
	(208.74,103.86) --
	(219.91, 96.65) --
	(231.08, 89.74) --
	(242.25, 82.16) --
	(253.42, 74.88) --
	(264.59, 67.13) --
	(275.76, 58.77) --
	(286.94, 50.32) --
	(298.11, 40.70) --
	(309.28, 31.68) --
	(320.45, 22.61);

\path[draw=drawColor,line width= 0.8pt,line join=round,line cap=round] ( 41.17,160.62) --
	( 52.34,135.06) --
	( 63.51, 98.92) --
	( 74.68, 68.09) --
	( 85.86, 39.06) --
	( 97.03,  9.81) --
	(102.11,  0.00);
\end{scope}
\begin{scope}
\path[clip] (  0.00,  0.00) rectangle (460.00,180.67);
\definecolor[named]{drawColor}{rgb}{0.00,0.00,0.00}
\definecolor[named]{fillColor}{rgb}{0.00,0.00,0.00}

\node[text=drawColor,anchor=base west,inner sep=0pt, outer sep=0pt, scale = 0.90] at (367.35,159.23) {$\widetilde{P}_n$ with {NPD}};
\path[draw=red,line width= 0.8pt,dash pattern=on 2pt off 2pt on 6pt off 2pt ,line join=round,line cap=round] (340.35,162.67) -- (358.35,162.67);

\node[text=drawColor,anchor=base west,inner sep=0pt, outer sep=0pt, scale = 0.90] at (367.35,147.23-5) {$Q_n$ with {NPD}};
\path[draw=drawColor,line width= 0.8pt,dash pattern=on 1pt off 3pt ,line join=round,line cap=round] (340.35,150.67-5) -- (358.35,150.67-5);

\node[text=drawColor,anchor=base west,inner sep=0pt, outer sep=0pt, scale = 0.90] at (367.35,135.23-10) {$\tau$ with {NPD}};
\path[draw=orange,line width= 0.8pt,dash pattern=on 4pt off 4pt ,line join=round,line cap=round] (340.35,138.67-10) -- (358.35,138.67-10);

\node[text=drawColor,anchor=base west,inner sep=0pt, outer sep=0pt, scale = 0.90] at (367.35,123.23-15) {$P_n$ with {NPD}};
\path[draw=blue,line width= 0.8pt,dash pattern=on 7pt off 3pt ,line join=round,line cap=round] (340.35,126.67-15) -- (358.35,126.67-15);

\node[text=drawColor,anchor=base west,inner sep=0pt, outer sep=0pt, scale = 0.90] at (367.35,111.23-20) {{MCD}};
\path[draw=drawColor,line width= 0.8pt,dash pattern=on 1pt off 3pt on 4pt off 3pt ,line join=round,line cap=round] (340.35,102.67-25) -- (358.35,102.67-25);

\node[text=drawColor,anchor=base west,inner sep=0pt, outer sep=0pt, scale = 0.90] at (367.35, 99.23-25)  {$P_n$ with {OGK}};
\path[draw=drawColor,line width= 0.8pt,line join=round,line cap=round] (340.35,114.67-20) -- (358.35,114.67-20);

\node[text=drawColor,anchor=base west,inner sep=0pt, outer sep=0pt, scale = 0.90] at (367.35, 87.23-30) {Classical};
\path[draw=drawColor,line width= 0.8pt,line join=round,line cap=round] (340.35, 90.67-30) -- (358.35, 90.67-30);
\path[draw=drawColor,line width= 0.8pt,line join=round,line cap=round,fill=fillColor] (349.35,162.67-72-30) circle (  1.50);
\node[text=drawColor,anchor=base,inner sep=0pt, outer sep=0pt, scale = 0.90] at (260.00,  160.00) {GLASSO, banded, $p=60$};
\end{scope}
\end{tikzpicture}
\caption{PRIAL results for a selection of estimators applied to data generated with a banded precision matrix  with extreme outliers for $p=60$ using CLIME (top), QUIC (middle) and GLASSO (bottom).}
\label{climequicglasso}
\end{figure}

The same pattern holds true when using the QUIC or the GLASSO regularisation routines. To demonstrate this consider Figure \ref{climequicglasso} where the PRIAL results are shown for CLIME, QUIC and the GLASSO under the banded precision matrix scenario with extreme outliers and $p=60$.  As would be expected the QUIC and GLASSO results are essentially identical, and largely consistent with the CLIME results in the top panel.  

Consulting the raw entropy loss numbers reveals that the CLIME method gives slightly lower average entropy loss measurements, particularly for very high levels of contamination.   In practice it does not matter what regularisation routine is used, the benefits of taking a pairwise approach to covariance estimation in the presence of cellwise contamination will still hold.

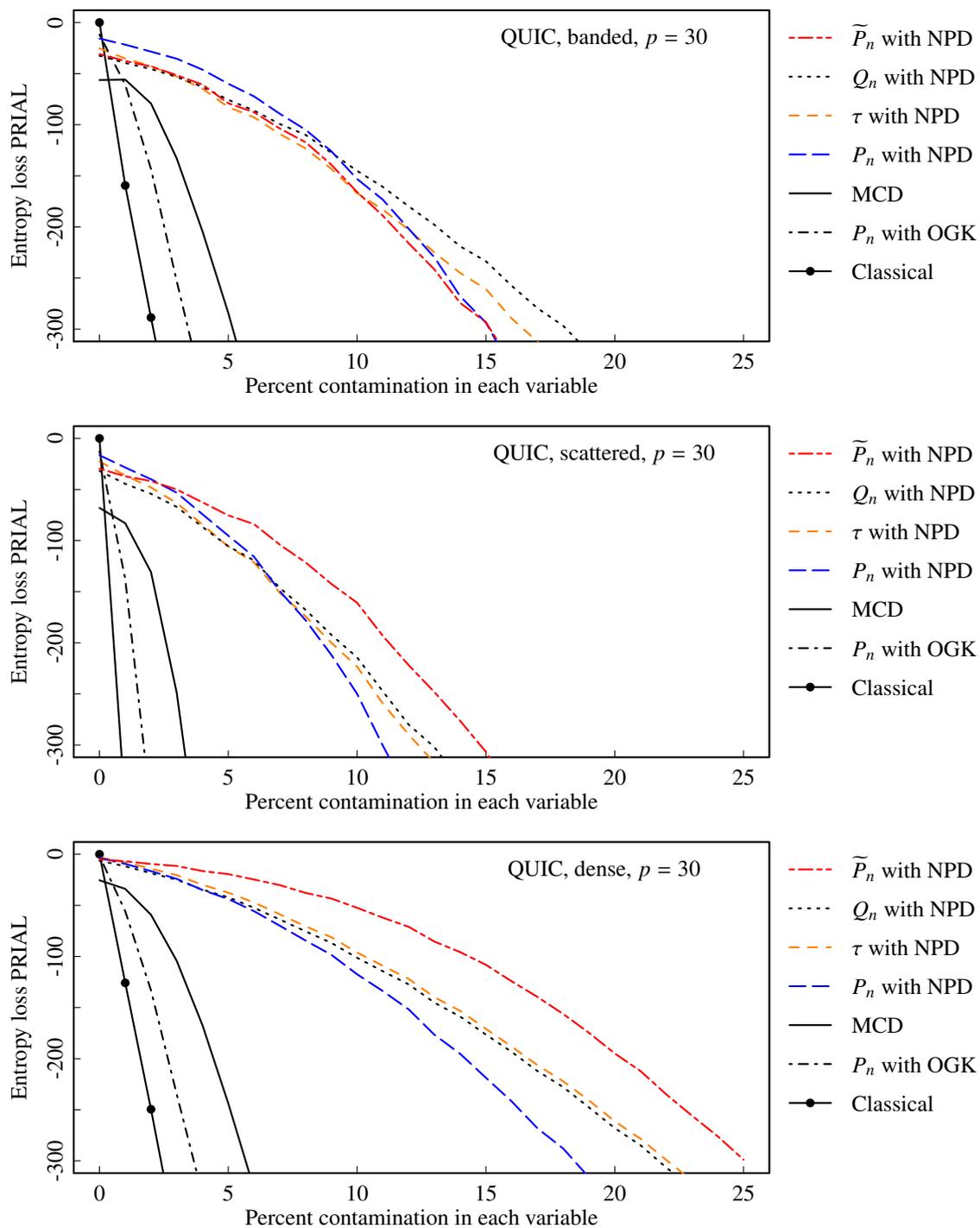
\begin{figure}[p]
\centering
\begin{tikzpicture}[x=1pt,y=1pt]
\definecolor[named]{fillColor}{rgb}{1.00,1.00,1.00}
\path[use as bounding box,fill=fillColor,fill opacity=0.00] (0,0) rectangle (433.62,180.67);
\begin{scope}
\path[clip] ( 30.00, 30.00) rectangle (331.62,174.67);
\definecolor[named]{drawColor}{rgb}{0.00,0.00,0.00}

\path[draw=drawColor,line width= 0.8pt,line join=round,line cap=round] ( 41.17,169.32) --
	( 52.34, 98.13) --
	( 63.51, 40.43) --
	( 71.53,  0.00);
\definecolor[named]{fillColor}{rgb}{0.00,0.00,0.00}

\path[draw=drawColor,line width= 0.8pt,line join=round,line cap=round,fill=fillColor] ( 41.17,169.32) circle (  1.50);

\path[draw=drawColor,line width= 0.8pt,line join=round,line cap=round,fill=fillColor] ( 52.34, 98.13) circle (  1.50);

\path[draw=drawColor,line width= 0.8pt,line join=round,line cap=round,fill=fillColor] ( 63.51, 40.43) circle (  1.50);
\end{scope}
\begin{scope}
\path[clip] (  0.00,  0.00) rectangle (460.00,180.67);
\definecolor[named]{drawColor}{rgb}{0.00,0.00,0.00}


\path[draw=drawColor,line width= 0.4pt,line join=round,line cap=round] ( 41.17, 30.00) -- ( 41.17, 32.89);

\path[draw=drawColor,line width= 0.4pt,line join=round,line cap=round] ( 97.03, 30.00) -- ( 97.03, 32.89);

\path[draw=drawColor,line width= 0.4pt,line join=round,line cap=round] (152.88, 30.00) -- (152.88, 32.89);

\path[draw=drawColor,line width= 0.4pt,line join=round,line cap=round] (208.74, 30.00) -- (208.74, 32.89);

\path[draw=drawColor,line width= 0.4pt,line join=round,line cap=round] (264.59, 30.00) -- (264.59, 32.89);

\path[draw=drawColor,line width= 0.4pt,line join=round,line cap=round] (320.45, 30.00) -- (320.45, 32.89);

0 at ( 41.17, 18.00) {0};

\node[text=drawColor,anchor=base,inner sep=0pt, outer sep=0pt, scale = 0.90] at ( 97.03, 18.00) {5};

\node[text=drawColor,anchor=base,inner sep=0pt, outer sep=0pt, scale = 0.90] at (152.88, 18.00) {10};

\node[text=drawColor,anchor=base,inner sep=0pt, outer sep=0pt, scale = 0.90] at (208.74, 18.00) {15};

\node[text=drawColor,anchor=base,inner sep=0pt, outer sep=0pt, scale = 0.90] at (264.59, 18.00) {20};

\node[text=drawColor,anchor=base,inner sep=0pt, outer sep=0pt, scale = 0.90] at (320.45, 18.00) {25};


\path[draw=drawColor,line width= 0.4pt,line join=round,line cap=round] ( 30.00, 35.36) -- ( 32.89, 35.36);

\path[draw=drawColor,line width= 0.4pt,line join=round,line cap=round] ( 30.00, 57.68) -- ( 32.89, 57.68);

\path[draw=drawColor,line width= 0.4pt,line join=round,line cap=round] ( 30.00, 80.01) -- ( 32.89, 80.01);

\path[draw=drawColor,line width= 0.4pt,line join=round,line cap=round] ( 30.00,102.34) -- ( 32.89,102.34);

\path[draw=drawColor,line width= 0.4pt,line join=round,line cap=round] ( 30.00,124.66) -- ( 32.89,124.66);

\path[draw=drawColor,line width= 0.4pt,line join=round,line cap=round] ( 30.00,146.99) -- ( 32.89,146.99);

\path[draw=drawColor,line width= 0.4pt,line join=round,line cap=round] ( 30.00,169.32) -- ( 32.89,169.32);

\node[text=drawColor,rotate= 90.00,anchor=base,inner sep=0pt, outer sep=0pt, scale = 0.90] at ( 25.20, 35.36) {-300};

\node[text=drawColor,rotate= 90.00,anchor=base,inner sep=0pt, outer sep=0pt, scale = 0.90] at ( 25.20, 80.01) {-200};

\node[text=drawColor,rotate= 90.00,anchor=base,inner sep=0pt, outer sep=0pt, scale = 0.90] at ( 25.20,124.66) {-100};

\node[text=drawColor,rotate= 90.00,anchor=base,inner sep=0pt, outer sep=0pt, scale = 0.90] at ( 25.20,169.32) {0};

\path[draw=drawColor,line width= 0.4pt,line join=round,line cap=round, thick] ( 30.00, 30.00) --
	(331.62, 30.00) --
	(331.62,174.67) --
	( 30.00,174.67) --
	( 30.00, 30.00);
\end{scope}
\begin{scope}
\path[clip] (  0.00,  0.00) rectangle (460.00,180.67);
\definecolor[named]{drawColor}{rgb}{0.00,0.00,0.00}

\node[text=drawColor,anchor=base,inner sep=0pt, outer sep=0pt, scale = 0.90] at (180.81,  2.40+5) {Percent contamination in each variable};

\node[text=drawColor,rotate= 90.00,anchor=base,inner sep=0pt, outer sep=0pt, scale = 0.90] at (  9.60,102.34) {Entropy loss {PRIAL}};
\end{scope}
\begin{scope}
\path[clip] ( 30.00, 30.00) rectangle (331.62,174.67);
\definecolor[named]{drawColor}{rgb}{0.00,0.00,0.00}

\path[draw=orange,line width= 0.8pt,dash pattern=on 4pt off 4pt ,line join=round,line cap=round] ( 41.17,158.02) --
	( 52.34,153.58) --
	( 63.51,150.30) --
	( 74.68,145.72) --
	( 85.86,140.42) --
	( 97.03,132.34) --
	(108.20,127.89) --
	(119.37,120.63) --
	(130.54,114.30) --
	(141.71,105.38) --
	(152.88, 94.79) --
	(164.05, 87.55) --
	(175.22, 78.82) --
	(186.40, 68.80) --
	(197.57, 59.91) --
	(208.74, 52.84) --
	(219.91, 40.03) --
	(231.08, 29.89) --
	(242.25, 21.89) --
	(253.42,  8.78) --
	(260.16,  0.00);

\path[draw=drawColor,line width= 0.8pt,dash pattern=on 1pt off 3pt ,line join=round,line cap=round] ( 41.17,154.81) --
	( 52.34,151.66) --
	( 63.51,149.02) --
	( 74.68,145.57) --
	( 85.86,141.21) --
	( 97.03,135.56) --
	(108.20,130.86) --
	(119.37,125.00) --
	(130.54,120.23) --
	(141.71,112.42) --
	(152.88,104.53) --
	(164.05, 97.50) --
	(175.22, 89.04) --
	(186.40, 80.96) --
	(197.57, 71.53) --
	(208.74, 65.00) --
	(219.91, 54.28) --
	(231.08, 44.52) --
	(242.25, 36.82) --
	(253.42, 25.30) --
	(264.59,  9.60) --
	(275.76,  4.14) --
	(281.10,  0.00);

\path[draw=drawColor,line width= 0.8pt,dash pattern=on 1pt off 3pt on 4pt off 3pt ,line join=round,line cap=round] ( 41.17,164.04) --
	( 52.34,142.17) --
	( 63.51,105.31) --
	( 74.68, 56.31) --
	( 85.86,  9.56) --
	( 88.64,  0.00);

\path[draw=blue,line width= 0.8pt,dash pattern=on 7pt off 3pt ,line join=round,line cap=round] ( 41.17,162.32) --
	( 52.34,159.64) --
	( 63.51,156.61) --
	( 74.68,153.50) --
	( 85.86,148.70) --
	( 97.03,142.57) --
	(108.20,137.04) --
	(119.37,129.46) --
	(130.54,122.47) --
	(141.71,113.06) --
	(152.88,100.97) --
	(164.05, 91.92) --
	(175.22, 79.24) --
	(186.40, 66.71) --
	(197.57, 49.75) --
	(208.74, 38.17) --
	(219.91, 18.06) --
	(231.08,  3.96) --
	(234.04,  0.00);

\path[draw=red,line width= 0.8pt,dash pattern=on 2pt off 2pt on 6pt off 2pt ,line join=round,line cap=round] ( 41.17,155.54) --
	( 52.34,152.51) --
	( 63.51,150.11) --
	( 74.68,146.32) --
	( 85.86,142.18) --
	( 97.03,134.05) --
	(108.20,130.11) --
	(119.37,123.01) --
	(130.54,117.01) --
	(141.71,107.19) --
	(152.88, 95.26) --
	(164.05, 84.92) --
	(175.22, 72.90) --
	(186.40, 61.41) --
	(197.57, 46.79) --
	(208.74, 38.40) --
	(219.91, 20.37) --
	(231.08,  7.15) --
	(236.78,  0.00);

\path[draw=drawColor,line width= 0.8pt,line join=round,line cap=round] ( 41.17,144.21) --
	( 52.34,144.38) --
	( 63.51,133.95) --
	( 74.68,109.93) --
	( 85.86, 78.05) --
	( 97.03, 42.43) --
	(108.20,  2.27) --
	(109.00,  0.00);
\end{scope}
\begin{scope}
\path[clip] (  0.00,  0.00) rectangle (460.00,180.67);
\definecolor[named]{drawColor}{rgb}{0.00,0.00,0.00}
\definecolor[named]{fillColor}{rgb}{0.00,0.00,0.00}

\node[text=drawColor,anchor=base west,inner sep=0pt, outer sep=0pt, scale = 0.90] at (367.35,159.23) {$\widetilde{P}_n$ with {NPD}};
\path[draw=red,line width= 0.8pt,dash pattern=on 2pt off 2pt on 6pt off 2pt ,line join=round,line cap=round] (340.35,162.67) -- (358.35,162.67);

\node[text=drawColor,anchor=base west,inner sep=0pt, outer sep=0pt, scale = 0.90] at (367.35,147.23-5) {$Q_n$ with {NPD}};
\path[draw=drawColor,line width= 0.8pt,dash pattern=on 1pt off 3pt ,line join=round,line cap=round] (340.35,150.67-5) -- (358.35,150.67-5);

\node[text=drawColor,anchor=base west,inner sep=0pt, outer sep=0pt, scale = 0.90] at (367.35,135.23-10) {$\tau$ with {NPD}};
\path[draw=orange,line width= 0.8pt,dash pattern=on 4pt off 4pt ,line join=round,line cap=round] (340.35,138.67-10) -- (358.35,138.67-10);

\node[text=drawColor,anchor=base west,inner sep=0pt, outer sep=0pt, scale = 0.90] at (367.35,123.23-15) {$P_n$ with {NPD}};
\path[draw=blue,line width= 0.8pt,dash pattern=on 7pt off 3pt ,line join=round,line cap=round] (340.35,126.67-15) -- (358.35,126.67-15);

\node[text=drawColor,anchor=base west,inner sep=0pt, outer sep=0pt, scale = 0.90] at (367.35,111.23-20) {{MCD}};
\path[draw=drawColor,line width= 0.8pt,dash pattern=on 1pt off 3pt on 4pt off 3pt ,line join=round,line cap=round] (340.35,102.67-25) -- (358.35,102.67-25);

\node[text=drawColor,anchor=base west,inner sep=0pt, outer sep=0pt, scale = 0.90] at (367.35, 99.23-25)  {$P_n$ with {OGK}};
\path[draw=drawColor,line width= 0.8pt,line join=round,line cap=round] (340.35,114.67-20) -- (358.35,114.67-20);

\node[text=drawColor,anchor=base west,inner sep=0pt, outer sep=0pt, scale = 0.90] at (367.35, 87.23-30) {Classical};
\path[draw=drawColor,line width= 0.8pt,line join=round,line cap=round] (340.35, 90.67-30) -- (358.35, 90.67-30);
\path[draw=drawColor,line width= 0.8pt,line join=round,line cap=round,fill=fillColor] (349.35,162.67-72-30) circle (  1.50);
\node[text=drawColor,anchor=base,inner sep=0pt, outer sep=0pt, scale = 0.90] at (260.00,  160.00) {QUIC, banded, $p=30$};
\end{scope}
\end{tikzpicture}

\begin{tikzpicture}[x=1pt,y=1pt]
\definecolor[named]{fillColor}{rgb}{1.00,1.00,1.00}
\path[use as bounding box,fill=fillColor,fill opacity=0.00] (0,0) rectangle (433.62,180.67);
\begin{scope}
\path[clip] ( 30.00, 30.00) rectangle (331.62,174.67);
\definecolor[named]{drawColor}{rgb}{0.00,0.00,0.00}

\path[draw=drawColor,line width= 0.8pt,line join=round,line cap=round] ( 41.17,169.32) --
	( 52.34,  7.76) --
	( 52.89,  0.00);
\definecolor[named]{fillColor}{rgb}{0.00,0.00,0.00}

\path[draw=drawColor,line width= 0.8pt,line join=round,line cap=round,fill=fillColor] ( 41.17,169.32) circle (  1.50);

\path[draw=drawColor,line width= 0.8pt,line join=round,line cap=round,fill=fillColor] ( 52.34,  7.76) circle (  1.50);
\end{scope}
\begin{scope}
\path[clip] (  0.00,  0.00) rectangle (460.00,180.67);
\definecolor[named]{drawColor}{rgb}{0.00,0.00,0.00}


\path[draw=drawColor,line width= 0.4pt,line join=round,line cap=round] ( 41.17, 30.00) -- ( 41.17, 32.89);

\path[draw=drawColor,line width= 0.4pt,line join=round,line cap=round] ( 97.03, 30.00) -- ( 97.03, 32.89);

\path[draw=drawColor,line width= 0.4pt,line join=round,line cap=round] (152.88, 30.00) -- (152.88, 32.89);

\path[draw=drawColor,line width= 0.4pt,line join=round,line cap=round] (208.74, 30.00) -- (208.74, 32.89);

\path[draw=drawColor,line width= 0.4pt,line join=round,line cap=round] (264.59, 30.00) -- (264.59, 32.89);

\path[draw=drawColor,line width= 0.4pt,line join=round,line cap=round] (320.45, 30.00) -- (320.45, 32.89);

\node[text=drawColor,anchor=base,inner sep=0pt, outer sep=0pt, scale = 0.90] at ( 41.17, 18.00) {0};

\node[text=drawColor,anchor=base,inner sep=0pt, outer sep=0pt, scale = 0.90] at ( 97.03, 18.00) {5};

\node[text=drawColor,anchor=base,inner sep=0pt, outer sep=0pt, scale = 0.90] at (152.88, 18.00) {10};

\node[text=drawColor,anchor=base,inner sep=0pt, outer sep=0pt, scale = 0.90] at (208.74, 18.00) {15};

\node[text=drawColor,anchor=base,inner sep=0pt, outer sep=0pt, scale = 0.90] at (264.59, 18.00) {20};

\node[text=drawColor,anchor=base,inner sep=0pt, outer sep=0pt, scale = 0.90] at (320.45, 18.00) {25};


\path[draw=drawColor,line width= 0.4pt,line join=round,line cap=round] ( 30.00, 35.36) -- ( 32.89, 35.36);

\path[draw=drawColor,line width= 0.4pt,line join=round,line cap=round] ( 30.00, 57.68) -- ( 32.89, 57.68);

\path[draw=drawColor,line width= 0.4pt,line join=round,line cap=round] ( 30.00, 80.01) -- ( 32.89, 80.01);

\path[draw=drawColor,line width= 0.4pt,line join=round,line cap=round] ( 30.00,102.34) -- ( 32.89,102.34);

\path[draw=drawColor,line width= 0.4pt,line join=round,line cap=round] ( 30.00,124.66) -- ( 32.89,124.66);

\path[draw=drawColor,line width= 0.4pt,line join=round,line cap=round] ( 30.00,146.99) -- ( 32.89,146.99);

\path[draw=drawColor,line width= 0.4pt,line join=round,line cap=round] ( 30.00,169.32) -- ( 32.89,169.32);

\node[text=drawColor,rotate= 90.00,anchor=base,inner sep=0pt, outer sep=0pt, scale = 0.90] at ( 25.20, 35.36) {-300};

\node[text=drawColor,rotate= 90.00,anchor=base,inner sep=0pt, outer sep=0pt, scale = 0.90] at ( 25.20, 80.01) {-200};

\node[text=drawColor,rotate= 90.00,anchor=base,inner sep=0pt, outer sep=0pt, scale = 0.90] at ( 25.20,124.66) {-100};

\node[text=drawColor,rotate= 90.00,anchor=base,inner sep=0pt, outer sep=0pt, scale = 0.90] at ( 25.20,169.32) {0};

\path[draw=drawColor,line width= 0.4pt,line join=round,line cap=round, thick] ( 30.00, 30.00) --
	(331.62, 30.00) --
	(331.62,174.67) --
	( 30.00,174.67) --
	( 30.00, 30.00);
\end{scope}
\begin{scope}
\path[clip] (  0.00,  0.00) rectangle (460.00,180.67);
\definecolor[named]{drawColor}{rgb}{0.00,0.00,0.00}

\node[text=drawColor,anchor=base,inner sep=0pt, outer sep=0pt, scale = 0.90] at (180.81,  2.40+5) {Percent contamination in each variable};

\node[text=drawColor,rotate= 90.00,anchor=base,inner sep=0pt, outer sep=0pt, scale = 0.90] at (  9.60,102.34) {Entropy loss {PRIAL}};
\end{scope}
\begin{scope}
\path[clip] ( 30.00, 30.00) rectangle (331.62,174.67);
\definecolor[named]{drawColor}{rgb}{0.00,0.00,0.00}

\path[draw=orange,line width= 0.8pt,dash pattern=on 4pt off 4pt ,line join=round,line cap=round] ( 41.17,159.25) --
	( 52.34,153.13) --
	( 63.51,147.89) --
	( 74.68,141.09) --
	( 85.86,131.62) --
	( 97.03,122.06) --
	(108.20,115.11) --
	(119.37,101.82) --
	(130.54, 91.58) --
	(141.71, 80.02) --
	(152.88, 69.62) --
	(164.05, 53.45) --
	(175.22, 39.97) --
	(186.40, 27.60) --
	(197.57, 14.59) --
	(207.16,  0.00);

\path[draw=drawColor,line width= 0.8pt,dash pattern=on 1pt off 3pt ,line join=round,line cap=round] ( 41.17,155.13) --
	( 52.34,149.46) --
	( 63.51,145.10) --
	( 74.68,139.37) --
	( 85.86,130.62) --
	( 97.03,122.18) --
	(108.20,115.96) --
	(119.37,103.96) --
	(130.54, 94.40) --
	(141.71, 83.49) --
	(152.88, 73.63) --
	(164.05, 58.80) --
	(175.22, 44.29) --
	(186.40, 34.43) --
	(197.57, 19.52) --
	(208.74,  4.53) --
	(211.92,  0.00);

\path[draw=drawColor,line width= 0.8pt,dash pattern=on 1pt off 3pt on 4pt off 3pt ,line join=round,line cap=round] ( 41.17,163.66) --
	( 52.34,107.85) --
	( 63.51,  6.46) --
	( 64.20,  0.00);

\path[draw=blue,line width= 0.8pt,dash pattern=on 7pt off 3pt ,line join=round,line cap=round] ( 41.17,161.95) --
	( 52.34,156.59) --
	( 63.51,151.49) --
	( 74.68,145.55) --
	( 85.86,136.14) --
	( 97.03,126.81) --
	(108.20,117.60) --
	(119.37,102.55) --
	(130.54, 89.87) --
	(141.71, 74.73) --
	(152.88, 57.74) --
	(164.05, 35.11) --
	(175.22, 13.66) --
	(182.81,  0.00);

\path[draw=red,line width= 0.8pt,dash pattern=on 2pt off 2pt on 6pt off 2pt ,line join=round,line cap=round] ( 41.17,156.10) --
	( 52.34,152.74) --
	( 63.51,150.67) --
	( 74.68,146.97) --
	( 85.86,141.37) --
	( 97.03,135.68) --
	(108.20,131.84) --
	(119.37,122.86) --
	(130.54,115.12) --
	(141.71,105.75) --
	(152.88, 97.49) --
	(164.05, 83.01) --
	(175.22, 70.26) --
	(186.40, 58.47) --
	(197.57, 45.88) --
	(208.74, 32.36) --
	(219.91, 13.75) --
	(227.13,  0.00);

\path[draw=drawColor,line width= 0.8pt,line join=round,line cap=round] ( 41.17,138.95) --
	( 52.34,132.25) --
	( 63.51,110.90) --
	( 74.68, 58.15) --
	( 82.56,  0.00);
\end{scope}
\begin{scope}
\path[clip] (  0.00,  0.00) rectangle (460.00,180.67);
\definecolor[named]{drawColor}{rgb}{0.00,0.00,0.00}
\definecolor[named]{fillColor}{rgb}{0.00,0.00,0.00}

\node[text=drawColor,anchor=base west,inner sep=0pt, outer sep=0pt, scale = 0.90] at (367.35,159.23) {$\widetilde{P}_n$ with {NPD}};
\path[draw=red,line width= 0.8pt,dash pattern=on 2pt off 2pt on 6pt off 2pt ,line join=round,line cap=round] (340.35,162.67) -- (358.35,162.67);

\node[text=drawColor,anchor=base west,inner sep=0pt, outer sep=0pt, scale = 0.90] at (367.35,147.23-5) {$Q_n$ with {NPD}};
\path[draw=drawColor,line width= 0.8pt,dash pattern=on 1pt off 3pt ,line join=round,line cap=round] (340.35,150.67-5) -- (358.35,150.67-5);

\node[text=drawColor,anchor=base west,inner sep=0pt, outer sep=0pt, scale = 0.90] at (367.35,135.23-10) {$\tau$ with {NPD}};
\path[draw=orange,line width= 0.8pt,dash pattern=on 4pt off 4pt ,line join=round,line cap=round] (340.35,138.67-10) -- (358.35,138.67-10);

\node[text=drawColor,anchor=base west,inner sep=0pt, outer sep=0pt, scale = 0.90] at (367.35,123.23-15) {$P_n$ with {NPD}};
\path[draw=blue,line width= 0.8pt,dash pattern=on 7pt off 3pt ,line join=round,line cap=round] (340.35,126.67-15) -- (358.35,126.67-15);

\node[text=drawColor,anchor=base west,inner sep=0pt, outer sep=0pt, scale = 0.90] at (367.35,111.23-20) {{MCD}};
\path[draw=drawColor,line width= 0.8pt,dash pattern=on 1pt off 3pt on 4pt off 3pt ,line join=round,line cap=round] (340.35,102.67-25) -- (358.35,102.67-25);

\node[text=drawColor,anchor=base west,inner sep=0pt, outer sep=0pt, scale = 0.90] at (367.35, 99.23-25)  {$P_n$ with {OGK}};
\path[draw=drawColor,line width= 0.8pt,line join=round,line cap=round] (340.35,114.67-20) -- (358.35,114.67-20);

\node[text=drawColor,anchor=base west,inner sep=0pt, outer sep=0pt, scale = 0.90] at (367.35, 87.23-30) {Classical};
\path[draw=drawColor,line width= 0.8pt,line join=round,line cap=round] (340.35, 90.67-30) -- (358.35, 90.67-30);
\path[draw=drawColor,line width= 0.8pt,line join=round,line cap=round,fill=fillColor] (349.35,162.67-72-30) circle (  1.50);
\node[text=drawColor,anchor=base,inner sep=0pt, outer sep=0pt, scale = 0.90] at (260.00,  160.00) {QUIC, scattered, $p=30$};
\end{scope}
\end{tikzpicture}

\begin{tikzpicture}[x=1pt,y=1pt]
\definecolor[named]{fillColor}{rgb}{1.00,1.00,1.00}
\path[use as bounding box,fill=fillColor,fill opacity=0.00] (0,0) rectangle (433.62,180.67);
\begin{scope}
\path[clip] ( 30.00, 30.00) rectangle (331.62,174.67);
\definecolor[named]{drawColor}{rgb}{0.00,0.00,0.00}

\path[draw=drawColor,line width= 0.8pt,line join=round,line cap=round] ( 41.17,169.32) --
	( 52.34,113.12) --
	( 63.51, 57.90) --
	( 74.63,  0.00);
\definecolor[named]{fillColor}{rgb}{0.00,0.00,0.00}

\path[draw=drawColor,line width= 0.8pt,line join=round,line cap=round,fill=fillColor] ( 41.17,169.32) circle (  1.50);

\path[draw=drawColor,line width= 0.8pt,line join=round,line cap=round,fill=fillColor] ( 52.34,113.12) circle (  1.50);

\path[draw=drawColor,line width= 0.8pt,line join=round,line cap=round,fill=fillColor] ( 63.51, 57.90) circle (  1.50);

\path[draw=drawColor,line width= 0.8pt,line join=round,line cap=round,fill=fillColor] ( 74.68, -0.29) circle (  1.50);
\end{scope}
\begin{scope}
\path[clip] (  0.00,  0.00) rectangle (460.00,180.67);
\definecolor[named]{drawColor}{rgb}{0.00,0.00,0.00}


\path[draw=drawColor,line width= 0.4pt,line join=round,line cap=round] ( 41.17, 30.00) -- ( 41.17, 32.89);

\path[draw=drawColor,line width= 0.4pt,line join=round,line cap=round] ( 97.03, 30.00) -- ( 97.03, 32.89);

\path[draw=drawColor,line width= 0.4pt,line join=round,line cap=round] (152.88, 30.00) -- (152.88, 32.89);

\path[draw=drawColor,line width= 0.4pt,line join=round,line cap=round] (208.74, 30.00) -- (208.74, 32.89);

\path[draw=drawColor,line width= 0.4pt,line join=round,line cap=round] (264.59, 30.00) -- (264.59, 32.89);

\path[draw=drawColor,line width= 0.4pt,line join=round,line cap=round] (320.45, 30.00) -- (320.45, 32.89);

\node[text=drawColor,anchor=base,inner sep=0pt, outer sep=0pt, scale = 0.90] at ( 41.17, 18.00) {0};

\node[text=drawColor,anchor=base,inner sep=0pt, outer sep=0pt, scale = 0.90] at ( 97.03, 18.00) {5};

\node[text=drawColor,anchor=base,inner sep=0pt, outer sep=0pt, scale = 0.90] at (152.88, 18.00) {10};

\node[text=drawColor,anchor=base,inner sep=0pt, outer sep=0pt, scale = 0.90] at (208.74, 18.00) {15};

\node[text=drawColor,anchor=base,inner sep=0pt, outer sep=0pt, scale = 0.90] at (264.59, 18.00) {20};

\node[text=drawColor,anchor=base,inner sep=0pt, outer sep=0pt, scale = 0.90] at (320.45, 18.00) {25};


\path[draw=drawColor,line width= 0.4pt,line join=round,line cap=round] ( 30.00, 35.36) -- ( 32.89, 35.36);

\path[draw=drawColor,line width= 0.4pt,line join=round,line cap=round] ( 30.00, 57.68) -- ( 32.89, 57.68);

\path[draw=drawColor,line width= 0.4pt,line join=round,line cap=round] ( 30.00, 80.01) -- ( 32.89, 80.01);

\path[draw=drawColor,line width= 0.4pt,line join=round,line cap=round] ( 30.00,102.34) -- ( 32.89,102.34);

\path[draw=drawColor,line width= 0.4pt,line join=round,line cap=round] ( 30.00,124.66) -- ( 32.89,124.66);

\path[draw=drawColor,line width= 0.4pt,line join=round,line cap=round] ( 30.00,146.99) -- ( 32.89,146.99);

\path[draw=drawColor,line width= 0.4pt,line join=round,line cap=round] ( 30.00,169.32) -- ( 32.89,169.32);

\node[text=drawColor,rotate= 90.00,anchor=base,inner sep=0pt, outer sep=0pt, scale = 0.90] at ( 25.20, 35.36) {-300};

\node[text=drawColor,rotate= 90.00,anchor=base,inner sep=0pt, outer sep=0pt, scale = 0.90] at ( 25.20, 80.01) {-200};

\node[text=drawColor,rotate= 90.00,anchor=base,inner sep=0pt, outer sep=0pt, scale = 0.90] at ( 25.20,124.66) {-100};

\node[text=drawColor,rotate= 90.00,anchor=base,inner sep=0pt, outer sep=0pt, scale = 0.90] at ( 25.20,169.32) {0};

\path[draw=drawColor,line width= 0.4pt,line join=round,line cap=round, thick] ( 30.00, 30.00) --
	(331.62, 30.00) --
	(331.62,174.67) --
	( 30.00,174.67) --
	( 30.00, 30.00);
\end{scope}
\begin{scope}
\path[clip] (  0.00,  0.00) rectangle (460.00,180.67);
\definecolor[named]{drawColor}{rgb}{0.00,0.00,0.00}

\node[text=drawColor,anchor=base,inner sep=0pt, outer sep=0pt, scale = 0.90] at (180.81,  2.40+5) {Percent contamination in each variable};

\node[text=drawColor,rotate= 90.00,anchor=base,inner sep=0pt, outer sep=0pt, scale = 0.90] at (  9.60,102.34) {Entropy loss {PRIAL}};
\end{scope}
\begin{scope}
\path[clip] ( 30.00, 30.00) rectangle (331.62,174.67);
\definecolor[named]{drawColor}{rgb}{0.00,0.00,0.00}

\path[draw=orange,line width= 0.8pt,dash pattern=on 4pt off 4pt ,line join=round,line cap=round] ( 41.17,167.97) --
	( 52.34,165.80) --
	( 63.51,162.98) --
	( 74.68,160.18) --
	( 85.86,155.95) --
	( 97.03,152.53) --
	(108.20,148.22) --
	(119.37,143.05) --
	(130.54,137.81) --
	(141.71,133.06) --
	(152.88,126.37) --
	(164.05,120.56) --
	(175.22,114.87) --
	(186.40,106.84) --
	(197.57,100.91) --
	(208.74, 93.02) --
	(219.91, 85.26) --
	(231.08, 77.18) --
	(242.25, 70.10) --
	(253.42, 61.78) --
	(264.59, 52.62) --
	(275.76, 45.03) --
	(286.94, 35.76) --
	(298.11, 26.71) --
	(309.28, 17.38) --
	(320.45,  7.57);

\path[draw=drawColor,line width= 0.8pt,dash pattern=on 1pt off 3pt ,line join=round,line cap=round] ( 41.17,166.49) --
	( 52.34,164.02) --
	( 63.51,161.03) --
	( 74.68,158.16) --
	( 85.86,153.84) --
	( 97.03,150.50) --
	(108.20,145.91) --
	(119.37,140.80) --
	(130.54,135.63) --
	(141.71,130.58) --
	(152.88,124.08) --
	(164.05,118.19) --
	(175.22,112.48) --
	(186.40,104.50) --
	(197.57, 98.40) --
	(208.74, 90.51) --
	(219.91, 82.97) --
	(231.08, 74.61) --
	(242.25, 67.59) --
	(253.42, 58.72) --
	(264.59, 49.66) --
	(275.76, 42.19) --
	(286.94, 32.41) --
	(298.11, 23.53) --
	(309.28, 14.94) --
	(320.45,  4.44);

\path[draw=drawColor,line width= 0.8pt,dash pattern=on 1pt off 3pt on 4pt off 3pt ,line join=round,line cap=round] ( 41.17,168.54) --
	( 52.34,144.40) --
	( 63.51,109.86) --
	( 74.68, 64.44) --
	( 85.86, 20.53) --
	( 91.44,  0.00);

\path[draw=blue,line width= 0.8pt,dash pattern=on 7pt off 3pt ,line join=round,line cap=round] ( 41.17,167.70) --
	( 52.34,165.09) --
	( 63.51,161.92) --
	( 74.68,158.56) --
	( 85.86,153.67) --
	( 97.03,149.75) --
	(108.20,144.43) --
	(119.37,138.07) --
	(130.54,131.60) --
	(141.71,125.38) --
	(152.88,116.87) --
	(164.05,109.52) --
	(175.22,101.57) --
	(186.40, 90.48) --
	(197.57, 82.12) --
	(208.74, 71.66) --
	(219.91, 61.33) --
	(231.08, 49.61) --
	(242.25, 40.75) --
	(253.42, 28.13) --
	(264.59, 15.90) --
	(275.76,  4.82) --
	(279.92,  0.00);

\path[draw=red,line width= 0.8pt,dash pattern=on 2pt off 2pt on 6pt off 2pt ,line join=round,line cap=round] ( 41.17,167.11) --
	( 52.34,166.23) --
	( 63.51,165.05) --
	( 74.68,164.18) --
	( 85.86,161.96) --
	( 97.03,160.63) --
	(108.20,158.26) --
	(119.37,155.81) --
	(130.54,152.41) --
	(141.71,149.94) --
	(152.88,145.92) --
	(164.05,141.59) --
	(175.22,137.62) --
	(186.40,131.17) --
	(197.57,126.55) --
	(208.74,121.00) --
	(219.91,113.79) --
	(231.08,106.98) --
	(242.25, 99.52) --
	(253.42, 91.31) --
	(264.59, 82.31) --
	(275.76, 74.59) --
	(286.94, 64.39) --
	(298.11, 55.08) --
	(309.28, 46.06) --
	(320.45, 35.86);

\path[draw=drawColor,line width= 0.8pt,line join=round,line cap=round] ( 41.17,157.93) --
	( 52.34,154.18) --
	( 63.51,142.87) --
	( 74.68,122.72) --
	( 85.86, 94.59) --
	( 97.03, 60.47) --
	(108.20, 23.13) --
	(115.60,  0.00);
\end{scope}
\begin{scope}
\path[clip] (  0.00,  0.00) rectangle (460.00,180.67);
\definecolor[named]{drawColor}{rgb}{0.00,0.00,0.00}
\definecolor[named]{fillColor}{rgb}{0.00,0.00,0.00}

\node[text=drawColor,anchor=base west,inner sep=0pt, outer sep=0pt, scale = 0.90] at (367.35,159.23) {$\widetilde{P}_n$ with {NPD}};
\path[draw=red,line width= 0.8pt,dash pattern=on 2pt off 2pt on 6pt off 2pt ,line join=round,line cap=round] (340.35,162.67) -- (358.35,162.67);

\node[text=drawColor,anchor=base west,inner sep=0pt, outer sep=0pt, scale = 0.90] at (367.35,147.23-5) {$Q_n$ with {NPD}};
\path[draw=drawColor,line width= 0.8pt,dash pattern=on 1pt off 3pt ,line join=round,line cap=round] (340.35,150.67-5) -- (358.35,150.67-5);

\node[text=drawColor,anchor=base west,inner sep=0pt, outer sep=0pt, scale = 0.90] at (367.35,135.23-10) {$\tau$ with {NPD}};
\path[draw=orange,line width= 0.8pt,dash pattern=on 4pt off 4pt ,line join=round,line cap=round] (340.35,138.67-10) -- (358.35,138.67-10);

\node[text=drawColor,anchor=base west,inner sep=0pt, outer sep=0pt, scale = 0.90] at (367.35,123.23-15) {$P_n$ with {NPD}};
\path[draw=blue,line width= 0.8pt,dash pattern=on 7pt off 3pt ,line join=round,line cap=round] (340.35,126.67-15) -- (358.35,126.67-15);

\node[text=drawColor,anchor=base west,inner sep=0pt, outer sep=0pt, scale = 0.90] at (367.35,111.23-20) {{MCD}};
\path[draw=drawColor,line width= 0.8pt,dash pattern=on 1pt off 3pt on 4pt off 3pt ,line join=round,line cap=round] (340.35,102.67-25) -- (358.35,102.67-25);

\node[text=drawColor,anchor=base west,inner sep=0pt, outer sep=0pt, scale = 0.90] at (367.35, 99.23-25)  {$P_n$ with {OGK}};
\path[draw=drawColor,line width= 0.8pt,line join=round,line cap=round] (340.35,114.67-20) -- (358.35,114.67-20);

\node[text=drawColor,anchor=base west,inner sep=0pt, outer sep=0pt, scale = 0.90] at (367.35, 87.23-30) {Classical};
\path[draw=drawColor,line width= 0.8pt,line join=round,line cap=round] (340.35, 90.67-30) -- (358.35, 90.67-30);
\path[draw=drawColor,line width= 0.8pt,line join=round,line cap=round,fill=fillColor] (349.35,162.67-72-30) circle (  1.50);
\node[text=drawColor,anchor=base,inner sep=0pt, outer sep=0pt, scale = 0.90] at (260.00,  160.00) {QUIC, dense, $p=30$};
\end{scope}
\end{tikzpicture}
\caption{PRIAL results for a selection of estimators applied to data generated with a banded precision matrix (top), scattered precision matrix (middle) and dense precision matrix (bottom)  with extreme outliers for $p=30$ using the QUIC routine.}
\label{scenarios}
\end{figure}

The NPD pairwise approach is a major improvement over standard robust estimators.  An example of this is given in Figure \ref{scenarios} where we present the average PRIAL results for the QUIC estimator with $p=30$ for the scenarios illustrated in Figure \ref{precisionscenarios}.   Across all scenarios the same general pattern holds, the classical method and the OGK and MCD methods fail quite rapidly whereas the NPD approach offers much greater resilience to the cellwise contamination.  

For the banded precision matrix scenario, top panel of Figure \ref{scenarios}, the NPD based methods under the various robust scale estimators give similar results with $P_n$ having a slight advantage over the others for low levels of contamination whereas $Q_n$ has an advantage for higher contamination proportions.  

For the scattered precision matrix and the dense precision matrix scenarios, $\widetilde{P}_n$ gives the best results.  The advantage of the adaptive trimming procedure is lost when the outliers are not so extreme, however, in such scenarios the adaptive trimming approach performs no worse than the other NPD methods.

We previously established that matrix norms are not a good performance measure for precision matrices. In terms of the other performance indicators, for all scenarios considered the log condition number remained bounded, suggesting that all three regularisation routines return well conditioned precision matrix estimates regardless of the level of contamination or the data generating process. 

The NPD also performed well in terms of the log determinant performance index.  As with the entropy loss, there appears to be an advantage to using $\widetilde{P}_n$ over the other scale estimators in each of the scenarios.  Unlike with the entropy loss, the advantage of the adaptive trimming procedure is still evident even when the contamination is less extreme.

\begin{figure}[htbp]
\centering
\input{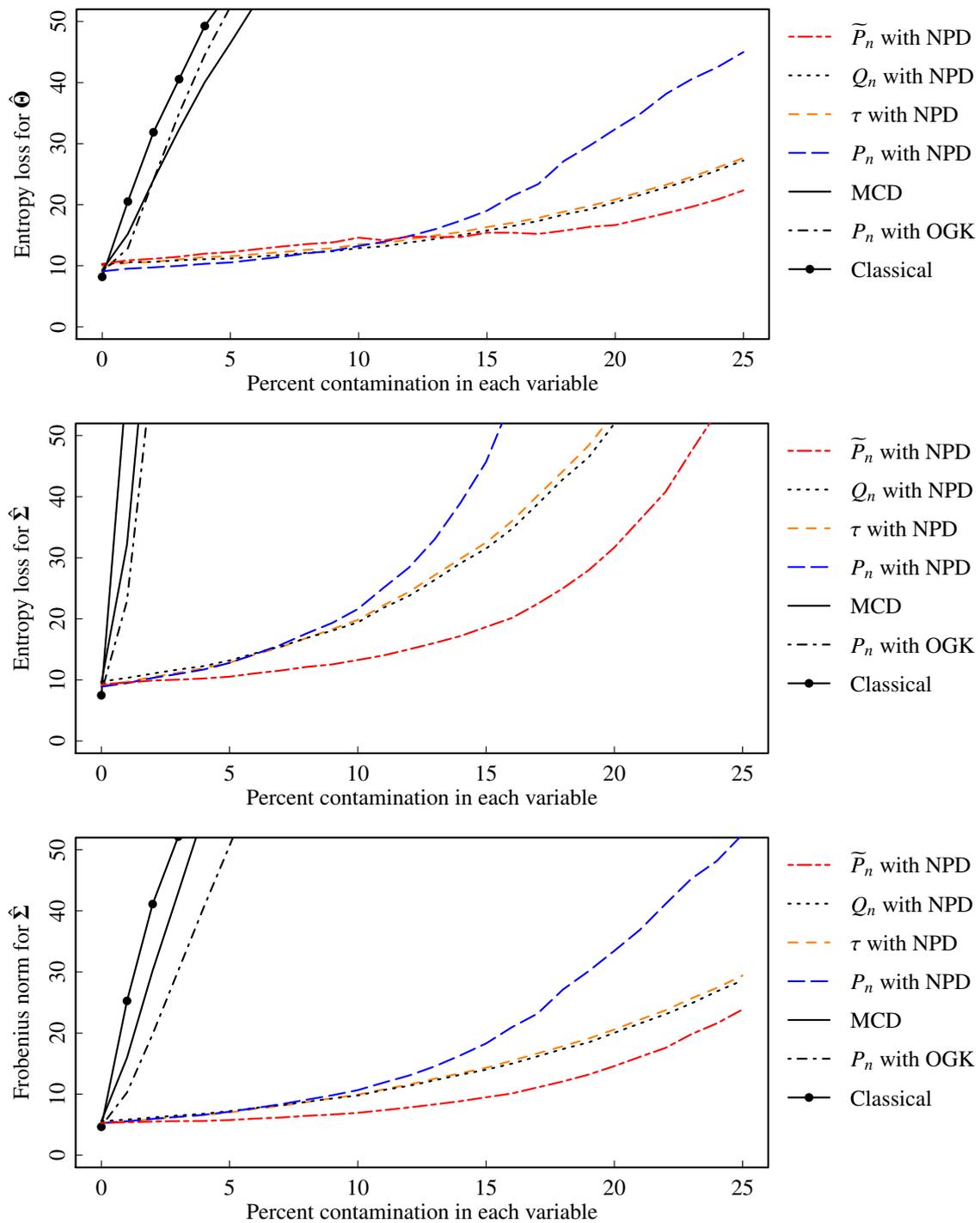}

\begin{tikzpicture}[x=1pt,y=1pt]
\definecolor[named]{fillColor}{rgb}{1.00,1.00,1.00}
\path[use as bounding box,fill=fillColor,fill opacity=0.00] (-2,0) rectangle (433.62,180.67);
\begin{scope}
\path[clip] ( 30.00, 30.00) rectangle (331.62,174.67);
\definecolor[named]{drawColor}{rgb}{0.00,0.00,0.00}

\path[draw=drawColor,line width= 0.8pt,line join=round,line cap=round] ( 41.17, 55.42) --
	( 51.26,180.67);
\definecolor[named]{fillColor}{rgb}{0.00,0.00,0.00}

\path[draw=drawColor,line width= 0.8pt,line join=round,line cap=round,fill=fillColor] ( 41.17, 55.42) circle (  1.50);
\end{scope}
\begin{scope}
\path[clip] (  0.00,  0.00) rectangle (433.62,180.67);
\definecolor[named]{drawColor}{rgb}{0.00,0.00,0.00}


\path[draw=drawColor,line width= 0.4pt,line join=round,line cap=round] ( 41.17, 30.00) -- ( 41.17, 32.89);

\path[draw=drawColor,line width= 0.4pt,line join=round,line cap=round] ( 97.03, 30.00) -- ( 97.03, 32.89);

\path[draw=drawColor,line width= 0.4pt,line join=round,line cap=round] (152.88, 30.00) -- (152.88, 32.89);

\path[draw=drawColor,line width= 0.4pt,line join=round,line cap=round] (208.74, 30.00) -- (208.74, 32.89);

\path[draw=drawColor,line width= 0.4pt,line join=round,line cap=round] (264.59, 30.00) -- (264.59, 32.89);

\path[draw=drawColor,line width= 0.4pt,line join=round,line cap=round] (320.45, 30.00) -- (320.45, 32.89);

\node[text=drawColor,anchor=base,inner sep=0pt, outer sep=0pt, scale=  0.90] at ( 41.17, 18.00) {0};

\node[text=drawColor,anchor=base,inner sep=0pt, outer sep=0pt, scale=  0.90] at ( 97.03, 18.00) {5};

\node[text=drawColor,anchor=base,inner sep=0pt, outer sep=0pt, scale=  0.90] at (152.88, 18.00) {10};

\node[text=drawColor,anchor=base,inner sep=0pt, outer sep=0pt, scale=  0.90] at (208.74, 18.00) {15};

\node[text=drawColor,anchor=base,inner sep=0pt, outer sep=0pt, scale=  0.90] at (264.59, 18.00) {20};

\node[text=drawColor,anchor=base,inner sep=0pt, outer sep=0pt, scale=  0.90] at (320.45, 18.00) {25};


\path[draw=drawColor,line width= 0.4pt,line join=round,line cap=round] ( 30.00, 35.36) -- ( 32.89, 35.36);

\path[draw=drawColor,line width= 0.4pt,line join=round,line cap=round] ( 30.00, 62.15) -- ( 32.89, 62.15);

\path[draw=drawColor,line width= 0.4pt,line join=round,line cap=round] ( 30.00, 88.94) -- ( 32.89, 88.94);

\path[draw=drawColor,line width= 0.4pt,line join=round,line cap=round] ( 30.00,115.73) -- ( 32.89,115.73);

\path[draw=drawColor,line width= 0.4pt,line join=round,line cap=round] ( 30.00,142.52) -- ( 32.89,142.52);

\path[draw=drawColor,line width= 0.4pt,line join=round,line cap=round] ( 30.00,169.32) -- ( 32.89,169.32);

\node[text=drawColor,rotate= 90.00,anchor=base,inner sep=0pt, outer sep=0pt, scale = 0.90] at ( 25.20, 35.36) {0};

\node[text=drawColor,rotate= 90.00,anchor=base,inner sep=0pt, outer sep=0pt, scale = 0.90] at ( 25.20, 62.15) {10};

\node[text=drawColor,rotate= 90.00,anchor=base,inner sep=0pt, outer sep=0pt, scale = 0.90] at ( 25.20, 88.94) {20};

\node[text=drawColor,rotate= 90.00,anchor=base,inner sep=0pt, outer sep=0pt, scale = 0.90] at ( 25.20,115.73) {30};

\node[text=drawColor,rotate= 90.00,anchor=base,inner sep=0pt, outer sep=0pt, scale = 0.90] at ( 25.20,142.52) {40};

\node[text=drawColor,rotate= 90.00,anchor=base,inner sep=0pt, outer sep=0pt, scale = 0.90] at ( 25.20,169.32) {50};

\path[draw=drawColor,line width= 0.4pt,line join=round,line cap=round, thick] ( 30.00, 30.00) --
	(331.62, 30.00) --
	(331.62,174.67) --
	( 30.00,174.67) --
	( 30.00, 30.00);
\end{scope}
\begin{scope}
\definecolor[named]{drawColor}{rgb}{0.00,0.00,0.00}

\node[text=drawColor,anchor=base,inner sep=0pt, outer sep=0pt, scale=  0.90] at (180.81,  2.40+5) {Percent contamination in each variable};

\node[text=drawColor,rotate= 90.00,anchor=base,inner sep=0pt, outer sep=0pt, scale = 0.90] at (  9.60,102.34) {Entropy loss for $\hat{\bm\Sigma}$};
\end{scope}
\begin{scope}
\path[clip] ( 30.00, 30.00) rectangle (331.62,174.67);
\definecolor[named]{drawColor}{rgb}{0.00,0.00,0.00}

\path[draw=orange,line width= 0.8pt,dash pattern=on 4pt off 4pt ,line join=round,line cap=round] ( 41.17, 59.60) --
	( 52.34, 61.38) --
	( 63.51, 63.39) --
	( 74.68, 65.32) --
	( 85.86, 67.31) --
	( 97.03, 69.75) --
	(108.20, 73.27) --
	(119.37, 76.53) --
	(130.54, 80.55) --
	(141.71, 84.42) --
	(152.88, 88.49) --
	(164.05, 94.86) --
	(175.22,100.83) --
	(186.40,108.12) --
	(197.57,115.35) --
	(208.74,122.44) --
	(219.91,131.73) --
	(231.08,142.86) --
	(242.25,153.77) --
	(253.42,165.01) --
	(264.59,179.16) --
	(265.71,180.67);

\path[draw=drawColor,line width= 0.8pt,dash pattern=on 1pt off 3pt ,line join=round,line cap=round] ( 41.17, 61.47) --
	( 52.34, 63.05) --
	( 63.51, 64.95) --
	( 74.68, 66.72) --
	( 85.86, 68.25) --
	( 97.03, 70.71) --
	(108.20, 73.86) --
	(119.37, 76.85) --
	(130.54, 80.54) --
	(141.71, 83.75) --
	(152.88, 87.39) --
	(164.05, 93.81) --
	(175.22, 99.05) --
	(186.40,106.03) --
	(197.57,113.34) --
	(208.74,119.78) --
	(219.91,128.29) --
	(231.08,139.19) --
	(242.25,150.20) --
	(253.42,159.81) --
	(264.59,174.61) --
	(269.26,180.67);

\path[draw=drawColor,line width= 0.8pt,dash pattern=on 1pt off 3pt on 4pt off 3pt ,line join=round,line cap=round] ( 41.17, 56.51) --
	( 52.34, 96.78) --
	( 61.43,180.67);

\path[draw=blue,line width= 0.8pt,dash pattern=on 7pt off 3pt ,line join=round,line cap=round] ( 41.17, 59.20) --
	( 52.34, 60.76) --
	( 63.51, 63.03) --
	( 74.68, 64.78) --
	( 85.86, 66.74) --
	( 97.03, 69.63) --
	(108.20, 73.50) --
	(119.37, 77.57) --
	(130.54, 82.54) --
	(141.71, 87.15) --
	(152.88, 93.26) --
	(164.05,102.52) --
	(175.22,111.50) --
	(186.40,123.92) --
	(197.57,139.83) --
	(208.74,157.83) --
	(218.10,180.67);

\path[draw=red,line width= 0.8pt,dash pattern=on 2pt off 2pt on 6pt off 2pt ,line join=round,line cap=round] ( 41.17, 60.31) --
	( 52.34, 60.98) --
	( 63.51, 61.91) --
	( 74.68, 62.29) --
	( 85.86, 62.79) --
	( 97.03, 63.58) --
	(108.20, 65.06) --
	(119.37, 66.29) --
	(130.54, 67.87) --
	(141.71, 68.97) --
	(152.88, 70.90) --
	(164.05, 72.91) --
	(175.22, 75.59) --
	(186.40, 78.39) --
	(197.57, 81.41) --
	(208.74, 85.38) --
	(219.91, 89.31) --
	(231.08, 95.48) --
	(242.25,102.38) --
	(253.42,110.30) --
	(264.59,120.27) --
	(275.76,132.46) --
	(286.94,144.73) --
	(298.11,162.46) --
	(309.28,179.58) --
	(309.87,180.67);

\path[draw=drawColor,line width= 0.8pt,line join=round,line cap=round] ( 41.17, 60.83) --
	( 52.34,121.55) --
	( 57.83,180.67);
\end{scope}
\begin{scope}
\path[clip] (  0.00,  0.00) rectangle (460.00,180.67);
\definecolor[named]{drawColor}{rgb}{0.00,0.00,0.00}
\definecolor[named]{fillColor}{rgb}{0.00,0.00,0.00}

\node[text=drawColor,anchor=base west,inner sep=0pt, outer sep=0pt, scale = 0.90] at (367.35,159.23) {$\widetilde{P}_n$ with {NPD}};
\path[draw=red,line width= 0.8pt,dash pattern=on 2pt off 2pt on 6pt off 2pt ,line join=round,line cap=round] (340.35,162.67) -- (358.35,162.67);

\node[text=drawColor,anchor=base west,inner sep=0pt, outer sep=0pt, scale = 0.90] at (367.35,147.23-5) {$Q_n$ with {NPD}};
\path[draw=drawColor,line width= 0.8pt,dash pattern=on 1pt off 3pt ,line join=round,line cap=round] (340.35,150.67-5) -- (358.35,150.67-5);

\node[text=drawColor,anchor=base west,inner sep=0pt, outer sep=0pt, scale = 0.90] at (367.35,135.23-10) {$\tau$ with {NPD}};
\path[draw=orange,line width= 0.8pt,dash pattern=on 4pt off 4pt ,line join=round,line cap=round] (340.35,138.67-10) -- (358.35,138.67-10);

\node[text=drawColor,anchor=base west,inner sep=0pt, outer sep=0pt, scale = 0.90] at (367.35,123.23-15) {$P_n$ with {NPD}};
\path[draw=blue,line width= 0.8pt,dash pattern=on 7pt off 3pt ,line join=round,line cap=round] (340.35,126.67-15) -- (358.35,126.67-15);

\node[text=drawColor,anchor=base west,inner sep=0pt, outer sep=0pt, scale = 0.90] at (367.35,111.23-20) {{MCD}};
\path[draw=drawColor,line width= 0.8pt,dash pattern=on 1pt off 3pt on 4pt off 3pt ,line join=round,line cap=round] (340.35,102.67-25) -- (358.35,102.67-25);

\node[text=drawColor,anchor=base west,inner sep=0pt, outer sep=0pt, scale = 0.90] at (367.35, 99.23-25)  {$P_n$ with {OGK}};
\path[draw=drawColor,line width= 0.8pt,line join=round,line cap=round] (340.35,114.67-20) -- (358.35,114.67-20);

\node[text=drawColor,anchor=base west,inner sep=0pt, outer sep=0pt, scale = 0.90] at (367.35, 87.23-30) {Classical};
\path[draw=drawColor,line width= 0.8pt,line join=round,line cap=round] (340.35, 90.67-30) -- (358.35, 90.67-30);
\path[draw=drawColor,line width= 0.8pt,line join=round,line cap=round,fill=fillColor] (349.35,162.67-72-30) circle (  1.50);
\end{scope}
\end{tikzpicture}

\begin{tikzpicture}[x=1pt,y=1pt]
\definecolor[named]{fillColor}{rgb}{1.00,1.00,1.00}
\path[use as bounding box,fill=fillColor,fill opacity=0.00] (-2,0) rectangle (433.62,180.67);
\begin{scope}
\path[clip] ( 30.00, 30.00) rectangle (331.62,174.67);
\definecolor[named]{drawColor}{rgb}{0.00,0.00,0.00}

\path[draw=drawColor,line width= 0.8pt,line join=round,line cap=round] ( 41.17, 47.81) --
	( 52.34,103.04) --
	( 63.51,145.50) --
	( 74.68,175.09) --
	( 76.19,180.67);
\definecolor[named]{fillColor}{rgb}{0.00,0.00,0.00}

\path[draw=drawColor,line width= 0.8pt,line join=round,line cap=round,fill=fillColor] ( 41.17, 47.81) circle (  1.50);

\path[draw=drawColor,line width= 0.8pt,line join=round,line cap=round,fill=fillColor] ( 52.34,103.04) circle (  1.50);

\path[draw=drawColor,line width= 0.8pt,line join=round,line cap=round,fill=fillColor] ( 63.51,145.50) circle (  1.50);

\path[draw=drawColor,line width= 0.8pt,line join=round,line cap=round,fill=fillColor] ( 74.68,175.09) circle (  1.50);
\end{scope}
\begin{scope}
\path[clip] (  0.00,  0.00) rectangle (460.00,180.67);
\definecolor[named]{drawColor}{rgb}{0.00,0.00,0.00}


\path[draw=drawColor,line width= 0.4pt,line join=round,line cap=round] ( 41.17, 30.00) -- ( 41.17, 32.89);

\path[draw=drawColor,line width= 0.4pt,line join=round,line cap=round] ( 97.03, 30.00) -- ( 97.03, 32.89);

\path[draw=drawColor,line width= 0.4pt,line join=round,line cap=round] (152.88, 30.00) -- (152.88, 32.89);

\path[draw=drawColor,line width= 0.4pt,line join=round,line cap=round] (208.74, 30.00) -- (208.74, 32.89);

\path[draw=drawColor,line width= 0.4pt,line join=round,line cap=round] (264.59, 30.00) -- (264.59, 32.89);

\path[draw=drawColor,line width= 0.4pt,line join=round,line cap=round] (320.45, 30.00) -- (320.45, 32.89);

\node[text=drawColor,anchor=base,inner sep=0pt, outer sep=0pt, scale = 0.90] at ( 41.17, 18.00) {0};

\node[text=drawColor,anchor=base,inner sep=0pt, outer sep=0pt, scale = 0.90] at ( 97.03, 18.00) {5};

\node[text=drawColor,anchor=base,inner sep=0pt, outer sep=0pt, scale = 0.90] at (152.88, 18.00) {10};

\node[text=drawColor,anchor=base,inner sep=0pt, outer sep=0pt, scale = 0.90] at (208.74, 18.00) {15};

\node[text=drawColor,anchor=base,inner sep=0pt, outer sep=0pt, scale = 0.90] at (264.59, 18.00) {20};

\node[text=drawColor,anchor=base,inner sep=0pt, outer sep=0pt, scale = 0.90] at (320.45, 18.00) {25};


\path[draw=drawColor,line width= 0.4pt,line join=round,line cap=round] ( 30.00, 35.36) -- ( 32.89, 35.36);

\path[draw=drawColor,line width= 0.4pt,line join=round,line cap=round] ( 30.00, 62.15) -- ( 32.89, 62.15);

\path[draw=drawColor,line width= 0.4pt,line join=round,line cap=round] ( 30.00, 88.94) -- ( 32.89, 88.94);

\path[draw=drawColor,line width= 0.4pt,line join=round,line cap=round] ( 30.00,115.73) -- ( 32.89,115.73);

\path[draw=drawColor,line width= 0.4pt,line join=round,line cap=round] ( 30.00,142.52) -- ( 32.89,142.52);

\path[draw=drawColor,line width= 0.4pt,line join=round,line cap=round] ( 30.00,169.32) -- ( 32.89,169.32);

\node[text=drawColor,rotate= 90.00,anchor=base,inner sep=0pt, outer sep=0pt, scale = 0.90] at ( 25.20, 35.36) {0};

\node[text=drawColor,rotate= 90.00,anchor=base,inner sep=0pt, outer sep=0pt, scale = 0.90] at ( 25.20, 62.15) {10};

\node[text=drawColor,rotate= 90.00,anchor=base,inner sep=0pt, outer sep=0pt, scale = 0.90] at ( 25.20, 88.94) {20};

\node[text=drawColor,rotate= 90.00,anchor=base,inner sep=0pt, outer sep=0pt, scale = 0.90] at ( 25.20,115.73) {30};

\node[text=drawColor,rotate= 90.00,anchor=base,inner sep=0pt, outer sep=0pt, scale = 0.90] at ( 25.20,142.52) {40};

\node[text=drawColor,rotate= 90.00,anchor=base,inner sep=0pt, outer sep=0pt, scale = 0.90] at ( 25.20,169.32) {50};

\path[draw=drawColor,line width= 0.4pt,line join=round,line cap=round, thick] ( 30.00, 30.00) --
	(331.62, 30.00) --
	(331.62,174.67) --
	( 30.00,174.67) --
	( 30.00, 30.00);
\end{scope}
\begin{scope}
\definecolor[named]{drawColor}{rgb}{0.00,0.00,0.00}

\node[text=drawColor,anchor=base,inner sep=0pt, outer sep=0pt, scale = 0.90] at (180.81,  2.40+5) {Percent contamination in each variable};

\node[text=drawColor,rotate= 90.00,anchor=base,inner sep=0pt, outer sep=0pt, scale = 0.90] at (  9.60,102.34) {Frobenius norm for $\hat{\bm\Sigma}$};
\end{scope}
\begin{scope}
\path[clip] ( 30.00, 30.00) rectangle (331.62,174.67);
\definecolor[named]{drawColor}{rgb}{0.00,0.00,0.00}

\path[draw=orange,line width= 0.8pt,dash pattern=on 4pt off 4pt ,line join=round,line cap=round] ( 41.17, 49.34) --
	( 52.34, 50.16) --
	( 63.51, 51.12) --
	( 74.68, 52.09) --
	( 85.86, 52.97) --
	( 97.03, 54.09) --
	(108.20, 55.67) --
	(119.37, 56.99) --
	(130.54, 58.69) --
	(141.71, 60.35) --
	(152.88, 61.90) --
	(164.05, 64.24) --
	(175.22, 66.36) --
	(186.40, 68.93) --
	(197.57, 71.30) --
	(208.74, 73.71) --
	(219.91, 76.80) --
	(231.08, 79.99) --
	(242.25, 83.17) --
	(253.42, 86.56) --
	(264.59, 90.48) --
	(275.76, 94.82) --
	(286.94, 98.98) --
	(298.11,104.12) --
	(309.28,108.77) --
	(320.45,114.16);

\path[draw=drawColor,line width= 0.8pt,dash pattern=on 1pt off 3pt ,line join=round,line cap=round] ( 41.17, 50.23) --
	( 52.34, 51.00) --
	( 63.51, 51.90) --
	( 74.68, 52.80) --
	( 85.86, 53.53) --
	( 97.03, 54.69) --
	(108.20, 56.08) --
	(119.37, 57.27) --
	(130.54, 58.80) --
	(141.71, 60.23) --
	(152.88, 61.56) --
	(164.05, 63.93) --
	(175.22, 65.81) --
	(186.40, 68.23) --
	(197.57, 70.67) --
	(208.74, 72.88) --
	(219.91, 75.51) --
	(231.08, 78.70) --
	(242.25, 81.96) --
	(253.42, 84.86) --
	(264.59, 89.04) --
	(275.76, 93.16) --
	(286.94, 97.19) --
	(298.11,101.92) --
	(309.28,107.22) --
	(320.45,112.01);

\path[draw=drawColor,line width= 0.8pt,dash pattern=on 1pt off 3pt on 4pt off 3pt ,line join=round,line cap=round] ( 41.17, 48.05) --
	( 52.34, 62.84) --
	( 63.51, 88.43) --
	( 74.68,116.33) --
	( 85.86,144.28) --
	( 97.03,171.08) --
	(100.88,180.67);

\path[draw=blue,line width= 0.8pt,dash pattern=on 7pt off 3pt ,line join=round,line cap=round] ( 41.17, 49.47) --
	( 52.34, 50.24) --
	( 63.51, 51.28) --
	( 74.68, 52.21) --
	( 85.86, 53.07) --
	( 97.03, 54.43) --
	(108.20, 56.05) --
	(119.37, 57.72) --
	(130.54, 59.69) --
	(141.71, 61.64) --
	(152.88, 63.91) --
	(164.05, 67.18) --
	(175.22, 70.33) --
	(186.40, 74.34) --
	(197.57, 79.16) --
	(208.74, 84.39) --
	(219.91, 91.45) --
	(231.08, 97.44) --
	(242.25,108.05) --
	(253.42,116.11) --
	(264.59,125.01) --
	(275.76,134.25) --
	(286.94,145.58) --
	(298.11,156.58) --
	(309.28,164.45) --
	(320.45,176.09);

\path[draw=red,line width= 0.8pt,dash pattern=on 2pt off 2pt on 6pt off 2pt ,line join=round,line cap=round] ( 41.17, 49.59) --
	( 52.34, 49.78) --
	( 63.51, 50.13) --
	( 74.68, 50.32) --
	( 85.86, 50.40) --
	( 97.03, 50.79) --
	(108.20, 51.40) --
	(119.37, 51.88) --
	(130.54, 52.63) --
	(141.71, 53.20) --
	(152.88, 53.94) --
	(164.05, 55.10) --
	(175.22, 56.30) --
	(186.40, 57.65) --
	(197.57, 59.07) --
	(208.74, 60.77) --
	(219.91, 62.43) --
	(231.08, 65.09) --
	(242.25, 67.73) --
	(253.42, 70.73) --
	(264.59, 74.45) --
	(275.76, 78.50) --
	(286.94, 82.46) --
	(298.11, 88.42) --
	(309.28, 93.21) --
	(320.45, 99.28);

\path[draw=drawColor,line width= 0.8pt,line join=round,line cap=round] ( 41.17, 50.28) --
	( 52.34, 78.10) --
	( 63.51,116.32) --
	( 74.68,150.55) --
	( 84.38,180.67);
\end{scope}
\begin{scope}
\path[clip] (  0.00,  0.00) rectangle (460.00,180.67);
\definecolor[named]{drawColor}{rgb}{0.00,0.00,0.00}
\definecolor[named]{fillColor}{rgb}{0.00,0.00,0.00}

\node[text=drawColor,anchor=base west,inner sep=0pt, outer sep=0pt, scale = 0.90] at (367.35,159.23) {$\widetilde{P}_n$ with {NPD}};
\path[draw=red,line width= 0.8pt,dash pattern=on 2pt off 2pt on 6pt off 2pt ,line join=round,line cap=round] (340.35,162.67) -- (358.35,162.67);

\node[text=drawColor,anchor=base west,inner sep=0pt, outer sep=0pt, scale = 0.90] at (367.35,147.23-5) {$Q_n$ with {NPD}};
\path[draw=drawColor,line width= 0.8pt,dash pattern=on 1pt off 3pt ,line join=round,line cap=round] (340.35,150.67-5) -- (358.35,150.67-5);

\node[text=drawColor,anchor=base west,inner sep=0pt, outer sep=0pt, scale = 0.90] at (367.35,135.23-10) {$\tau$ with {NPD}};
\path[draw=orange,line width= 0.8pt,dash pattern=on 4pt off 4pt ,line join=round,line cap=round] (340.35,138.67-10) -- (358.35,138.67-10);

\node[text=drawColor,anchor=base west,inner sep=0pt, outer sep=0pt, scale = 0.90] at (367.35,123.23-15) {$P_n$ with {NPD}};
\path[draw=blue,line width= 0.8pt,dash pattern=on 7pt off 3pt ,line join=round,line cap=round] (340.35,126.67-15) -- (358.35,126.67-15);

\node[text=drawColor,anchor=base west,inner sep=0pt, outer sep=0pt, scale = 0.90] at (367.35,111.23-20) {{MCD}};
\path[draw=drawColor,line width= 0.8pt,dash pattern=on 1pt off 3pt on 4pt off 3pt ,line join=round,line cap=round] (340.35,102.67-25) -- (358.35,102.67-25);

\node[text=drawColor,anchor=base west,inner sep=0pt, outer sep=0pt, scale = 0.90] at (367.35, 99.23-25)  {$P_n$ with {OGK}};
\path[draw=drawColor,line width= 0.8pt,line join=round,line cap=round] (340.35,114.67-20) -- (358.35,114.67-20);

\node[text=drawColor,anchor=base west,inner sep=0pt, outer sep=0pt, scale = 0.90] at (367.35, 87.23-30) {Classical};
\path[draw=drawColor,line width= 0.8pt,line join=round,line cap=round] (340.35, 90.67-30) -- (358.35, 90.67-30);
\path[draw=drawColor,line width= 0.8pt,line join=round,line cap=round,fill=fillColor] (349.35,162.67-72-30) circle (  1.50);
\end{scope}
\end{tikzpicture}
\caption{Average entropy loss results for the precision matrices resulting from the CLIME procedure (top) and average entropy loss and Frobenius norm results for the resulting covariance matrix estimates (middle and bottom) for $p=60$ with scattered sparsity and extreme outliers.}
\label{clime.covp60normFexpt2k100}
\end{figure}

It is also constructive to see how $\hat{\bm{\Sigma}}=\hat{\bm{\Theta}}^{-1}$, the inverse of the estimated regularised precision matrix,  compares with the true covariance matrix $\bm\Sigma$.  Figure \ref{clime.covp60normFexpt2k100} presents the average entropy loss and Frobenius norm results for the resulting estimated covariance matrices after regularisation using the CLIME procedure.  We see similar trends to those outlined earlier.  While using $P_n$ alone does not perform well when the amount of contamination in each variable is large, the adaptive trimming procedure gives excellent results.  The other pairwise methods also perform quite well.  However, as we would expect, the classical method and standard robust techniques, MCD and OGK fail quite rapidly.  In general, the patterns for the matrix norms applied to $\hat{\bm\Sigma}$ are very similar to those of the entropy loss for $\hat{\bm\Theta}$.   That is, the robust covariance matrices that are obtained at the end of the proposed procedure perform similarly well to the robust regularised precision matrices. 

There are also important differences between $\hat{\bm\Sigma}$ and the initial pairwise covariance matrix obtained after applying the NPD procedure.  While the matrix norms for the the initial pairwise matrices were comparable to those for $\hat{\bm\Sigma}$, the entropy loss and log determinant results for the initial pairwise covariance matrices were much worse due to small eigenvalues resulting from the NPD method.  As such, we would not recommend simply applying the NPD procedure to a pairwise covariance matrix without further regularisation. 

\subsection{Gaussian graphical discovery rates}

Another way to analyse the performance of a precision matrix estimator is through the lens of a Gaussian graphical model.  When the data follow a multivariate Gaussian distribution, pairwise conditional independence between variables $X_j$ and $X_k$ holds if and only if $\theta_{jk}=0$, therefore inferring linkages between variables corresponds to identifying the nonzero elements of $\bm{\Theta}=(\theta_{jk})$, see \citet{Lauritzen:1996} for further details.  Hence, rather than focussing on overall measures of similarity between the estimated precision and the true precision matrix, it can be informative to see how often the estimated precision matrix identifies the correct non-zero elements from the true precision matrix.  

\begin{figure}[p]
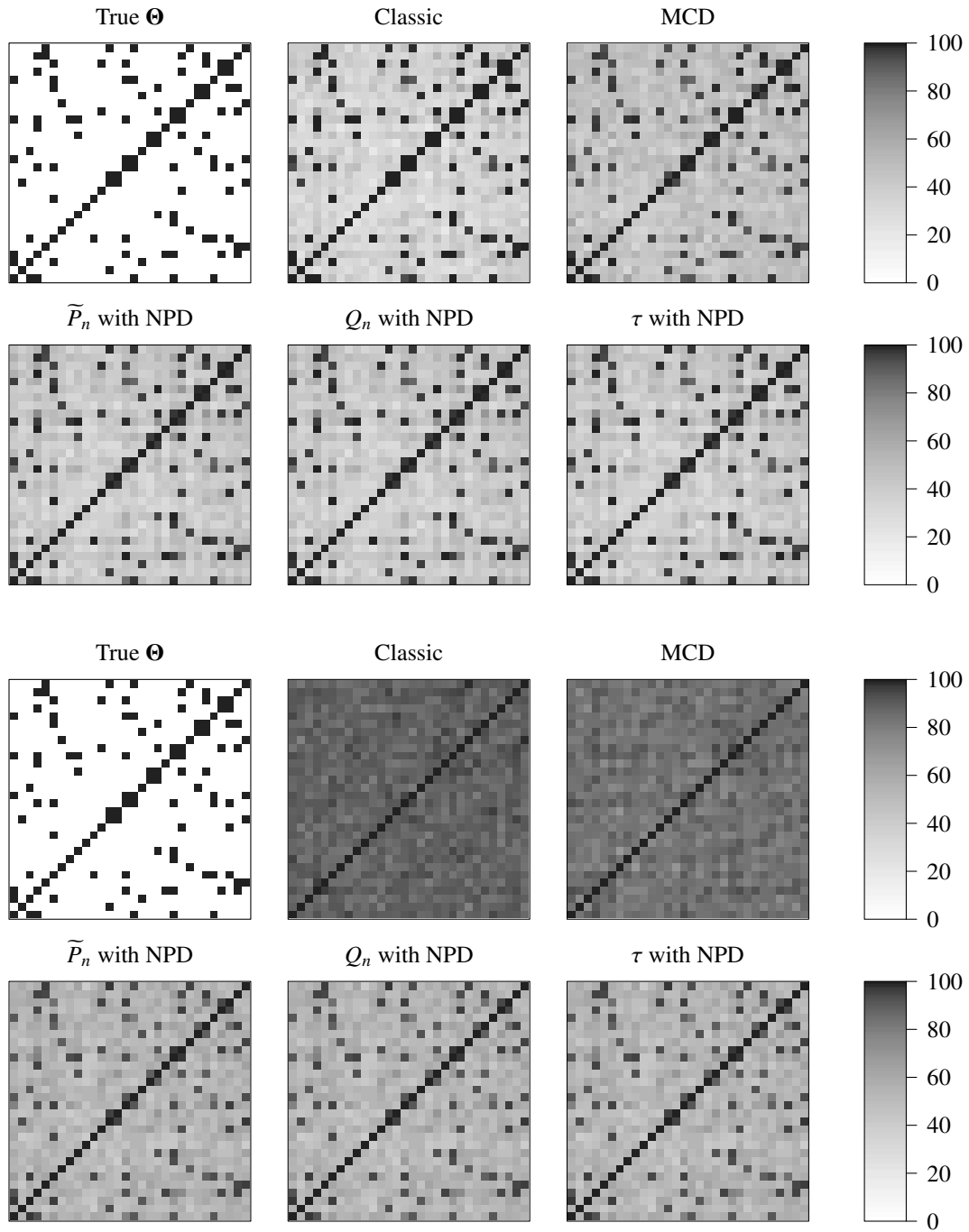

\centering
\input{TR}
\input{A}
\input{B}

\input{C}
\input{D}
\input{E}

\

\input{TR}
\input{A10}
\input{B10}

\input{C10}
\input{D10}
\input{E10}

\caption{Heat maps showing how often each element in the precision matrix is identified as being non-zero using the QUIC routine over 100 replications.  The top half have no contamination and the bottom half have 10\% extreme contamination.}
\label{fig:heuristic}
\end{figure}

Figure \ref{fig:heuristic} shows visually how well the QUIC estimator performs in the presence of cellwise contamination.  When there is no contamination, all methods appear to perform similarly well in terms of their ability to correctly identify the true non-zero elements in the precision matrix.  However, in the presence of 10\% extreme contamination, the classical covariance and the MCD approach both fail to identify any structure as they tend to return overly dense precision matrices.  On the other hand, the pairwise robust methods are, on average, still able to identify the underlying structure.

In the machine learning literature, the Matthews correlation coefficient (MCC), also known as the $\phi$ coefficient in the statistics literature, is often used to assess the ability of an estimator to identify the true non-zero elements in a precision matrix \citep{Matthews:1975}.  It takes into account the number of true positives (TP), false positives (FP), true negatives (TN) and false negatives (FN),
$$\operatorname{MCC} = \frac{\mathrm{TP}\times \mathrm{TN} - \mathrm{FP}\times\mathrm{FN}}{\sqrt{(\mathrm{TP}+\mathrm{FP})(\mathrm{TP}+\mathrm{FN})(\mathrm{TN}+\mathrm{FP})(\mathrm{TN}+\mathrm{FN})}}.$$

\begin{figure}[htbp]
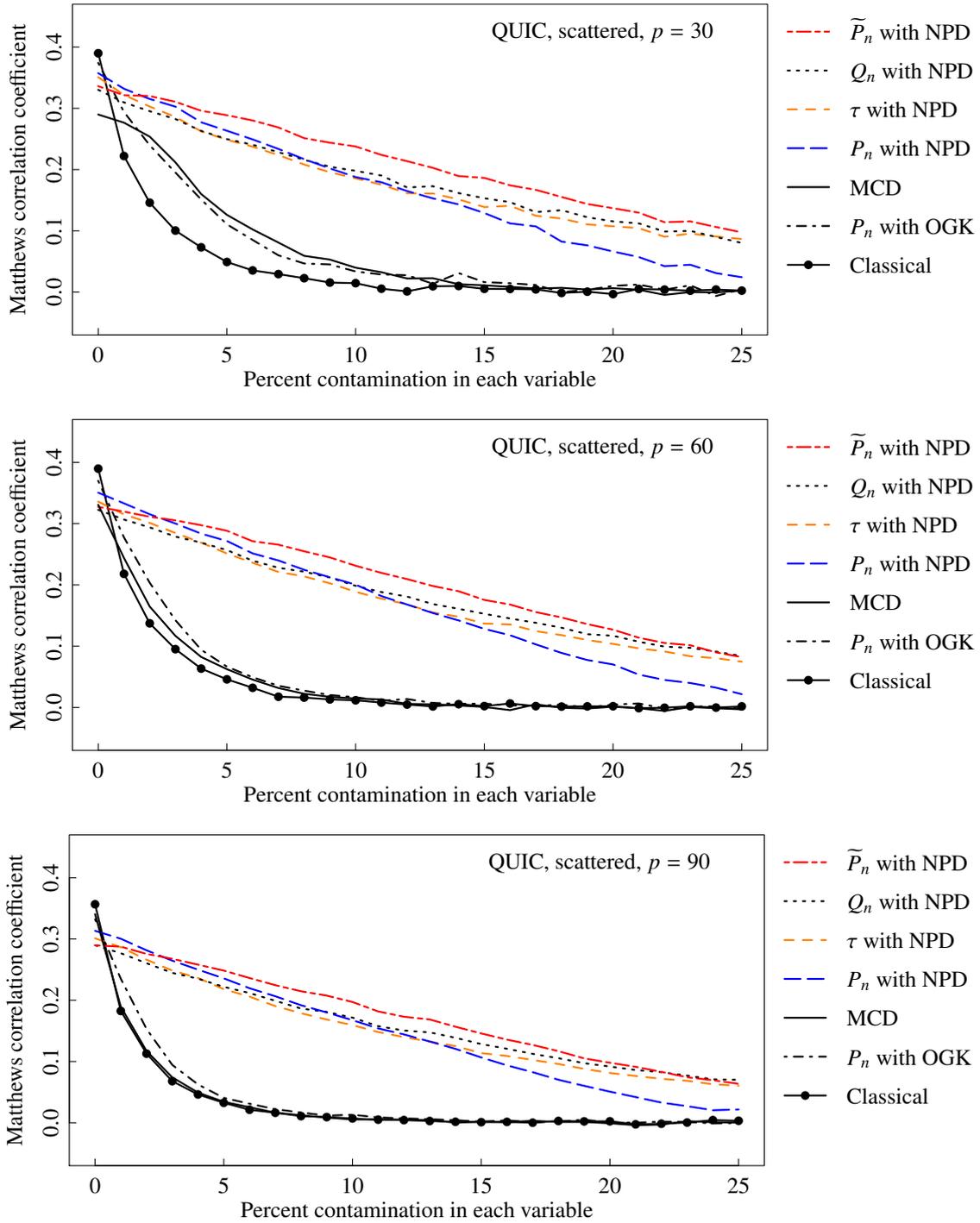

\centering
\input{quicp30MCCexpt2k100}
\input{quicp60MCCexpt2k100}
\input{quicp90MCCexpt2k100}
\caption{Matthews correlation coefficient results for the QUIC procedure with extreme outliers, $p=30$ (top), $p=60$ (middle) and $p=90$ (bottom) in the scattered sparsity precision matrix scenario.}
\label{fig:MCC}
\end{figure}

Typical MCC results are given in Figure \ref{fig:MCC} for the QUIC procedure.  When cellwise contamination is introduced, the MCD, OGK and classical covariance approaches lose their ability to identify the true structure in the precision matrix quite quickly.  The pairwise methods are much more resilient.  As the number of contaminated observations in each variable increases, the ability of the pairwise methods to identify the true structure decreases gradually.  When there is no contamination and $p=60$, the classical method has an MCC of 0.39 compared with an MCC of 0.35 for the pairwise method based on $P_{n}$.  When there is 5\% extreme contamination in each variable, the MCC for the $P_{n}$ based method is still at 0.29, while the classical approach is at 0.05.  This pattern of results is virtually unchanged for the pairwise methods when the contamination is less extreme.

\section{Simulation study for $p>n$}
\label{MVsimn50}

Of particular interest in a data mining context is the case when the number of variables $p$, is larger than the number of observations $n$.  In order to perform simulations in a reasonable amount of time we reformulated the simulation settings in Section \ref{MVsim} such that samples of size $n=50$ were drawn with dimension $p=15$, $30$, $60$ and $90$.  The same types of precision matrices were considered as in the previous section, though we also looked at the case where the condition number of the sparse precision matrix was much larger than the dimension.  Given the similar performance of the various regularisation routines found in the previous section, we restricted attention to the QUIC routine.

\begin{figure}[htbp]
\centering
\input{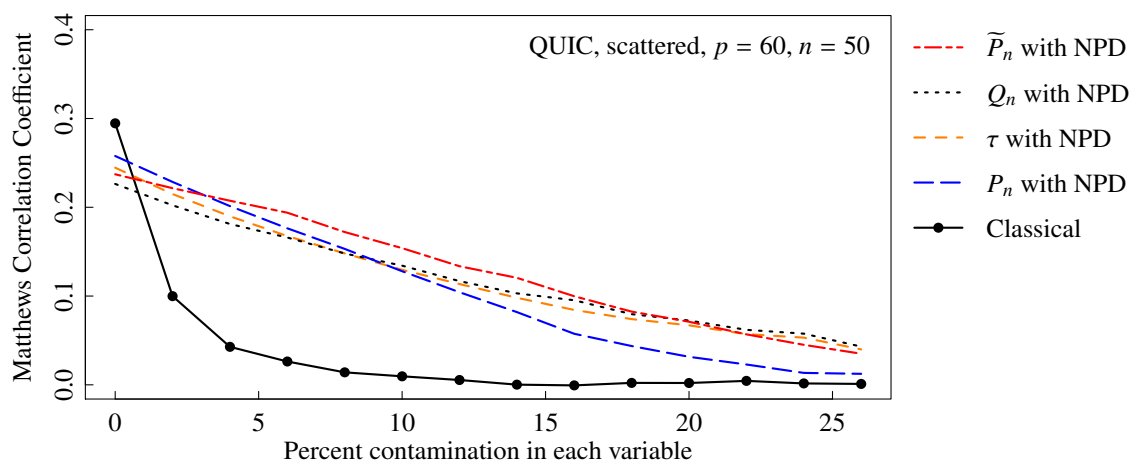}
\caption{Matthews correlation coefficients results for the QUIC procedure for a scattered precision matrix with $p=60$, $n=50$, and extreme outliers.}
\label{fig:p60MCC}
\end{figure}

\begin{figure}[htbp]
\centering
\begin{tikzpicture}[x=1pt,y=1pt]
\definecolor[named]{fillColor}{rgb}{1.00,1.00,1.00}
\path[use as bounding box,fill=fillColor,fill opacity=0.00] (0,0) rectangle (433.62,180.67);
\begin{scope}
\path[clip] ( 30.00, 30.00) rectangle (331.62,174.67);
\definecolor[named]{drawColor}{rgb}{0.00,0.00,0.00}

\path[draw=drawColor,line width= 0.8pt,line join=round,line cap=round] ( 41.17,169.32) --
	( 62.65, 72.85) --
	( 84.14,  6.76) --
	( 86.95,  0.00);
\definecolor[named]{fillColor}{rgb}{0.00,0.00,0.00}

\path[draw=drawColor,line width= 0.8pt,line join=round,line cap=round,fill=fillColor] ( 41.17,169.32) circle (  1.50);

\path[draw=drawColor,line width= 0.8pt,line join=round,line cap=round,fill=fillColor] ( 62.65, 72.85) circle (  1.50);

\path[draw=drawColor,line width= 0.8pt,line join=round,line cap=round,fill=fillColor] ( 84.14,  6.76) circle (  1.50);
\end{scope}
\begin{scope}
\path[clip] (  0.00,  0.00) rectangle (433.62,180.67);
\definecolor[named]{drawColor}{rgb}{0.00,0.00,0.00}

\path[draw=drawColor,line width= 0.4pt,line join=round,line cap=round] ( 41.17, 30.00) -- (309.71, 30.00);

\path[draw=drawColor,line width= 0.4pt,line join=round,line cap=round] ( 41.17, 30.00) -- ( 41.17, 32.89);

\path[draw=drawColor,line width= 0.4pt,line join=round,line cap=round] ( 94.88, 30.00) -- ( 94.88, 32.89);

\path[draw=drawColor,line width= 0.4pt,line join=round,line cap=round] (148.59, 30.00) -- (148.59, 32.89);

\path[draw=drawColor,line width= 0.4pt,line join=round,line cap=round] (202.29, 30.00) -- (202.29, 32.89);

\path[draw=drawColor,line width= 0.4pt,line join=round,line cap=round] (256.00, 30.00) -- (256.00, 32.89);

\path[draw=drawColor,line width= 0.4pt,line join=round,line cap=round] (309.71, 30.00) -- (309.71, 32.89);

\node[text=drawColor,anchor=base,inner sep=0pt, outer sep=0pt, scale=  0.90] at ( 41.17, 18.00) {0};

\node[text=drawColor,anchor=base,inner sep=0pt, outer sep=0pt, scale=  0.90] at ( 94.88, 18.00) {5};

\node[text=drawColor,anchor=base,inner sep=0pt, outer sep=0pt, scale=  0.90] at (148.59, 18.00) {10};

\node[text=drawColor,anchor=base,inner sep=0pt, outer sep=0pt, scale=  0.90] at (202.29, 18.00) {15};

\node[text=drawColor,anchor=base,inner sep=0pt, outer sep=0pt, scale=  0.90] at (256.00, 18.00) {20};

\node[text=drawColor,anchor=base,inner sep=0pt, outer sep=0pt, scale=  0.90] at (309.71, 18.00) {25};

\path[draw=drawColor,line width= 0.4pt,line join=round,line cap=round] ( 30.00, 35.36) -- ( 30.00,169.32);

\path[draw=drawColor,line width= 0.4pt,line join=round,line cap=round] ( 30.00, 35.36) -- ( 32.89, 35.36);

\path[draw=drawColor,line width= 0.4pt,line join=round,line cap=round] ( 30.00, 57.68) -- ( 32.89, 57.68);

\path[draw=drawColor,line width= 0.4pt,line join=round,line cap=round] ( 30.00, 80.01) -- ( 32.89, 80.01);

\path[draw=drawColor,line width= 0.4pt,line join=round,line cap=round] ( 30.00,102.34) -- ( 32.89,102.34);

\path[draw=drawColor,line width= 0.4pt,line join=round,line cap=round] ( 30.00,124.66) -- ( 32.89,124.66);

\path[draw=drawColor,line width= 0.4pt,line join=round,line cap=round] ( 30.00,146.99) -- ( 32.89,146.99);

\path[draw=drawColor,line width= 0.4pt,line join=round,line cap=round] ( 30.00,169.32) -- ( 32.89,169.32);

\node[text=drawColor,rotate= 90.00,anchor=base,inner sep=0pt, outer sep=0pt, scale = 0.90] at ( 25.20, 35.36) {-300};

\node[text=drawColor,rotate= 90.00,anchor=base,inner sep=0pt, outer sep=0pt, scale = 0.90] at ( 25.20, 80.01) {-200};

\node[text=drawColor,rotate= 90.00,anchor=base,inner sep=0pt, outer sep=0pt, scale = 0.90] at ( 25.20,124.66) {-100};

\node[text=drawColor,rotate= 90.00,anchor=base,inner sep=0pt, outer sep=0pt, scale = 0.90] at ( 25.20,169.32) {0};

\path[draw=drawColor,line width= 0.4pt,line join=round,line cap=round] ( 30.00, 30.00) --
	(331.62, 30.00) --
	(331.62,174.67) --
	( 30.00,174.67) --
	( 30.00, 30.00);
\end{scope}
\begin{scope}
\path[clip] (  0.00,  0.00) rectangle (433.62,180.67);
\definecolor[named]{drawColor}{rgb}{0.00,0.00,0.00}

\node[text=drawColor,anchor=base,inner sep=0pt, outer sep=0pt, scale=  0.90] at (180.81,  2.40+5) {Percent contamination in each variable};

\node[text=drawColor,rotate= 90.00,anchor=base,inner sep=0pt, outer sep=0pt, scale = 0.90] at (  9.60,102.34) {Entropy loss PRIAL};
\end{scope}
\begin{scope}
\path[clip] ( 30.00, 30.00) rectangle (331.62,174.67);
\definecolor[named]{drawColor}{rgb}{0.00,0.00,0.00}

\path[draw=orange,line width= 0.8pt,dash pattern=on 4pt off 4pt ,line join=round,line cap=round] ( 41.17,164.85) --
	( 62.65,159.63) --
	( 84.14,153.43) --
	(105.62,145.43) --
	(127.10,136.80) --
	(148.59,127.05) --
	(170.07,116.14) --
	(191.55,104.19) --
	(213.03, 91.43) --
	(234.52, 78.85) --
	(256.00, 64.83) --
	(277.48, 50.10) --
	(298.97, 34.60) --
	(320.45, 17.29);

\path[draw=drawColor,line width= 0.8pt,dash pattern=on 1pt off 3pt ,line join=round,line cap=round] ( 41.17,161.75) --
	( 62.65,156.41) --
	( 84.14,150.33) --
	(105.62,142.58) --
	(127.10,134.61) --
	(148.59,125.15) --
	(170.07,114.63) --
	(191.55,103.15) --
	(213.03, 90.96) --
	(234.52, 78.34) --
	(256.00, 64.75) --
	(277.48, 50.19) --
	(298.97, 35.58) --
	(320.45, 18.85);

\path[draw=blue,line width= 0.8pt,dash pattern=on 7pt off 3pt ,line join=round,line cap=round] ( 41.17,165.53) --
	( 62.65,159.86) --
	( 84.14,152.52) --
	(105.62,142.89) --
	(127.10,130.35) --
	(148.59,115.69) --
	(170.07, 97.19) --
	(191.55, 78.67) --
	(213.03, 58.14) --
	(234.52, 37.26) --
	(256.00, 13.23) --
	(268.58,  0.00);

\path[draw=red,line width= 0.8pt,dash pattern=on 2pt off 2pt on 6pt off 2pt,line join=round,line cap=round] ( 41.17,163.39) --
	( 62.65,160.61) --
	( 84.14,157.66) --
	(105.62,154.05) --
	(127.10,148.85) --
	(148.59,141.94) --
	(170.07,133.90) --
	(191.55,123.63) --
	(213.03,110.55) --
	(234.52, 96.22) --
	(256.00, 78.40) --
	(277.48, 59.87) --
	(298.97, 40.09) --
	(320.45, 19.74);
\end{scope}

\begin{scope}
\path[clip] (  0.00,  0.00) rectangle (460.00,180.67);
\definecolor[named]{drawColor}{rgb}{0.00,0.00,0.00}
\definecolor[named]{fillColor}{rgb}{0.00,0.00,0.00}

\node[text=drawColor,anchor=base west,inner sep=0pt, outer sep=0pt, scale = 0.90] at (367.35,159.23) {$\widetilde{P}_n$ with {NPD}};
\path[draw=red,line width= 0.8pt,dash pattern=on 2pt off 2pt on 6pt off 2pt ,line join=round,line cap=round] (340.35,162.67) -- (358.35,162.67);

\node[text=drawColor,anchor=base west,inner sep=0pt, outer sep=0pt, scale = 0.90] at (367.35,147.23-5) {$Q_n$ with {NPD}};
\path[draw=drawColor,line width= 0.8pt,dash pattern=on 1pt off 3pt ,line join=round,line cap=round] (340.35,150.67-5) -- (358.35,150.67-5);

\node[text=drawColor,anchor=base west,inner sep=0pt, outer sep=0pt, scale = 0.90] at (367.35,135.23-10) {$\tau$ with {NPD}};
\path[draw=orange,line width= 0.8pt,dash pattern=on 4pt off 4pt ,line join=round,line cap=round] (340.35,138.67-10) -- (358.35,138.67-10);

\node[text=drawColor,anchor=base west,inner sep=0pt, outer sep=0pt, scale = 0.90] at (367.35,123.23-15) {$P_n$ with {NPD}};
\path[draw=blue,line width= 0.8pt,dash pattern=on 7pt off 3pt ,line join=round,line cap=round] (340.35,126.67-15) -- (358.35,126.67-15);
%
%

\node[text=drawColor,anchor=base west,inner sep=0pt, outer sep=0pt, scale = 0.90] at (367.35,111.23-20) {Classical};
\path[draw=drawColor,line width= 0.8pt,line join=round,line cap=round] (340.35,111.23-20 +3.4) -- (358.35,111.23-20 +3.4);
\path[draw=drawColor,line width= 0.8pt,line join=round,line cap=round,fill=fillColor] (349.35,111.23-20 +3.4) circle (  1.50);


\end{scope}
%
%
%
%
%
%
%
%
%
%
%
\end{tikzpicture}
\begin{tikzpicture}[x=1pt,y=1pt]
\definecolor[named]{fillColor}{rgb}{1.00,1.00,1.00}
\path[use as bounding box,fill=fillColor,fill opacity=0.00] (0,0) rectangle (433.62,180.67);
\begin{scope}
\path[clip] ( 30.00, 30.00) rectangle (331.62,174.67);
\definecolor[named]{drawColor}{rgb}{0.00,0.00,0.00}

\path[draw=drawColor,line width= 0.8pt,line join=round,line cap=round] ( 41.17,169.32) --
	( 62.65, 82.81) --
	( 84.14, 26.67) --
	( 95.56,  0.00);
\definecolor[named]{fillColor}{rgb}{0.00,0.00,0.00}

\path[draw=drawColor,line width= 0.8pt,line join=round,line cap=round,fill=fillColor] ( 41.17,169.32) circle (  1.50);

\path[draw=drawColor,line width= 0.8pt,line join=round,line cap=round,fill=fillColor] ( 62.65, 82.81) circle (  1.50);

\path[draw=drawColor,line width= 0.8pt,line join=round,line cap=round,fill=fillColor] ( 84.14, 26.67) circle (  1.50);
\end{scope}
\begin{scope}
\path[clip] (  0.00,  0.00) rectangle (433.62,180.67);
\definecolor[named]{drawColor}{rgb}{0.00,0.00,0.00}

\path[draw=drawColor,line width= 0.4pt,line join=round,line cap=round] ( 41.17, 30.00) -- (309.71, 30.00);

\path[draw=drawColor,line width= 0.4pt,line join=round,line cap=round] ( 41.17, 30.00) -- ( 41.17, 32.89);

\path[draw=drawColor,line width= 0.4pt,line join=round,line cap=round] ( 94.88, 30.00) -- ( 94.88, 32.89);

\path[draw=drawColor,line width= 0.4pt,line join=round,line cap=round] (148.59, 30.00) -- (148.59, 32.89);

\path[draw=drawColor,line width= 0.4pt,line join=round,line cap=round] (202.29, 30.00) -- (202.29, 32.89);

\path[draw=drawColor,line width= 0.4pt,line join=round,line cap=round] (256.00, 30.00) -- (256.00, 32.89);

\path[draw=drawColor,line width= 0.4pt,line join=round,line cap=round] (309.71, 30.00) -- (309.71, 32.89);

\node[text=drawColor,anchor=base,inner sep=0pt, outer sep=0pt, scale=  0.90] at ( 41.17, 18.00) {0};

\node[text=drawColor,anchor=base,inner sep=0pt, outer sep=0pt, scale=  0.90] at ( 94.88, 18.00) {5};

\node[text=drawColor,anchor=base,inner sep=0pt, outer sep=0pt, scale=  0.90] at (148.59, 18.00) {10};

\node[text=drawColor,anchor=base,inner sep=0pt, outer sep=0pt, scale=  0.90] at (202.29, 18.00) {15};

\node[text=drawColor,anchor=base,inner sep=0pt, outer sep=0pt, scale=  0.90] at (256.00, 18.00) {20};

\node[text=drawColor,anchor=base,inner sep=0pt, outer sep=0pt, scale=  0.90] at (309.71, 18.00) {25};

\path[draw=drawColor,line width= 0.4pt,line join=round,line cap=round] ( 30.00, 35.36) -- ( 30.00,169.32);

\path[draw=drawColor,line width= 0.4pt,line join=round,line cap=round] ( 30.00, 35.36) -- ( 32.89, 35.36);

\path[draw=drawColor,line width= 0.4pt,line join=round,line cap=round] ( 30.00, 57.68) -- ( 32.89, 57.68);

\path[draw=drawColor,line width= 0.4pt,line join=round,line cap=round] ( 30.00, 80.01) -- ( 32.89, 80.01);

\path[draw=drawColor,line width= 0.4pt,line join=round,line cap=round] ( 30.00,102.34) -- ( 32.89,102.34);

\path[draw=drawColor,line width= 0.4pt,line join=round,line cap=round] ( 30.00,124.66) -- ( 32.89,124.66);

\path[draw=drawColor,line width= 0.4pt,line join=round,line cap=round] ( 30.00,146.99) -- ( 32.89,146.99);

\path[draw=drawColor,line width= 0.4pt,line join=round,line cap=round] ( 30.00,169.32) -- ( 32.89,169.32);

\node[text=drawColor,rotate= 90.00,anchor=base,inner sep=0pt, outer sep=0pt, scale = 0.90] at ( 25.20, 35.36) {-300};

\node[text=drawColor,rotate= 90.00,anchor=base,inner sep=0pt, outer sep=0pt, scale = 0.90] at ( 25.20, 80.01) {-200};

\node[text=drawColor,rotate= 90.00,anchor=base,inner sep=0pt, outer sep=0pt, scale = 0.90] at ( 25.20,124.66) {-100};

\node[text=drawColor,rotate= 90.00,anchor=base,inner sep=0pt, outer sep=0pt, scale = 0.90] at ( 25.20,169.32) {0};

\path[draw=drawColor,line width= 0.4pt,line join=round,line cap=round] ( 30.00, 30.00) --
	(331.62, 30.00) --
	(331.62,174.67) --
	( 30.00,174.67) --
	( 30.00, 30.00);
\end{scope}
\begin{scope}
\path[clip] (  0.00,  0.00) rectangle (433.62,180.67);
\definecolor[named]{drawColor}{rgb}{0.00,0.00,0.00}

\node[text=drawColor,anchor=base,inner sep=0pt, outer sep=0pt, scale=  0.90] at (180.81,  2.40+5) {Percent contamination in each variable};

\node[text=drawColor,rotate= 90.00,anchor=base,inner sep=0pt, outer sep=0pt, scale = 0.90] at (  9.60,102.34) {Entropy loss PRIAL};
\end{scope}
\begin{scope}
\path[clip] ( 30.00, 30.00) rectangle (331.62,174.67);
\definecolor[named]{drawColor}{rgb}{0.00,0.00,0.00}

\path[draw=orange,line width= 0.8pt,dash pattern=on 4pt off 4pt ,line join=round,line cap=round] ( 41.17,159.92) --
	( 62.65,153.47) --
	( 84.14,146.27) --
	(105.62,138.36) --
	(127.10,128.91) --
	(148.59,118.27) --
	(170.07,107.83) --
	(191.55, 96.55) --
	(213.03, 84.73) --
	(234.52, 72.83) --
	(256.00, 59.78) --
	(277.48, 46.07) --
	(298.97, 32.69) --
	(320.45, 18.38);

\path[draw=drawColor,line width= 0.8pt,dash pattern=on 1pt off 3pt ,line join=round,line cap=round] ( 41.17,153.97) --
	( 62.65,148.27) --
	( 84.14,141.89) --
	(105.62,135.01) --
	(127.10,126.53) --
	(148.59,116.22) --
	(170.07,106.64) --
	(191.55, 95.90) --
	(213.03, 83.99) --
	(234.52, 72.70) --
	(256.00, 60.03) --
	(277.48, 46.14) --
	(298.97, 33.11) --
	(320.45, 18.96);

\path[draw=blue,line width= 0.8pt,dash pattern=on 7pt off 3pt ,line join=round,line cap=round] ( 41.17,161.74) --
	( 62.65,155.20) --
	( 84.14,147.81) --
	(105.62,137.78) --
	(127.10,126.08) --
	(148.59,111.21) --
	(170.07, 95.40) --
	(191.55, 78.36) --
	(213.03, 61.00) --
	(234.52, 39.91) --
	(256.00, 20.82) --
	(276.79,  0.00);

\path[draw=red,line width= 0.8pt,dash pattern=on 2pt off 2pt on 6pt off 2pt,line join=round,line cap=round] ( 41.17,157.74) --
	( 62.65,153.67) --
	( 84.14,149.45) --
	(105.62,143.75) --
	(127.10,137.27) --
	(148.59,128.59) --
	(170.07,119.62) --
	(191.55,109.19) --
	(213.03, 96.31) --
	(234.52, 83.93) --
	(256.00, 68.54) --
	(277.48, 52.62) --
	(298.97, 35.02) --
	(320.45, 18.88);
\end{scope}

\begin{scope}
\path[clip] (  0.00,  0.00) rectangle (460.00,180.67);
\definecolor[named]{drawColor}{rgb}{0.00,0.00,0.00}
\definecolor[named]{fillColor}{rgb}{0.00,0.00,0.00}

\node[text=drawColor,anchor=base west,inner sep=0pt, outer sep=0pt, scale = 0.90] at (367.35,159.23) {$\widetilde{P}_n$ with {NPD}};
\path[draw=red,line width= 0.8pt,dash pattern=on 2pt off 2pt on 6pt off 2pt ,line join=round,line cap=round] (340.35,162.67) -- (358.35,162.67);

\node[text=drawColor,anchor=base west,inner sep=0pt, outer sep=0pt, scale = 0.90] at (367.35,147.23-5) {$Q_n$ with {NPD}};
\path[draw=drawColor,line width= 0.8pt,dash pattern=on 1pt off 3pt ,line join=round,line cap=round] (340.35,150.67-5) -- (358.35,150.67-5);

\node[text=drawColor,anchor=base west,inner sep=0pt, outer sep=0pt, scale = 0.90] at (367.35,135.23-10) {$\tau$ with {NPD}};
\path[draw=orange,line width= 0.8pt,dash pattern=on 4pt off 4pt ,line join=round,line cap=round] (340.35,138.67-10) -- (358.35,138.67-10);

\node[text=drawColor,anchor=base west,inner sep=0pt, outer sep=0pt, scale = 0.90] at (367.35,123.23-15) {$P_n$ with {NPD}};
\path[draw=blue,line width= 0.8pt,dash pattern=on 7pt off 3pt ,line join=round,line cap=round] (340.35,126.67-15) -- (358.35,126.67-15);
%
%

\node[text=drawColor,anchor=base west,inner sep=0pt, outer sep=0pt, scale = 0.90] at (367.35,111.23-20) {Classical};
\path[draw=drawColor,line width= 0.8pt,line join=round,line cap=round] (340.35,111.23-20 +3.4) -- (358.35,111.23-20 +3.4);
\path[draw=drawColor,line width= 0.8pt,line join=round,line cap=round,fill=fillColor] (349.35,111.23-20 +3.4) circle (  1.50);


\end{scope}
%
%
%
%
%
%
%
%
%
%
%
\end{tikzpicture}
\caption{Entropy loss PRIAL results for the QUIC procedure with a scattered precision matrix, $p=60$, $n=50$ and extreme outliers.  The condition number of the underlying precision matrix is $60$ in the top panel and $1000$ in the bottom panel.}
\label{fig:p60}
\end{figure}
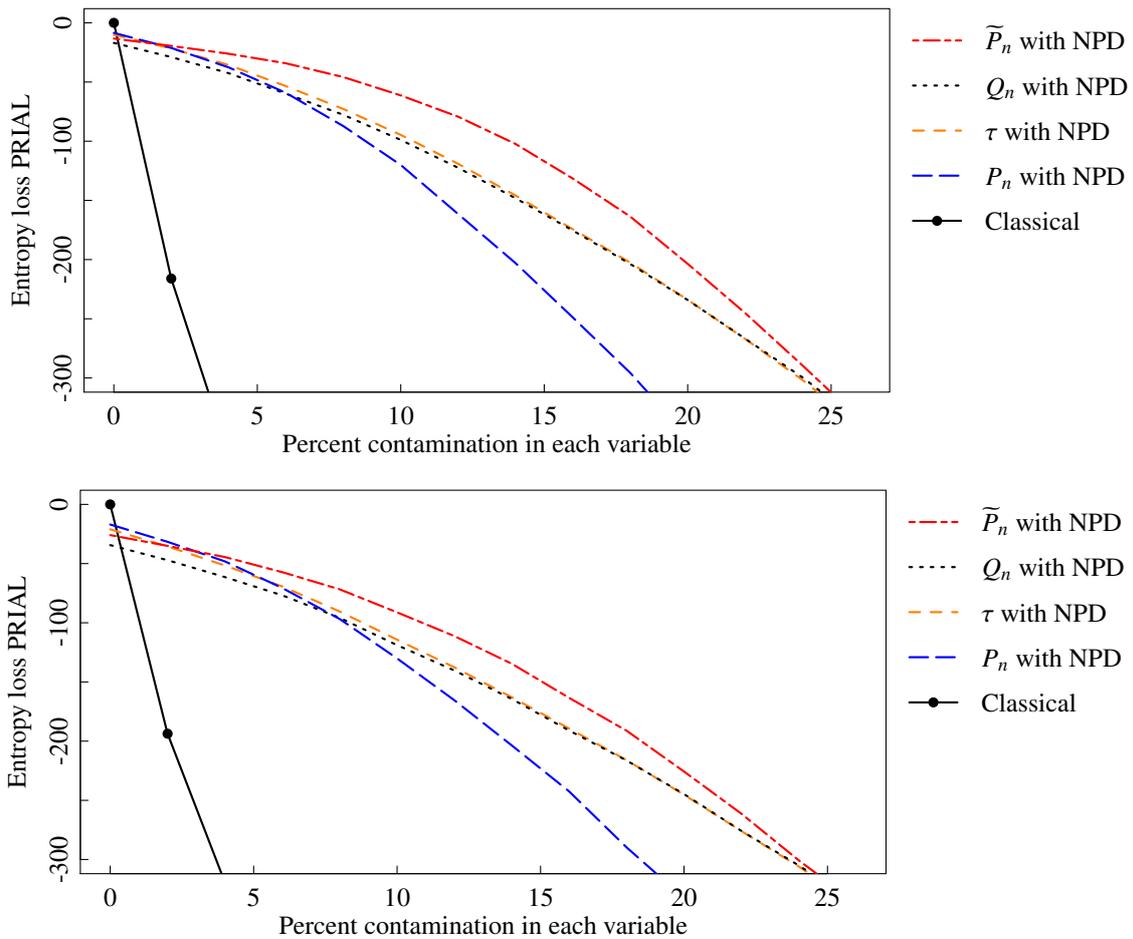

The results are similar to those found in the $p<n$ setting.  Figure \ref{fig:p60MCC} presents the MCC results for the case when $p=60$ and $n=50$.  We can draw a direct comparison between Figure \ref{fig:p60MCC} and the middle panel of Figure \ref{fig:MCC} where $p=60$ but $n=100$.  An important difference is that when $p>n$, the MCC values are lower across all levels of contamination, indicating that it is more difficult to recover the support of a Gaussian graphical model.  For example, the classical approach had a MCC of 0.39 when $n=100$, but only 0.29 when $n=50$.  Figure \ref{fig:p60} allow us to compare the relative performance of the pairwise techniques as the condition number of the true precision matrix increases from 60 to 1000.  The baseline value for the entropy loss is 8.8 for the classical approach when the condition number is 60, which increases slightly to 10.0 when the condition number is 1000.  We observe that the performance of the various pairwise estimators decreases slightly as the condition number increases.  For example when there is 8\% contamination in each variable, the adaptively trimmed $P_{n}$ estimator, has a PRIAL of -46\% when the condition number is 60 (top panel), which decreases to -72\% when the condition number is 1000 (bottom panel).

\begin{figure}[htbp]
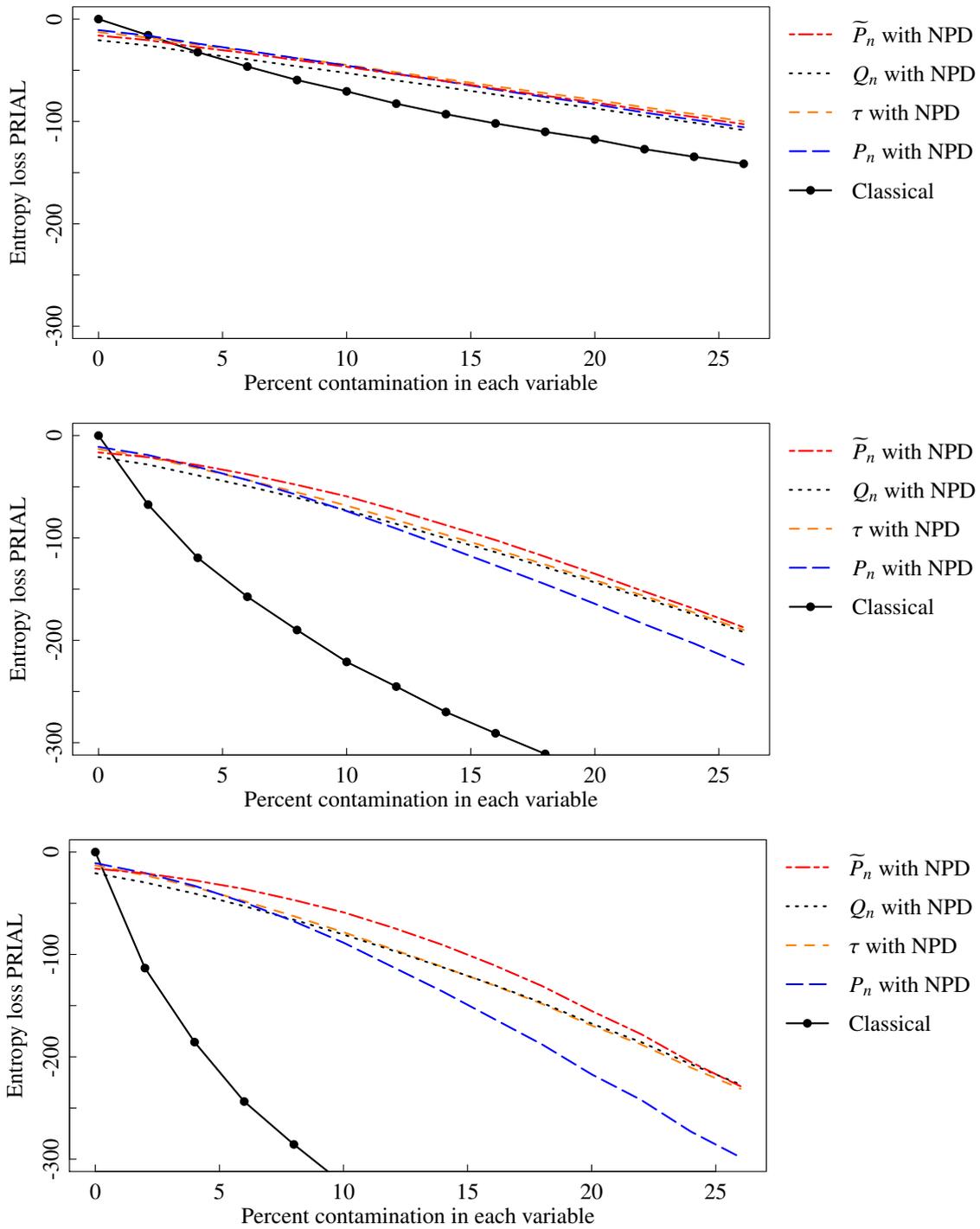

\centering
\input{quicp90loss1expt1k10n50rel}
\input{quicp90loss1expt1k50n50rel}
\input{quicp90loss1expt1k100n50rel}
\caption{Results for the QUIC procedure for a banded precision matrix with $p=90$ and $n=50$.   The outlier generating distribution is a multivariate $t_{10}$ distribution with scale matrix $10\,\mathbf{I}_{90}$ (top), $50\,\mathbf{I}_{90}$ (middle) and $100\,\mathbf{I}_{90}$ (bottom).}
\label{fig:p90}
\end{figure}

Figure \ref{fig:p90} demonstrates the impact of changing the extremity of the outliers when $p=90$ and $n=50$.  As described in Section \ref{MVsim}, the outliers are generated from a multivariate $t_{10}$ distribution with scale matrix $k\,\mathbf{I}_{p}$, for $k=10$, $50$ and $100$. With moderate outliers, as shown in the top panel of Figure \ref{fig:p90}, all the robust procedures perform comparably and the classical approach is not affected too badly.  As the extremity of the outliers increases, shown in the bottom two panels of Figure \ref{fig:p90} the performance of the classical approach deteriorates quickly.  Between the middle and bottom panels, there is little difference in the performance of the high-breakdown value robust estimators, indicating that they have stabilised and are likely to continue giving the same result even if the existing outliers were moved further away.  It is clear that the method based on $P_{n}$, with its lower breakdown value, continues to be affected by the size of contamination when there is a large proportion of contamination in each variable and we would expect its performance to continue to deteriorate if the outlier generating distribution was even more extreme.

Note that $\widetilde{P}_{n}$, the adaptively trimmed $P_{n}$, remains the most stable as the extremity of the outliers increases.  In fact, when there is 10\% cellwise contamination the PRIAL remains at -59\% whether $k=50$ or $100$, which is down from -47\% when $k=10$.  This can be attributed to the adaptive trimming correctly identifying the vast majority of the contaminated bivariate observations when the outliers are so extreme.

\section{Conclusion} \label{MVconclusion}

A pairwise approach to covariance estimation has a natural resilience to the type of cellwise contamination seen in high dimensional scenarios where classical robust procedures, such as the MCD, $M$-estimators, Quadrant Covariance and OGK, tend to fail.  

We considered a broad range of scenarios: from dense precision matrices, as typically found in standard analyses with $n\gg p$; to banded precision matrices that often occur in time series settings and may also be representative of scenarios with block diagonal precision matrices; as well as scattered sparsity, where the linkages between variables are not known beforehand and can show up anywhere within the precision matrix, as is often found in settings where $p>n$.  

After careful consideration of the various performance indices available in the multivariate setting, outlined in the supplementary material, our primary choice of performance measure was the entropy loss.  When appropriate, we showed that the entropy loss returned similar conclusions to other performance indices, such as the Frobenius norm and log determinant. 

We have shown that combining robust pairwise covariance estimation with the NPD method and regularisation techniques such as the CLIME, QUIC or GLASSO yield precision matrices that are robust to cellwise contamination.  The additional advantages of the regularisation techniques, such as the promotion of sparsity also carry through.  While it was expected, given that they are solving the same minimisation problem, it is reassuring to find that the QUIC estimates are virtually indistinguishable from the standard GLASSO approach in all scenarios considered here.  Furthermore, it did not appear to matter which of the three considered regularisation routines was applied, as all gave broadly similar results in the various scenarios considered.  This is comforting given the current pace of research in this area, with new procedures being suggested frequently.

We demonstrated that the proposed approach maintains its ability to identify the true precision matrix structure, as measured by the Matthews correlation coefficient, under moderate levels of contamination.   

We also investigated what happens when the resulting precision matrix is inverted to find the corresponding covariance matrix estimate.   Applying the same performance indices to the resulting covariance matrices, we found they perform similarly well to the underlying precision matrix.

The simulation study allowed for quite high levels of arbitrary contamination in multivariate data sets. As such, the pairwise techniques based on the standard $P_n$ estimator unsurprisingly did not perform as well as $Q_n$ and $\tau$-scale estimators, however, the adaptively trimmed $P_n$, $\widetilde{P}_n$ with trimming parameter $d=3$ typically performed extremely well, due to its ability to detect and trim extreme outliers in bivariate space.  Finally, we showed that the performance of the proposed technique continues to perform well even when $p$ is moderately larger than $n$.

\section*{Acknowledgements}

The authors would like to thank two anonymous referees for their helpful comments on an earlier draft.

\bibliographystyle{model2-names}
\bibliography{InvCovMat}

\begin{thebibliography}{54}
\expandafter\ifx\csname natexlab\endcsname\relax\def\natexlab#1{#1}\fi
\providecommand{\url}[1]{\texttt{#1}}
\providecommand{\href}[2]{#2}
\providecommand{\path}[1]{#1}
\providecommand{\DOIprefix}{doi:}
\providecommand{\ArXivprefix}{arXiv:}
\providecommand{\URLprefix}{URL: }
\providecommand{\Pubmedprefix}{pmid:}
\providecommand{\doi}[1]{\href{http://dx.doi.org/#1}{\path{#1}}}
\providecommand{\Pubmed}[1]{\href{pmid:#1}{\path{#1}}}
\providecommand{\bibinfo}[2]{#2}
\ifx\xfnm\relax \def\xfnm[#1]{\unskip,\space#1}\fi
\bibitem[{Agostinelli et~al.(2014)Agostinelli, Leung, Yohai and
  Zamar}]{Agostinelli:2014}
\bibinfo{author}{Agostinelli, C.}, \bibinfo{author}{Leung, A.},
  \bibinfo{author}{Yohai, V.}, \bibinfo{author}{Zamar, R.},
  \bibinfo{year}{2014}.
\newblock \bibinfo{title}{Robust estimation of multivariate location and
  scatter in the presence of cellwise and casewise contamination}.
\newblock \bibinfo{type}{Technical Report}. arXiv:1406.6031.
\bibitem[{Alqallaf et~al.(2009)Alqallaf, Van~Aelst, Yohai and
  Zamar}]{Alqallaf:2009}
\bibinfo{author}{Alqallaf, F.}, \bibinfo{author}{Van~Aelst, S.},
  \bibinfo{author}{Yohai, V.J.}, \bibinfo{author}{Zamar, R.H.},
  \bibinfo{year}{2009}.
\newblock \bibinfo{title}{Propagation of outliers in multivariate data}.
\newblock \bibinfo{journal}{The Annals of Statistics} \bibinfo{volume}{37},
  \bibinfo{pages}{311--331}.
\newblock \DOIprefix\doi{10.1214/07-AOS588}.
\bibitem[{Alqallaf et~al.(2002)Alqallaf, Konis, Martin and
  Zamar}]{Alqallaf:2002}
\bibinfo{author}{Alqallaf, F.A.}, \bibinfo{author}{Konis, K.P.},
  \bibinfo{author}{Martin, R.D.}, \bibinfo{author}{Zamar, R.H.},
  \bibinfo{year}{2002}.
\newblock \bibinfo{title}{Scalable robust covariance and correlation estimates
  for data mining}, in: \bibinfo{booktitle}{Proceedings of the Eighth ACM
  SIGKDD International Conference on Knowledge Discovery and Data Mining},
  \bibinfo{publisher}{ACM}, \bibinfo{address}{New York, NY, USA}. pp.
  \bibinfo{pages}{14--23}.
\newblock \DOIprefix\doi{10.1145/775047.775050}.
\bibitem[{Barnett and Lewis(1994)}]{Barnett:1994}
\bibinfo{author}{Barnett, V.}, \bibinfo{author}{Lewis, T.},
  \bibinfo{year}{1994}.
\newblock \bibinfo{title}{Outliers in statistical data}.
\newblock Wiley Series in Probability and Mathematical Statistics: Applied
  Probability and Statistics. \bibinfo{edition}{3rd} ed.,
  \bibinfo{publisher}{Wiley}, \bibinfo{address}{New York}.
\bibitem[{Bieroza et~al.(2011)Bieroza, Baker and Bridgeman}]{Bieroza:2011}
\bibinfo{author}{Bieroza, M.}, \bibinfo{author}{Baker, A.},
  \bibinfo{author}{Bridgeman, J.}, \bibinfo{year}{2011}.
\newblock \bibinfo{title}{Classification and calibration of organic matter
  fluorescence data with multiway analysis methods and artificial neural
  networks: an operational tool for improved drinking water treatment}.
\newblock \bibinfo{journal}{Environmetrics} \bibinfo{volume}{22},
  \bibinfo{pages}{256--270}.
\newblock \DOIprefix\doi{10.1002/env.1045}.
\bibitem[{Cai et~al.(2011)Cai, Liu and Luo}]{Cai:2011}
\bibinfo{author}{Cai, T.}, \bibinfo{author}{Liu, W.}, \bibinfo{author}{Luo,
  X.}, \bibinfo{year}{2011}.
\newblock \bibinfo{title}{A constrained {$\ell_1$} minimization approach to
  sparse precision matrix estimation}.
\newblock \bibinfo{journal}{Journal of the American Statistical Association}
  \bibinfo{volume}{106}, \bibinfo{pages}{594--607}.
\newblock \DOIprefix\doi{10.1198/jasa.2011.tm10155}.
\bibitem[{Cai et~al.(2012)Cai, Liu and Luo}]{Cai:2012}
\bibinfo{author}{Cai, T.}, \bibinfo{author}{Liu, W.}, \bibinfo{author}{Luo,
  X.}, \bibinfo{year}{2012}.
\newblock \bibinfo{title}{clime: Constrained {$\ell_1$-minimization} for
  inverse (covariance) matrix estimation}.
\newblock \URLprefix \url{http://CRAN.R-project.org/package=clime}.
  \bibinfo{note}{{R} package version 0.4.1}.
\bibitem[{Cator and Lopuha{\"a}(2010)}]{Cator:2010}
\bibinfo{author}{Cator, E.A.}, \bibinfo{author}{Lopuha{\"a}, H.P.},
  \bibinfo{year}{2010}.
\newblock \bibinfo{title}{Asymptotic expansion of the minimum covariance
  determinant estimators}.
\newblock \bibinfo{journal}{Journal of Multivariate Analysis}
  \bibinfo{volume}{101}, \bibinfo{pages}{2372--2388}.
\newblock \DOIprefix\doi{10.1016/j.jmva.2010.06.009}.
\bibitem[{Cator and Lopuha{\"a}(2012)}]{Cator:2012}
\bibinfo{author}{Cator, E.A.}, \bibinfo{author}{Lopuha{\"a}, H.P.},
  \bibinfo{year}{2012}.
\newblock \bibinfo{title}{Central limit theorem and influence function for the
  {MCD} estimators at general multivariate distributions}.
\newblock \bibinfo{journal}{Bernoulli} \bibinfo{volume}{18},
  \bibinfo{pages}{520--551}.
\newblock \DOIprefix\doi{10.3150/11-BEJ353}.
\bibitem[{Croux et~al.(2003)Croux, Filzmoser, Pison and Rousseeuw}]{Croux:2003}
\bibinfo{author}{Croux, C.}, \bibinfo{author}{Filzmoser, P.},
  \bibinfo{author}{Pison, G.}, \bibinfo{author}{Rousseeuw, P.J.},
  \bibinfo{year}{2003}.
\newblock \bibinfo{title}{{Fitting multiplicative models by robust alternating
  regressions}}.
\newblock \bibinfo{journal}{Statistics and Computing} \bibinfo{volume}{13},
  \bibinfo{pages}{23--36}.
\newblock \DOIprefix\doi{10.1023/A:1021979409012}.
\bibitem[{Dempster(1972)}]{Dempster:1972}
\bibinfo{author}{Dempster, A.P.}, \bibinfo{year}{1972}.
\newblock \bibinfo{title}{Covariance selection}.
\newblock \bibinfo{journal}{Biometrics} \bibinfo{volume}{28},
  \bibinfo{pages}{157--175}.
\newblock \DOIprefix\doi{10.2307/2528966}.
\bibitem[{Dey and Srinivasan(1985)}]{Dey:1985}
\bibinfo{author}{Dey, D.K.}, \bibinfo{author}{Srinivasan, C.},
  \bibinfo{year}{1985}.
\newblock \bibinfo{title}{Estimation of a covariance matrix under {Stein's}
  loss}.
\newblock \bibinfo{journal}{The Annals of Statistics} \bibinfo{volume}{13},
  \bibinfo{pages}{1581--1591}.
\newblock \DOIprefix\doi{10.1214/aos/1176349756}.
\bibitem[{Farcomeni(2014)}]{Farcomeni:2014}
\bibinfo{author}{Farcomeni, A.}, \bibinfo{year}{2014}.
\newblock \bibinfo{title}{Robust constrained clustering in presence of
  entry-wise outliers}.
\newblock \bibinfo{journal}{Technometrics} \bibinfo{volume}{56},
  \bibinfo{pages}{102--111}.
\newblock \DOIprefix\doi{10.1080/00401706.2013.826148}.
\bibitem[{Filzmoser et~al.(2014)Filzmoser, Ruiz-Gazen and
  Thomas-Agnan}]{Filzmoser:2013}
\bibinfo{author}{Filzmoser, P.}, \bibinfo{author}{Ruiz-Gazen, A.},
  \bibinfo{author}{Thomas-Agnan, C.}, \bibinfo{year}{2014}.
\newblock \bibinfo{title}{Identification of local multivariate outliers}.
\newblock \bibinfo{journal}{Statistical Papers} \bibinfo{volume}{55},
  \bibinfo{pages}{29--47}.
\newblock \DOIprefix\doi{10.1007/s00362-013-0524-z}.
\bibitem[{Friedman et~al.(2008)Friedman, Hastie and Tibshirani}]{Friedman:2008}
\bibinfo{author}{Friedman, J.H.}, \bibinfo{author}{Hastie, T.},
  \bibinfo{author}{Tibshirani, R.}, \bibinfo{year}{2008}.
\newblock \bibinfo{title}{Sparse inverse covariance estimation with the
  graphical lasso}.
\newblock \bibinfo{journal}{Biostatistics} \bibinfo{volume}{9},
  \bibinfo{pages}{432--441}.
\newblock \DOIprefix\doi{10.1093/biostatistics/kxm045}.
\bibitem[{Gales and van Dalen(2007)}]{Gales:2007}
\bibinfo{author}{Gales, M.J.F.}, \bibinfo{author}{van Dalen, R.C.},
  \bibinfo{year}{2007}.
\newblock \bibinfo{title}{Predictive linear transforms for noise robust speech
  recognition}, in: \bibinfo{booktitle}{IEEE Workshop on Automatic Speech
  Recognition \& Understanding}, \bibinfo{organization}{IEEE}. pp.
  \bibinfo{pages}{59--64}.
\bibitem[{Gentle(2007)}]{Gentle:2007}
\bibinfo{author}{Gentle, J.}, \bibinfo{year}{2007}.
\newblock \bibinfo{title}{Matrix Algebra Theory, Computations and Applications
  in Statistics}.
\newblock \bibinfo{publisher}{Springer}, \bibinfo{address}{New York}.
\bibitem[{Gnanadesikan and Kettenring(1972)}]{Gnanadesikan:1972}
\bibinfo{author}{Gnanadesikan, R.}, \bibinfo{author}{Kettenring, J.R.},
  \bibinfo{year}{1972}.
\newblock \bibinfo{title}{Robust estimates, residuals, and outlier detection
  with multiresponse data}.
\newblock \bibinfo{journal}{Biometrics} \bibinfo{volume}{28},
  \bibinfo{pages}{81--124}.
\newblock \DOIprefix\doi{10.2307/2528963}.
\bibitem[{Gupta and Srivastava(2010)}]{Gupta:2010}
\bibinfo{author}{Gupta, M.}, \bibinfo{author}{Srivastava, S.},
  \bibinfo{year}{2010}.
\newblock \bibinfo{title}{Parametric {Bayesian} estimation of differential
  entropy and relative entropy}.
\newblock \bibinfo{journal}{Entropy} \bibinfo{volume}{12},
  \bibinfo{pages}{818--843}.
\newblock \DOIprefix\doi{10.3390/e12040818}.
\bibitem[{Higham(2002)}]{Higham:2002}
\bibinfo{author}{Higham, N.J.}, \bibinfo{year}{2002}.
\newblock \bibinfo{title}{Computing the nearest correlation matrix -- a problem
  from finance}.
\newblock \bibinfo{journal}{IMA Journal of Numerical Analysis}
  \bibinfo{volume}{22}, \bibinfo{pages}{329--343}.
\newblock \DOIprefix\doi{10.1093/imanum/22.3.329}.
\bibitem[{Hsieh et~al.(2011)Hsieh, Dhillon, Ravikumar and Sustik}]{Hsieh:2011}
\bibinfo{author}{Hsieh, C.J.}, \bibinfo{author}{Dhillon, I.S.},
  \bibinfo{author}{Ravikumar, P.K.}, \bibinfo{author}{Sustik, M.A.},
  \bibinfo{year}{2011}.
\newblock \bibinfo{title}{Sparse inverse covariance matrix estimation using
  quadratic approximation}, in: \bibinfo{editor}{Shawe-Taylor, J.},
  \bibinfo{editor}{Zemel, R.}, \bibinfo{editor}{Bartlett, P.},
  \bibinfo{editor}{Pereira, F.}, \bibinfo{editor}{Weinberger, K.} (Eds.),
  \bibinfo{booktitle}{Advances in Neural Information Processing Systems}, pp.
  \bibinfo{pages}{2330--2338}.
\bibitem[{Hsieh et~al.(2013)Hsieh, Sustik, Dhillon, Ravikumar and
  Poldrack}]{Hsieh:2013}
\bibinfo{author}{Hsieh, C.J.}, \bibinfo{author}{Sustik, M.A.},
  \bibinfo{author}{Dhillon, I.S.}, \bibinfo{author}{Ravikumar, P.K.},
  \bibinfo{author}{Poldrack, R.}, \bibinfo{year}{2013}.
\newblock \bibinfo{title}{{BIG \& QUIC: Sparse} inverse covariance estimation
  for a million variables}, in: \bibinfo{editor}{Burges, C.},
  \bibinfo{editor}{Bottou, L.}, \bibinfo{editor}{Welling, M.},
  \bibinfo{editor}{Ghahramani, Z.}, \bibinfo{editor}{Weinberger, K.} (Eds.),
  \bibinfo{booktitle}{Advances in Neural Information Processing Systems}, pp.
  \bibinfo{pages}{3165--3173}.
\bibitem[{Huber(1964)}]{Huber:1964}
\bibinfo{author}{Huber, P.J.}, \bibinfo{year}{1964}.
\newblock \bibinfo{title}{Robust estimation of a location parameter}.
\newblock \bibinfo{journal}{Annals of Mathematical Statistics}
  \bibinfo{volume}{35}, \bibinfo{pages}{73--101}.
\newblock \DOIprefix\doi{10.1214/aoms/1177703732}.
\bibitem[{Hubert et~al.(2014)Hubert, Rousseeuw and Vakili}]{Hubert:2013}
\bibinfo{author}{Hubert, M.}, \bibinfo{author}{Rousseeuw, P.J.},
  \bibinfo{author}{Vakili, K.}, \bibinfo{year}{2014}.
\newblock \bibinfo{title}{Shape bias of robust covariance estimators: an
  empirical study}.
\newblock \bibinfo{journal}{Statistical Papers} \bibinfo{volume}{55},
  \bibinfo{pages}{15--28}.
\newblock \DOIprefix\doi{10.1007/s00362-013-0544-8}.
\bibitem[{Hubert et~al.(2008)Hubert, Rousseeuw and Van~Aelst}]{Hubert:2008}
\bibinfo{author}{Hubert, M.}, \bibinfo{author}{Rousseeuw, P.J.},
  \bibinfo{author}{Van~Aelst, S.}, \bibinfo{year}{2008}.
\newblock \bibinfo{title}{High-breakdown robust multivariate methods}.
\newblock \bibinfo{journal}{Statistical Science} \bibinfo{volume}{23},
  \bibinfo{pages}{92--119}.
\newblock \DOIprefix\doi{10.1214/088342307000000087}.
\bibitem[{James and Stein(1961)}]{James:1961}
\bibinfo{author}{James, W.}, \bibinfo{author}{Stein, C.}, \bibinfo{year}{1961}.
\newblock \bibinfo{title}{Estimation with quadratic loss}, in:
  \bibinfo{editor}{Neyman, J.} (Ed.), \bibinfo{booktitle}{Proceedings of the
  Fourth Berkeley Symposium on Mathematical Statististics and Probability},
  \bibinfo{publisher}{University of California Press},
  \bibinfo{address}{Berkeley}. pp. \bibinfo{pages}{361--379}.
\bibitem[{Johnstone(2001)}]{Johnstone:2001}
\bibinfo{author}{Johnstone, I.M.}, \bibinfo{year}{2001}.
\newblock \bibinfo{title}{On the distribution of the largest eigenvalue in
  principal components analysis}.
\newblock \bibinfo{journal}{The Annals of Statistics} \bibinfo{volume}{29},
  \bibinfo{pages}{295--327}.
\newblock \DOIprefix\doi{10.1214/aos/1009210544}.
\bibitem[{Lauritzen(1996)}]{Lauritzen:1996}
\bibinfo{author}{Lauritzen, S.}, \bibinfo{year}{1996}.
\newblock \bibinfo{title}{Graphical Models}.
\newblock \bibinfo{publisher}{Oxford University Press}, \bibinfo{address}{New
  York}.
\bibitem[{Ledoit and Wolf(2004)}]{Ledoit:2004}
\bibinfo{author}{Ledoit, O.}, \bibinfo{author}{Wolf, M.}, \bibinfo{year}{2004}.
\newblock \bibinfo{title}{{A well-conditioned estimator for large-dimensional
  covariance matrices}}.
\newblock \bibinfo{journal}{Journal of Multivariate Analysis}
  \bibinfo{volume}{88}, \bibinfo{pages}{365--411}.
\newblock \DOIprefix\doi{10.1016/S0047-259X(03)00096-4}.
\bibitem[{Li(2004)}]{Li:2004}
\bibinfo{author}{Li, Y.}, \bibinfo{year}{2004}.
\newblock \bibinfo{title}{On incremental and robust subspace learning}.
\newblock \bibinfo{journal}{Pattern Recognition} \bibinfo{volume}{37},
  \bibinfo{pages}{1509 -- 1518}.
\newblock \DOIprefix\doi{10.1016/j.patcog.2003.11.010}.
\bibitem[{Lin and Perlman(1985)}]{Lin:1985}
\bibinfo{author}{Lin, S.P.}, \bibinfo{author}{Perlman, M.D.},
  \bibinfo{year}{1985}.
\newblock \bibinfo{title}{A {Monte Carlo} comparison of four estimators of a
  covariance matrix}, in: \bibinfo{editor}{Krishnaiah, P.R.} (Ed.),
  \bibinfo{booktitle}{Proceedings of the Sixth International Symposium on
  Multivariate Analysis}, \bibinfo{publisher}{North-Holland},
  \bibinfo{address}{Amsterdam}. pp. \bibinfo{pages}{411--429}.
\bibitem[{Little and Rubin(2002)}]{Little:2002}
\bibinfo{author}{Little, R.}, \bibinfo{author}{Rubin, D.},
  \bibinfo{year}{2002}.
\newblock \bibinfo{title}{Statistical Analysis with Missing Data}.
\newblock \bibinfo{edition}{2nd} ed., \bibinfo{publisher}{Wiley},
  \bibinfo{address}{Hoboken}.
\bibitem[{Liu et~al.(2003)Liu, Hawkins, Ghosh and Young}]{Liu:2003}
\bibinfo{author}{Liu, L.}, \bibinfo{author}{Hawkins, D.M.},
  \bibinfo{author}{Ghosh, S.}, \bibinfo{author}{Young, S.S.},
  \bibinfo{year}{2003}.
\newblock \bibinfo{title}{Robust singular value decomposition analysis of
  microarray data}.
\newblock \bibinfo{journal}{Proceedings of the National Academy of Sciences of
  the United States of America} \bibinfo{volume}{100},
  \bibinfo{pages}{13167--13172}.
\newblock \DOIprefix\doi{10.1073/pnas.1733249100}.
\bibitem[{L{\o}land et~al.(2013)L{\o}land, Huseby, Hjort and
  Frigessi}]{Loland:2013}
\bibinfo{author}{L{\o}land, A.}, \bibinfo{author}{Huseby, R.B.},
  \bibinfo{author}{Hjort, N.L.}, \bibinfo{author}{Frigessi, A.},
  \bibinfo{year}{2013}.
\newblock \bibinfo{title}{Statistical corrections of invalid correlation
  matrices}.
\newblock \bibinfo{journal}{Scandinavian Journal of Statistics}
  \bibinfo{volume}{40}, \bibinfo{pages}{807--824}.
\newblock \DOIprefix\doi{10.1111/sjos.12035}.
\bibitem[{Ma and Genton(2001)}]{Ma:2001}
\bibinfo{author}{Ma, Y.}, \bibinfo{author}{Genton, M.G.}, \bibinfo{year}{2001}.
\newblock \bibinfo{title}{Highly robust estimation of dispersion matrices}.
\newblock \bibinfo{journal}{Journal of Multivariate Analysis}
  \bibinfo{volume}{78}, \bibinfo{pages}{11--36}.
\newblock \DOIprefix\doi{10.1006/jmva.2000.1942}.
\bibitem[{Mardia et~al.(1979)Mardia, Kent and Bibby}]{Mardia:1979}
\bibinfo{author}{Mardia, K.V.}, \bibinfo{author}{Kent, J.T.},
  \bibinfo{author}{Bibby, J.M.}, \bibinfo{year}{1979}.
\newblock \bibinfo{title}{Multivariate Analysis}.
\newblock Probability and Mathematical Statistics, \bibinfo{publisher}{Academic
  Press}, \bibinfo{address}{London}.
\bibitem[{Maronna et~al.(2006)Maronna, Martin and Yohai}]{Maronna:2006}
\bibinfo{author}{Maronna, R.A.}, \bibinfo{author}{Martin, R.D.},
  \bibinfo{author}{Yohai, V.J.}, \bibinfo{year}{2006}.
\newblock \bibinfo{title}{Robust Statistics}.
\newblock \bibinfo{publisher}{Wiley}, \bibinfo{address}{London}.
\bibitem[{Maronna and Yohai(2008)}]{Maronna:2008}
\bibinfo{author}{Maronna, R.A.}, \bibinfo{author}{Yohai, V.J.},
  \bibinfo{year}{2008}.
\newblock \bibinfo{title}{Robust low-rank approximation of data matrices with
  elementwise contamination}.
\newblock \bibinfo{journal}{Technometrics} \bibinfo{volume}{50},
  \bibinfo{pages}{295--304}.
\newblock \DOIprefix\doi{10.1198/004017008000000190}.
\bibitem[{Maronna and Zamar(2002)}]{Maronna:2002}
\bibinfo{author}{Maronna, R.A.}, \bibinfo{author}{Zamar, R.H.},
  \bibinfo{year}{2002}.
\newblock \bibinfo{title}{Robust estimates of location and dispersion for
  high-dimensional datasets}.
\newblock \bibinfo{journal}{Technometrics} \bibinfo{volume}{44},
  \bibinfo{pages}{307--317}.
\newblock \DOIprefix\doi{10.1198/004017002188618509}.
\bibitem[{Matthews(1975)}]{Matthews:1975}
\bibinfo{author}{Matthews, B.W.}, \bibinfo{year}{1975}.
\newblock \bibinfo{title}{Comparison of the predicted and observed secondary
  structure of {T4} phage lysozyme}.
\newblock \bibinfo{journal}{Biochimica et Biophysica Acta}
  \bibinfo{volume}{405}, \bibinfo{pages}{442--451}.
\newblock \DOIprefix\doi{10.1016/0005-2795(75)90109-9}.
\bibitem[{Mavroeidis and Marchiori(2014)}]{Mavroeidis:2014}
\bibinfo{author}{Mavroeidis, D.}, \bibinfo{author}{Marchiori, E.},
  \bibinfo{year}{2014}.
\newblock \bibinfo{title}{Feature selection for k-means clustering stability:
  theoretical analysis and an algorithm}.
\newblock \bibinfo{journal}{Data Mining and Knowledge Discovery}
  \bibinfo{volume}{28}, \bibinfo{pages}{918--960}.
\newblock \DOIprefix\doi{10.1007/s10618-013-0320-3}.
\bibitem[{Rousseeuw and Croux(1993)}]{Rousseeuw:1993}
\bibinfo{author}{Rousseeuw, P.J.}, \bibinfo{author}{Croux, C.},
  \bibinfo{year}{1993}.
\newblock \bibinfo{title}{Alternatives to the median absolute deviation}.
\newblock \bibinfo{journal}{Journal of the American Statistical Association}
  \bibinfo{volume}{88}, \bibinfo{pages}{1273--1283}.
\newblock \DOIprefix\doi{10.1080/01621459.1993.10476408}.
\bibitem[{Rousseeuw et~al.(2013)Rousseeuw, Croux, Todorov, Ruckstuhl,
  Salibian-Barrera, Verbeke, Koller and Maechler}]{robustbase}
\bibinfo{author}{Rousseeuw, P.J.}, \bibinfo{author}{Croux, C.},
  \bibinfo{author}{Todorov, V.}, \bibinfo{author}{Ruckstuhl, A.},
  \bibinfo{author}{Salibian-Barrera, M.}, \bibinfo{author}{Verbeke, T.},
  \bibinfo{author}{Koller, M.}, \bibinfo{author}{Maechler, M.},
  \bibinfo{year}{2013}.
\newblock \bibinfo{title}{{robustbase}: Basic Robust Statistics}.
\newblock \URLprefix \url{http://CRAN.R-project.org/package=robustbase}.
  \bibinfo{note}{{R} package version 0.9-10}.
\bibitem[{Rousseeuw and Molenberghs(1993)}]{Rousseeuw:1993b}
\bibinfo{author}{Rousseeuw, P.J.}, \bibinfo{author}{Molenberghs, G.},
  \bibinfo{year}{1993}.
\newblock \bibinfo{title}{Transformation of non positive semidefinite
  correlation matrices}.
\newblock \bibinfo{journal}{Communications in Statistics -- Theory and Methods}
  \bibinfo{volume}{22}, \bibinfo{pages}{965--984}.
\newblock \DOIprefix\doi{10.1080/03610928308831068}.
\bibitem[{Stein(1956)}]{Stein:1956}
\bibinfo{author}{Stein, C.}, \bibinfo{year}{1956}.
\newblock \bibinfo{title}{Some Problems in Multivariate Analysis, {Part I}}.
\newblock \bibinfo{type}{Technical Report} \bibinfo{number}{6}. Stanford
  University. \bibinfo{address}{Stanford}.
\bibitem[{Tarr et~al.(2012)Tarr, M{\"u}ller and Weber}]{Tarr:2012}
\bibinfo{author}{Tarr, G.}, \bibinfo{author}{M{\"u}ller, S.},
  \bibinfo{author}{Weber, N.C.}, \bibinfo{year}{2012}.
\newblock \bibinfo{title}{A robust scale estimator based on pairwise means}.
\newblock \bibinfo{journal}{Journal of Nonparametric Statistics}
  \bibinfo{volume}{24}, \bibinfo{pages}{187--199}.
\newblock \DOIprefix\doi{10.1080/10485252.2011.621424}.
\bibitem[{Tarr et~al.(2014)Tarr, M{\"u}ller and Weber}]{Tarr:2014}
\bibinfo{author}{Tarr, G.}, \bibinfo{author}{M{\"u}ller, S.},
  \bibinfo{author}{Weber, N.C.}, \bibinfo{year}{2014}.
\newblock \bibinfo{title}{Robust estimation of precision matrices under
  cellwise contamination}.
\newblock \bibinfo{journal}{Under review} .
\bibitem[{De~la Torre and Black(2001)}]{DelaTorre:2001}
\bibinfo{author}{De~la Torre, F.}, \bibinfo{author}{Black, M.J.},
  \bibinfo{year}{2001}.
\newblock \bibinfo{title}{Robust principal component analysis for computer
  vision}, in: \bibinfo{booktitle}{International Conference on Computer
  Vision}, \bibinfo{publisher}{IEEE}, \bibinfo{address}{Vancouver}. pp.
  \bibinfo{pages}{362--369}.
\newblock \DOIprefix\doi{10.1109/ICCV.2001.937541}.
\bibitem[{Tukey(1962)}]{Tukey:1962}
\bibinfo{author}{Tukey, J.W.}, \bibinfo{year}{1962}.
\newblock \bibinfo{title}{The future of data analysis}.
\newblock \bibinfo{journal}{Annals of Mathematical Statistics}
  \bibinfo{volume}{33}, \bibinfo{pages}{1--67}.
\newblock \DOIprefix\doi{doi:10.1214/aoms/1177704711}.
\bibitem[{Van~Aelst et~al.(2012)Van~Aelst, Vandervieren and
  Willems}]{VanAelst:2012}
\bibinfo{author}{Van~Aelst, S.}, \bibinfo{author}{Vandervieren, E.},
  \bibinfo{author}{Willems, G.}, \bibinfo{year}{2012}.
\newblock \bibinfo{title}{A {Stahel-Donoho} estimator based on {Huberized}
  outlyingness}.
\newblock \bibinfo{journal}{Computational Statistics \& Data Analysis}
  \bibinfo{volume}{56}, \bibinfo{pages}{531 -- 542}.
\newblock \DOIprefix\doi{10.1016/j.csda.2011.08.014}.
\bibitem[{Wilks(1932)}]{Wilks:1932}
\bibinfo{author}{Wilks, S.S.}, \bibinfo{year}{1932}.
\newblock \bibinfo{title}{Certain generalizations in the analysis of variance}.
\newblock \bibinfo{journal}{Biometrika} \bibinfo{volume}{24},
  \bibinfo{pages}{471--494}.
\newblock \DOIprefix\doi{10.2307/2331979}.
\bibitem[{Won et~al.(2013)Won, Lim, Kim and Rajaratnam}]{Won:2013}
\bibinfo{author}{Won, J.H.}, \bibinfo{author}{Lim, J.}, \bibinfo{author}{Kim,
  S.J.}, \bibinfo{author}{Rajaratnam, B.}, \bibinfo{year}{2013}.
\newblock \bibinfo{title}{Condition-number-regularized covariance estimation}.
\newblock \bibinfo{journal}{Journal of the Royal Statistical Society: Series B
  (Statistical Methodology)} \bibinfo{volume}{75}, \bibinfo{pages}{427--450}.
\newblock \DOIprefix\doi{10.1111/j.1467-9868.2012.01049.x}.
\bibitem[{Yang and Berger(1994)}]{Yang:1994}
\bibinfo{author}{Yang, R.}, \bibinfo{author}{Berger, J.O.},
  \bibinfo{year}{1994}.
\newblock \bibinfo{title}{Estimation of a covariance matrix using the reference
  prior}.
\newblock \bibinfo{journal}{The Annals of Statistics} \bibinfo{volume}{22},
  \bibinfo{pages}{1195--1211}.
\newblock \DOIprefix\doi{10.1214/aos/1176325625}.
\bibitem[{Yuan and Lin(2007)}]{Yuan:2007}
\bibinfo{author}{Yuan, M.}, \bibinfo{author}{Lin, Y.}, \bibinfo{year}{2007}.
\newblock \bibinfo{title}{Model selection and estimation in the {Gaussian}
  graphical model}.
\newblock \bibinfo{journal}{Biometrika} \bibinfo{volume}{94},
  \bibinfo{pages}{19--35}.
\newblock \DOIprefix\doi{10.1093/biomet/asm018}.

\end{thebibliography}

\newpage
\begin{frontmatter}
\title{Supplementary material for: Robust estimation of precision matrices under cellwise contamination}

\begin{abstract}
This supplementary material gives a summary of the performance indices used in the article ``Robust estimation of covariance and precision matrices under cellwise contamination'' by \citet{Tarr:2014}.
\end{abstract}
\end{frontmatter}

\setcounter{section}{0}
\setcounter{figure}{0}
\section{Performance indices} \label{performanceindices}

We require a way to assess the performance of various covariance and precision matrix estimators under cellwise contamination.  There are a range of possible ways to measure how close an estimated matrix is to the true value.  In order to assess the performance of our proposed estimators, we first need to identify which performance indices are appropriate.  One class of performance indices considered are matrix norms which measure the size of a matrix.  The second class looks at how closely the estimated precision (or covariance) matrix reflects the nature of the theoretical precision (or covariance) matrix, through either the determinant, condition number or an overall entropy loss index.  Let $ \bm{\Sigma}$ denote the true covariance matrix and $ \bm{\Theta}= \bm{\Sigma}^{-1}$ denote the true precision matrix.  In this section we define the performance measures and consider their appropriateness in the context of cellwise contamination.

\subsection{Measures of performance}

\subsubsection{Matrix norms}

The Frobenius norm is perhaps the most common matrix norm, it is an element-wise norm, the Euclidean norm of $\mathbf{A}$ treated as if it were a vector of length $p^2$, $||\mathbf{A}||_{F}=\sqrt{\operatorname{tr}(\mathbf{A}' \mathbf{A})}$.  An alternative way of constructing a matrix norm is to take a vector norm and use it to generate a matrix norm of the form $||\mathbf{A}|| = \sup_{||\mathbf{x}||=1}  ||\mathbf{A}\mathbf{x}||$, where $||\cdot||$ on the left is the induced (or operator) norm and $||\cdot||$ on the right is a vector norm.  Examples of induced norms are the $L_p$ norms.  For example, the one norm, $||\mathbf{A}||_{1} = \max_{j} \sum_{i=1}^{n}|a_{ij}|$; the infinity norm, $||\mathbf{A}||_{\infty} = \max_{i} \sum_{j=1}^{n}|a_{ij}|$; and the spectral norm, $||\mathbf{A}||_{2} = \sigma_\text{max}(\mathbf{A})$, where $\sigma_\text{max}(\mathbf{A})$ is the largest singular value of $\mathbf{A}$.    When $\mathbf{A}$ is nonsingular $||\mathbf{A}^{-1}||_2 =1/\sigma_\text{min}(\mathbf{A})$ where $\sigma_\text{min}(\mathbf{A})$ is the smallest singular value of $\mathbf{A}$.

In our experiments, we apply the matrix norms to $\mathbf{A}= \bm{\Theta}_0 - \mathbf{I}$ where $ \bm{\Theta}_0 =  \bm{\Theta}^{-1}\hat{ \bm{\Theta}}$. While $\bm{\Theta}$ and $\hat{\bm{\Theta}}$ are symmetric, it is not the case that the product of two symmetric matrices yields a symmetric matrix, hence in general $\bm{\Theta}_0 \neq  \bm{\Theta}_0' $,  so in practice the one norm and the infinity norm may yield different results.  

For each norm, one may na\"ively assume that the closer to zero the better, however, in Section \ref{indexbehav} we demonstrate that this is not always the case, particularly when estimating precision matrices in the presence of outliers.

Aside from matrix norms, there are a few other commonly employed performance indices.  

\subsubsection{Entropy loss}

The entropy loss, as suggested by \citet{Stein:1956} and featured in \citet{James:1961} and also \citet{Dey:1985}, when applied to precision matrices is defined as,
\begin{align*}
L_E( \bm{\Theta},\hat{ \bm{\Theta}}) & = \operatorname{tr}( \bm{\Theta} ^{-1}\hat{ \bm{\Theta}}) - \log\det( \bm{\Theta}^{-1}\hat{ \bm{\Theta}}) - p \\
& = \sum_{i=1}^p \left( \lambda_i - \log\lambda_i\right) - p,
\end{align*}
where $\lambda_i$, $i=1,\ldots,p$, are the eigenvalues of $ \bm{\Theta}_0$.  \citet{Stein:1956} notes that this function is ``somewhat arbitrary'' but it is convex in $\hat{ \bm{\Theta}}$ and assuming that $ \bm{\Theta}$ and $\hat{ \bm{\Theta}}$ are positive semidefinite, $L_E( \bm{\Theta},\hat{ \bm{\Theta}})\geq0$ with equality if and only if $ \bm{\Theta} = \hat{ \bm{\Theta}}$. 

However, the entropy loss is not as ``arbitrary'' as it may seem at first.  Note that the Kullback-Leibler divergence from $\mathcal{N} _1(\bm{\mu}, \bm{\Sigma}_1)$ to $\mathcal{N} _2(\bm{\mu}, \bm{\Sigma}_2)$ is, $D_\text{KL}(\mathcal{N} _1,\mathcal{N} _2) = \frac{1}{2} L_E( \bm{\Sigma}_1, \bm{\Sigma}_2)$.
The close link between the entropy loss and the Kullback-Leibler or Bregman divergence loss is shown in a Bayesian context by \citet{Gupta:2010}.  

There is also a clear link between the entropy loss and the likelihood ratio test for $H_0:  \bm{\Sigma} =  \bm{\Sigma}^\star$ with unknown mean assuming the data come from a Gaussian distribution, $-2(l_{1} - l_0) = n L_E( \bm{\Sigma}^\star,\mathbf{S})$,
where $l_0$ and $l_{1}$  are the log-likelihoods under the null and alternative hypotheses, see, for example \citet[p.\ 126]{Mardia:1979}.

The entropy loss is used extensively as a basis for developing and assessing improved precision and covariance matrix estimators, for example in \citet{Lin:1985}, \citet{Yang:1994} and more recently, \citet{Won:2013}.  

\subsubsection{Log determinant}

In the multivariate Gaussian setting, \citet{Wilks:1932} names the determinant of the covariance matrix, $\det(\bm{\Sigma})$, the generalised variance.   The generalised precision is similarly defined as the determinant of the precision matrix, $\det( \bm{\Theta})$.  This idea can be used as the basis for a performance index. Consider the log of the determinant of the standardised covariance or precision matrix, $L_D( \bm{\Theta}_0 ) = \log \det( \bm{\Theta}_0 ) = \sum_{i=1}^p \log \lambda_i$.  The determinant of an identity matrix is 1, so the optimal value of $L_D( \bm{\Theta}_0)$ is 0.  Positive (negative) log determinant results indicate that the generalised variance or precision is being over (under) estimated.  Note that, 
$L_D( \bm{\Theta}_0)  = -L_D( \bm{\Sigma}_0)$. Thus, methods that underestimate the generalised variance will overestimate the generalised precision.

The log determinant is a very crude performance index which can be dominated by one eigenvalue that is very close to zero. Furthermore, it is incorporated as part of the entropy loss so there is little need to focus on it in the results of the simulation studies.

\subsubsection{Log condition number}
Formally, the condition number of a square matrix is the product of the norm of the matrix and the norm of its inverse, $\kappa( \bm{\Theta}_0) = || \bm{\Theta}_0^{-1}||\cdot || \bm{\Theta}_0||$,
and hence depends on the kind of matrix-norm.  It is common to use the spectral norm, in which case the condition number is the ratio of the largest to the smallest non-zero singular value of the matrix.  The condition number associated with the systems of equations, $\mathbf{A} \mathbf{x} = \mathbf{b}$, gives a bound on how inaccurate the solution may be.  A system is said to be ill-conditioned if small changes in the inputs, $\mathbf{A}$ and $\mathbf{b}$, result in large changes in the solution, $\mathbf{x}$.    

Consider the performance index defined as the log of the condition number of $ \bm{\Theta}_0$,
\begin{align*}
L_\kappa( \bm{\Theta}_0) = \log\kappa( \bm{\Theta}_0) = \log (\sigma_{\max}( \bm{\Theta}_0)) - \log(\sigma_{\min}( \bm{\Theta}_0)).
\end{align*}

The condition number for an identity matrix is 1 and the condition number for a singular matrix is infinity, so the $0\leq L_\kappa( \bm{\Theta}_0)\leq \infty$.

\citet{Gentle:2007} notes that while the condition number of a matrix provides a useful indication of its ability to solve linear equations accurately, it can be misleading at times when the rows (or columns) of the matrix have very different scales.  That is, the condition number can be changed by simply scaling the rows or columns which does not actually make a linear system of equations any better or worse conditioned.  This is known as artificial ill-conditioning.  

\begin{figure}
\centering
\input{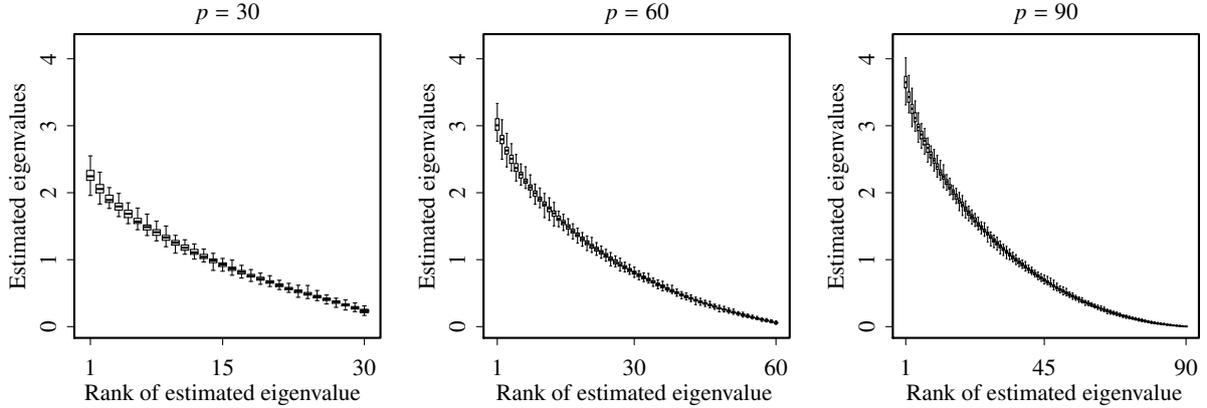}
\caption{A series of boxplots showing the distribution of the largest through to the smallest eigenvalues of an estimated covariance matrix, $\mathbf{S}$, over $N=100$ samples from $\mathcal{N} (\mathbf{0},\mathbf{I})$ with $n=100$ and $p=30$, $60$ and $90$.}
\label{eigendistortion}
\end{figure}

In the context of the sample covariance matrix, $\mathbf{S}$, \citet{Ledoit:2004} note that ``when the ratio $p/n$ is less than one but not negligible, the sample covariance matrix is invertible but numerically ill-conditioned, which means that inverting it amplifies estimation error dramatically.''     \citet{Won:2013} go further stating that ``the eigenstructure [of $\mathbf{S}$] tends to be systematically distorted unless $p/n$ is extremely small, resulting in numerically ill-conditioned estimators for $ \bm{\Sigma}$.''  Figure \ref{eigendistortion} demonstrates the systematic deterioration in the eigenstructure as  $p/n\rightarrow 1$.  The eigenvalues of the true covariance matrix are all identically 1, however this is not reflected in the eigenvalues of the estimated sample covariance matrices.
 
As with the log determinant, the log condition number is not a particularly discerning  performance index. To assess whether a robust estimator provides reasonable estimates, the most that it can contribute is whether or not the log condition number remains bounded.

\subsubsection{Quadratic loss}

Another index that is frequently used in the literature to assess the performance of covariance matrix estimators is the quadratic loss.  The exact specification varies from paper to paper, for example \citet{Ledoit:2004} define it as, $||\hat{ \bm{\Sigma}} -  \bm{\Sigma}||^2_F$.  An alternative specification of the quadratic loss, more in line with the entropy loss, is used in \citet{Won:2013}, $||\hat{ \bm{\Sigma}} \bm{\Sigma}^{-1} - \mathbf{I}||^2_F$.  It is obvious that the quadratic loss is intrinsically linked to the Frobenius norm.

\subsection{Behaviour of performance indices} 
\label{indexbehav}

The performance indices outlined in this section are typically used to compare competing estimators in uncontaminated data sets.  Contaminated data will have potentially severe implications for structure and size of the estimated precision and covariance matrices, and it is not clear how these indices will behave in such settings.  As such, we begin our investigation by exploring how these indices react to the presence of gross outliers in a data set.

The model used to assess the behaviour of the various performance indices is typical of that  used in the simulation study, $n=100$ observations drawn from the standard multivariate Gaussian distribution, $\mathcal{N} (\mathbf{0}, \mathbf{I})$, with $p=30$.  

\begin{figure}[p] 
\centering
\input{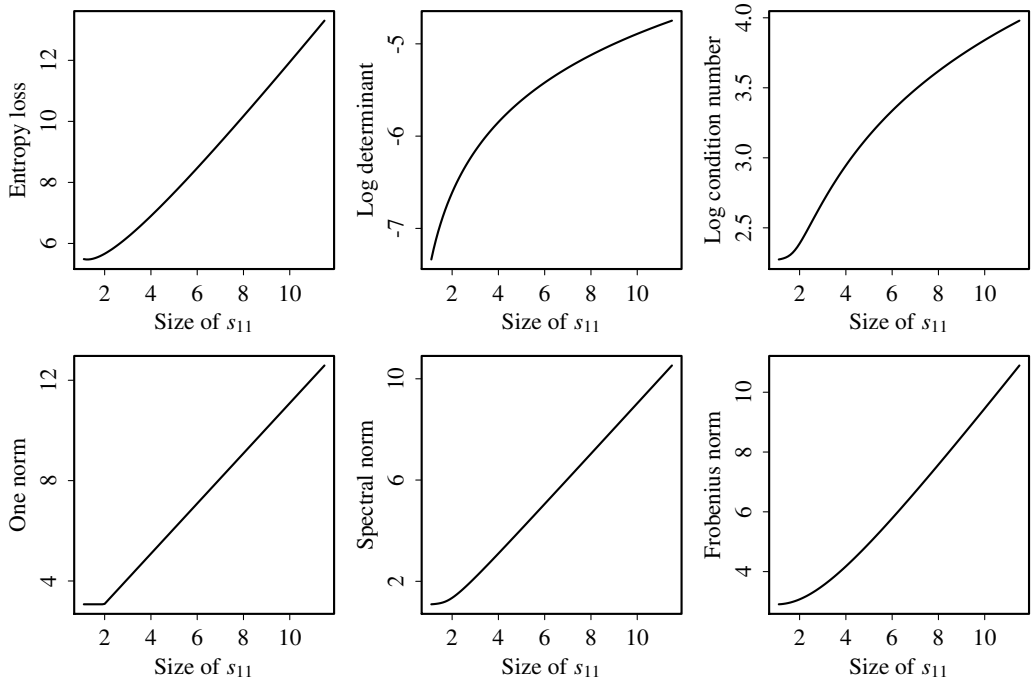}
\caption{The impact of artificially inflating the size of the top left element of the sample covariance matrix, $s_{11}$, on each of the performance indices when applied to the resulting covariance matrix.}
\label{InflateVar}
\end{figure}

\begin{figure}[p]
\centering
\input{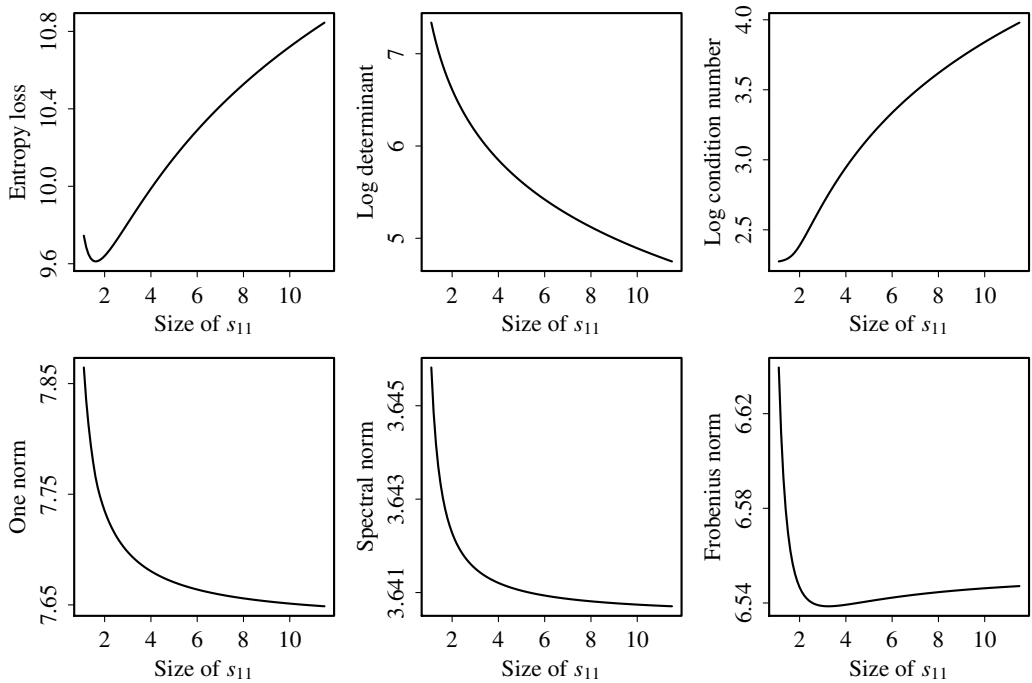}
\caption{The impact of artificially inflating the size of the top left element of the sample covariance matrix, $s_{11}$, on each of the performance indices when applied to the resulting precision matrix.}
\label{InflateVarTheta}
\end{figure}

\subsubsection{Inflated variances}

This section explores how the various performance indices react if we artificially inflate the variance of the first variable, i.e.\ increase the value of $s_{11}$ in the sample covariance matrix, $\mathbf{S}=(s_{ij})$, based on a single sample of uncontaminated data.  In this simple case, where $ \bm{\Sigma} =  \bm{\Theta} = \mathbf{I}$, the matrix norms are applied to $\mathbf{S} -\mathbf{I}$ or $\mathbf{S}^{-1}-\mathbf{I}$ and the entropy loss, log determinant and log condition number are simply applied to $\mathbf{S}$ or $\mathbf{S}^{-1}$.  Note that in this setting the one norm and the infinity norm will give identical results and so only the results for the one norm are shown.

Figure \ref{InflateVar} shows the behaviour of the various indices when applied to these adjusted covariance matrices and Figure \ref{InflateVarTheta} presents the same for the resulting precision matrices, $\hat{ \bm{\Theta}} = \mathbf{S}^{-1}$. The horizontal axis shows the size of $s_{11}$, the artificially inflated variance of the first variable in the sample covariance matrix.  

In Figure \ref{InflateVar}, the majority of the performance indices behave similarly in the covariance case -- there is an overall positive trend as the variance of the first variable increases.  The spectral norm and Frobenius norm both increase uniformly with $s_{11}$.  The one norm, and correspondingly the infinity norm (not shown), remains flat as the sum of the absolute value of the elements in another column (or row) remains larger than the first column (row) up until the point where $s_{11}\approx 2$, at which point the one norm (and infinity norm) increase linearly with $s_{11}$.   For large $s_{11}$, it is clear that all considered performance indices register that the adjusted $\mathbf{S}$ matrix is no longer close to the true value, $\mathbf{I}$.
 
\begin{figure}[tbp]
\centering
\input{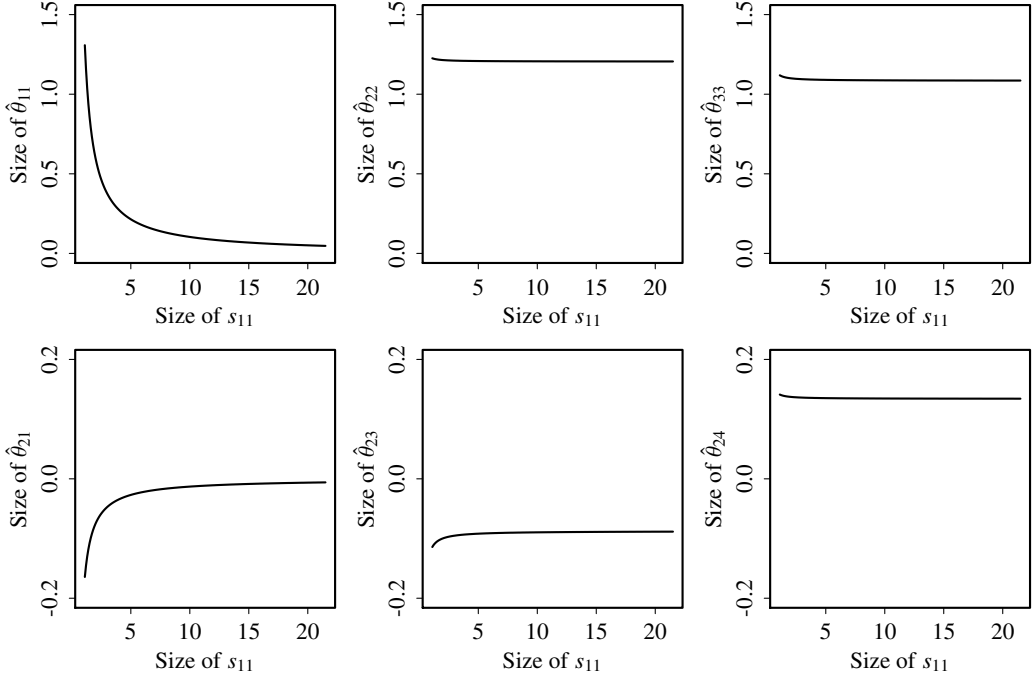}
\caption{The change in few elements of $\hat{ \bm{\Theta}} = {\mathbf{S}}^{-1}$ as $s_{11}$ is artificially inflated.  Main diagonal elements are in the top row and the off diagonal elements are in the bottom row.  Note that ${\mathbf{S}}=(s_{ij})$ and $\hat{ \bm{\Theta}}=(\hat{\theta}_{ij})$.}
\label{InvConvergence}
\end{figure}
 
In stark contrast, Figure \ref{InflateVarTheta} shows the indices when applied to the resulting precision matrix (after the first entry in the covariance matrix has been artificially inflated).  As expected, the condition number of the resulting inverse is identical to that of the original covariance matrix and $L_D(\mathbf{S}^{-1})=-L_D(\mathbf{S})$, however the other indices exhibit somewhat different behaviour.    In particular the matrix norms tend to decrease, rather than increase.  This is explained by noting that as the first element in the covariance matrix is artificially inflated, the first row and column of the precision matrix decay towards zero while the other elements are more or less constant.  This is demonstrated in Figure \ref{InvConvergence} where $\hat{\theta}_{11}$ and $\hat{\theta}_{21}$ both tend towards zero whereas the other main diagonal and off diagonal elements remain quite stable as the level of contamination increases. 

In Figure \ref{InflateVarTheta}, the Frobenius norm, an elementwise norm, exhibits a minimum turning point before levelling off. This is due to the first row and column of the precision matrix converging rapidly to zero, often from relatively large starting points, whereas the convergence of the other elements is not as drastic and not necessarily shrinking towards zero, hence the upward trend.  

The entropy loss broadly exhibits similar behaviour in both Figures \ref{InflateVar} and \ref{InflateVarTheta}.  The minimum turning point in Figure \ref{InflateVarTheta} is somewhat similar to that of the Frobenius norm and is explained by noting that this is due to the sum of the eigenvalues decaying quite quickly before levelling off as $s_{11}$ increases, whereas the sum of the logs of the eigenvalues decays much more slowly. Regardless, it is clear that the entropy loss tends to reflect the impact of the inflated variance in both the covariance matrix and the resulting precision matrix.

\subsubsection{Contamination in the data}

Instead of directly manipulating the estimated covariance matrix, consider introducing contamination into the original data set and observing what effect that has on the performance metrics applied to the covariance and resulting precision matrix.  For each level of contamination we take $N=1000$ samples from $\mathcal{N} (\mathbf{0},\mathbf{I})$.  For each sample we estimate the classical sample covariance matrix, $\mathbf{S}$, and take the inverse to obtain $\hat{ \bm{\Theta}} = {\mathbf{S}}^{-1}$.  

\begin{figure}[p]
\centering
\input{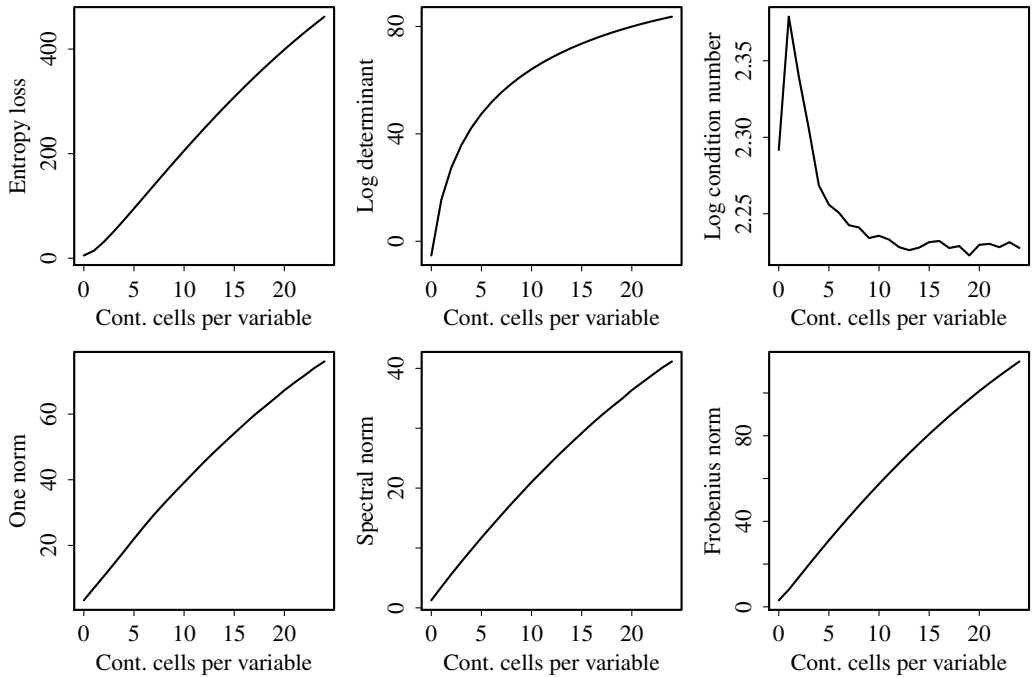}
\caption{The impact of randomly contaminating a certain number of cells in each variable of a $100\times30$ data matrix on the various performance indices when applied to the resulting covariance matrix.}
\label{CovScatteredCont}
\end{figure}

\begin{figure}[p]
\centering
\input{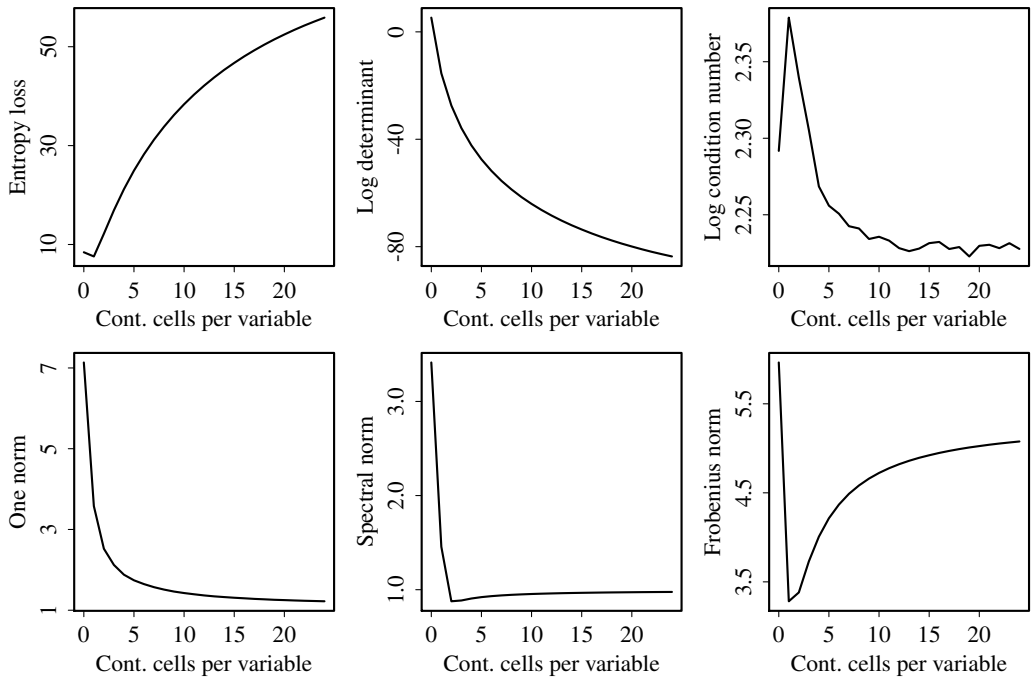}
\caption{The impact of randomly contaminating a certain number of cells in each variable of a $100\times30$ data matrix on the various performance indices when applied to the resulting precision matrix.}
\label{InvScatteredCont}
\end{figure}

Figures \ref{CovScatteredCont} and \ref{InvScatteredCont} show the behaviour of the various loss indices over $N=1000$ replications.  The horizontal axis represents the number of contaminated observations within each of the $p=30$ variables.  Starting with an uncontaminated multivariate Gaussian distribution, we then progressively add one contaminated observation to each variable until there are 24 contaminated observations within each variable.  The contamination is performed by assigning to each randomly selected cell the value of 10.

In general, cellwise outlying contamination will destroy any existing dependence structure and inflate the main diagonal of the covariance matrix, resulting in increases for the entropy loss and matrix norms applied to the covariance as seen in Figure \ref{CovScatteredCont}.  The log determinant of the estimated covariance matrix trends upwards, demonstrating the over estimated generalised variance with increasing levels of contamination.  Apart from a relatively minor spike when there is only one contaminated observation in each variable, the condition number is not adversely affected by increasing levels of contamination, reflecting the stabilised eigenvalues of the resulting covariance matrix.  This demonstrates that in this setting, the condition number is not an appropriate index against which to compare the performance of competing robust estimators.  

The interpretation of the performance indices when they are applied to the resulting precision matrix is more complicated. We see in Figure \ref{InvScatteredCont} that there is still structure present in the precision matrix, in the sense that there is a main diagonal behaving distinctly from the off diagonal elements. However, all the elements tend to shrink towards zero.  Hence, for large amounts of contamination, $\hat{ \bm{\Theta}} - \mathbf{I} \approx -\mathbf{I}$ and so the matrix norms tend to converge to $||-\mathbf{I}||$.  

As in the previous scenario, the Frobenius norm exhibits a minimal turning point before plateauing.  This is explained by noting that while the introduction of contamination has an immediate shrinkage effect on the main diagonal of the precision matrix, including one or two influential observations in each variable induces an artificially high level of correlation between some variables. Hence, it can take some time for the off diagonal elements to stabilise.  When more than a few outlying cells are present in each variable, the artificial correlation structure wanes and hence the Frobenius norm applied to the resulting precision matrix trends towards $||\mathbf{I}||_F = \sqrt{p}$.  Hence, in this contamination setting the matrix norms appear to be useful only when applied to the covariance matrix, not the precision matrix.

As in the previous scenario, the entropy loss behaves consistently for both the covariance and precision matrix.    Similarly to Figure \ref{InflateVarTheta}, it exhibits a slight drop when only one cell in each variable is contaminated after which it increases as the proportion of contaminated cells grows. As such, the entropy loss is the preferred performance index when looking across both covariance and precision matrix estimators.

\end{document}